\documentclass[iop]{emulateapj}

\newcommand{\ts}{\thinspace}
\newcommand{\etal}{\mbox{et~al.}}

\shorttitle{70\ts${\mu}$m Sources: Morphology}
\shortauthors{Kartaltepe \etal}
\slugcomment{Accepted: June 22, 2010} 

\begin{document}

\title{A Multiwavelength Study of a Sample of \lowercase{70\ts${\mu}$m} Selected Galaxies in the COSMOS Field II:  The Role of Mergers in Galaxy Evolution }

\author{Jeyhan S. Kartaltepe\altaffilmark{1,2}, D. B. Sanders\altaffilmark{1}, E. Le Floc'h\altaffilmark{1,4}, 
D. T. Frayer\altaffilmark{3},
H. Aussel\altaffilmark{4},
S. Arnouts\altaffilmark{5},
O. Ilbert\altaffilmark{6},
M. Salvato\altaffilmark{7},
N. Z. Scoville\altaffilmark{7},
J. Surace\altaffilmark{8},
L. Yan\altaffilmark{8} ,
P. Capak\altaffilmark{7},
K. Caputi\altaffilmark{10},
C. M. Carollo\altaffilmark{10},
P. Cassata\altaffilmark{6},
F. Civano\altaffilmark{11},
G. Hasinger\altaffilmark{9}, 
A. M. Koekemoer\altaffilmark{12},
O. Le F{\`e}vre\altaffilmark{6},
S. Lilly\altaffilmark{10},
C. T. Liu\altaffilmark{13},
H. J. McCracken\altaffilmark{14},
E. Schinnerer\altaffilmark{15},
V. Smol{\v c}i{\' c}\altaffilmark{7},
Y. Taniguchi\altaffilmark{16},
D. J. Thompson\altaffilmark{17},
J. Trump\altaffilmark{18},
V. F. Baldassare\altaffilmark{19}, and
S. L. Fiorenza\altaffilmark{13}
}

\altaffiltext{$\star$}{Based on observations with the NASA/ESA {\em Hubble Space Telescope}, obtained at the Space Telescope Science Institute, which is operated by AURA Inc, under NASA contract NAS 5-26555; also based on data collected at : the Subaru Telescope, which is operated by the National Astronomical Observatory of Japan; the XMM-Newton, an ESA science mission with
instruments and contributions directly funded by ESA Member States and NASA; the European Southern Observatory under Large Program 175.A-0839, Chile; the National Radio Astronomy Observatory which is a facility of the National Science Foundation operated under cooperative agreement by Associated Universities, Inc ;  and and the Canada-France-Hawaii Telescope with MegaPrime/MegaCam operated as a joint project by the CFHT Corporation, CEA/DAPNIA, the National Research Council of Canada, the Canadian Astronomy Data Centre, the Centre National de la Recherche Scientifique de France, TERAPIX and the University of Hawaii.}  

\altaffiltext{1}{Institute for Astronomy, 2680 Woodlawn Dr., University of Hawaii, Honolulu, Hawaii, 96822}

\altaffiltext{2}{Current address: National Optical Astronomy Observatory, 950 N. Cherry Ave., Tucson, AZ, 85719, email: jeyhan@noao.edu}

\altaffiltext{3}{Infrared Processing and Analysis Center, California Institute of Technology 100-22, Pasadena, CA 91125, USA }

\altaffiltext{4}{CNRS, AIM-Unit«e Mixte de Recherche CEA-CNRS-Universit«e Paris VII-UMR 7158, F-91191 Gif-sur-Yvette, France. }

\altaffiltext{5}{Canada France Hawaii telescope corporation, 65-1238 Mamalahoa Hwy, Kamuela, Hawaii 96743, USA }

\altaffiltext{6}{Laboratoire d'Astrophysique de Marseille, BP 8, Traverse du Siphon, 13376 Marseille Cedex 12, France}

\altaffiltext{7}{California Institute of Technology, MC 105-24, 1200 East California Boulevard, Pasadena, CA 91125}

\altaffiltext{8}{Spitzer Science Center, California Institute of Technology, Pasadena, CA 91125}

\altaffiltext{9}{Max Planck Institut f\"ur Extraterrestrische Physik,  D-85478 Garching, Germany}

\altaffiltext{10}{Department of Physics, ETH Zurich, CH-8093 Zurich, Switzerland}

\altaffiltext{11}{Harvard-Smithsonian Center for Astrophysics, 60 Garden Street, Cambridge, MA 02138}

\altaffiltext{12}{Space Telescope Science Institute, 3700 San Martin Drive, Baltimore, MD 21218}

\altaffiltext{13}{Astrophysical Observatory, City University of New York, College of Staten Island, 2800 Victory Blvd, Staten Island, NY  10314}

\altaffiltext{14}{Institut d'Astrophysique de Paris, UMR7095 CNRS, Universit\`e Pierre et Marie Curie, 98 bis Boulevard Arago, 75014 Paris, France}

\altaffiltext{15}{Max Planck Institut f\"ur Astronomie, K\"onigstuhl 17, Heidelberg, D-69117, Germany}

\altaffiltext{16}{Physics Department, Graduate School of Science, Ehime University, 2-5 Bunkyou, Matuyama, 790-8577, Japan}

\altaffiltext{17}{Large Binocular Telescope Observatory, University of Arizona, 933 N. Cherry Ave. Tucson, AZ  85721-0065,   USA}

\altaffiltext{18}{Steward Observatory, University of Arizona, 933 North Cherry Avenue, Tucson, AZ 85721}

\altaffiltext{19}{Dept of Physics and Astronomy, City University of New York, Hunter College, 695 Park Ave., New York, NY 10065}

\begin{abstract}

We analyze the morphological properties of a large sample of 1503 70\ts$\mu$m selected galaxies in the COSMOS field spanning the redshift range $0.01<z<3.5$ with a median redshift of 0.5 and an infrared luminosity range of $10^{8} < L_{\rm IR} (8-1000\ts{\mu}\rm m) < 10^{14}\ts L_{\odot}$ with a median luminosity of $10^{11.4}\ts L_{\odot}$. In general these galaxies are massive, with a stellar mass range of $10^{10}-10^{12}\ts M_{\odot}$, and luminous, with $-25<M_{\rm K}<-20$.  We find a strong correlation between the fraction of major mergers and $L_{\rm IR}$, with the fraction at the highest luminosity ($L_{\rm IR}>10^{12}\ts L_{\odot}$) being up to $\sim 50\%$. We also find that the fraction of spirals drops dramatically with $L_{\rm IR}$. Minor mergers likely play a role in boosting the infrared luminosity for sources with low luminosities ($L_{\rm IR}<10^{11.5}\ts L_{\odot}$). The precise fraction of mergers in any given $L_{\rm IR}$ bin varies by redshift due to sources at $z>1$ being difficult to classify and subject to the effects of band pass shifting, therefore, these numbers can only be considered lower limits. At $z<1$, where the morphological classifications are most robust, major mergers clearly dominate the ULIRG population ($\sim 50-80\%$) and are important for the LIRG population ($\sim 25-40\%$). At $z>1$, the fraction of major mergers are lower, but are at least $30-40\%$ for ULIRGs. In a comparison of our visual classifications with several automated classification techniques we find general agreement, however, the fraction of identified mergers is underestimated due to automated classification methods being sensitive to only certain timescales of a major merger. Although the general morphological trends agree with what has been observed for local (U)LIRGs, the fraction of major mergers is slightly lower than seen locally. This is in part due to the difficulty of identifying merger signatures at high redshift. The distribution of the $U-V$ color of the galaxies in our sample peaks in the green valley ($<U-V>=1.1$) with a large spread at bluer and redder colors and with the major mergers peaking more strongly in the green valley than the rest of the morphological classes. We argue that given the number of major gas-rich mergers observed and the relatively short timescale that they would be observable in the (U)LIRG phase that it is plausible for the observed red sequence of massive ellipticals ($<10^{12}\ts M_{\odot}$) to have been formed entirely by gas-rich major mergers.

\end{abstract}

\keywords{cosmology: observations --- galaxies: active  --- galaxies: evolution --- galaxies: high-redshift --- infrared: galaxies --- surveys }

\section{Introduction}

One of the fundamental premises of hierarchical galaxy evolution models is that galaxy mergers are one of the primary drivers in galaxy evolution.  Observations have suggested that mergers between galaxies may account for many of the changes seen over the history of the universe, from morphological transformations to the assembly of mass in galaxies. The merger driven evolutionary sequence initially proposed by \citet{Sanders:1988p1635} suggests that the merger of two gas rich disk galaxies can eventually lead to the formation of an elliptical galaxy after passing through an infrared luminous and QSO phase \citep{Schweizer:1992p4259,Toomre:1972p4262}. This scenario can be tested by observing galaxies at these phases to see if they do indeed represent a transition between disk galaxies and ellipticals.

Studies of luminous and ultraluminous infrared galaxies (LIRGs, $L_{\rm IR}  (8-1000\ts{\mu}\rm m) = 10^{11}-10^{12}\ts L_{\odot}$ and ULIRGs, $L_{\rm IR} = 10^{12}-10^{13}\ts L_{\odot}$) are an excellent way to probe this aspect of the cosmological relevance of galaxy mergers. These objects are rare in the local universe but were once much more common \citep[e.g.,][]{Sanders:1996p1630} and even dominate the cosmic star formation rate at $z>1$  \citep{LeFloch:2005p2544}. Soon after their discovery it quickly became apparent that local ULIRGs are dominated by strong interactions and mergers \citep[e.g.,][]{Kleinmann:1987p6225, Sanders:1987p6296, Sanders:1988p1639}. Subsequent studies of large samples detected by the {\it Infrared Astronomy Satellite} (IRAS) (e.g., the Bright Galaxy Sample (BGS): \citealt{Soifer:1989p2523}; the Revised Bright Galaxy Sample (RBGS): \citealt{Sanders:2003p1575}; 1\ts Jy ULIRG Survey: \citealt{Kim:1998p3280}) revealed that the fraction of interactions and mergers among LIRGs and ULIRGs is strongly correlated with luminosity \citep[e.g.,][]{Kim:1995p3875, Veilleux:2002p920,Ishida:2004p3661,Goto:2005p4302}.

In contrast to the wealth of information known about (U)LIRGs locally, little is known about their high redshift counterparts. It has only recently become possible to detect large samples of distant (U)LIRGs in the mid- and far-infrared with the {\it Spitzer Space Telescope} and many of the morphological studies to date have found contradictory results  \citep[e.g.,][]{Melbourne:2005p1175,Melbourne:2008p944,Lotz:2008p347,Dasyra:2008p1540,Shi:2009p4334}. Most of these samples are based on detections at 24\ts${\mu}$m where the MIPS detector is most sensitive. However, even though many sources are detected at 24\ts${\mu}$m it is difficult to place constraints on the infrared luminosity, particularly at high redshift, with only a detection at this wavelength. The luminosities obtained from 24\ts${\mu}$m detections alone can be off by as much as $0.5-1$\ts dex compared to those obtained with a detection at 70\ts${\mu}$m as well (\citealt{Kartaltepe:2010p6994}; hereafter Paper I). This uncertainty makes it very difficult to look for trends as a function of $L_{\rm IR}$ and many of these samples are potentially dominated by sources at the lower luminosity end thereby biasing the results.

Identifying galaxy mergers at high redshift can also be problematic. High resolution imaging is necessary, either with imaging from space (with the {\it Hubble Space Telescope} (HST)) or with adaptive optics (AO) from the ground, which can be observationally expensive for large samples. The effects of surface brightness dimming and band pass shifting also make it difficult to identify merging signatures, such as tidal tails and surrounding debris, consistently over a wide redshift range \citep{Abraham:1999p4420}. Finally, it is time consuming to morphologically classify large samples of galaxies visually, yet automated classification techniques are only sensitive to certain stages of the merger process \citep[e.g.,][]{Conselice:2003p4555,Conselice:2006p7000,Lotz:2008p4477}) and can suffer from contamination by non-merging systems as the rest-frame wavelength observed approaches the near-UV \citep[e.g.,][]{Jogee:2009p7014}. All of these factors need to be taken into consideration for morphological studies at high redshift.

The large area and wide redshift range observed by the Cosmic Evolution Survey (COSMOS: \citealt{Scoville:2007p1776}) makes it ideal for studying the morphological properties of (U)LIRGs. The COSMOS field has been observed by both HST (ACS F814W images, \citealt{Koekemoer:2007p2299}) and {\it Spitzer} (in all of the IRAC and MIPS bands). The depth of these observations allow us to identify a large sample of objects at 70\ts${\mu}$m over a wide range in infrared luminosity and analyze their morphological properties. This paper is the second in a two part series on a large sample of $\sim$1500 70\ts${\mu}$m selected galaxies in the COSMOS field. The first paper (Paper I) described the selection of our sample and its general properties, then presented the spectral energy distributions (SEDs) and total infrared luminosity ($L_{\rm IR}$) of each source with an average uncertainty of 0.2 dex.

This paper is organized as follows: \S2 briefly introduces the data sets used and \S3 describes our morphological classifications and compares them to those obtained by automated methods. \S4 presents the color-magnitude diagrams for our sample as a function of luminosity and redshift. We discuss our results in \S5 and summarize our findings in \S6. Throughout this paper we assume a  $\Lambda$CDM cosmology with $\rm H_0=70\ts \rm km\ts s^{-1} \ts Mpc^{-1}$, $\Omega_{\Lambda}=0.7$, and $\Omega_{m}=0.3$. All magnitudes are in the AB system unless otherwise stated.

\section{The 70\ts${\mu}$m Sample}

\subsection{The Data Sets}

The 70\ts${\mu}$m sample used in this study was selected from the Spitzer-MIPS coverage of the COSMOS field taken during cycles 2 and 3. The details of the Spitzer IRAC and MIPS data acquisition, reduction, and catalogs can be found in \cite{Sanders:2007p11}, \cite{LeFloch:2009p6953}, and \cite{Frayer:2009p6986}. The main optical imaging data used for this study were obtained as a part of the COSMOS Treasury project with HST using the Advanced Camera for Surveys (ACS) and the F814W filter \citep{Koekemoer:2007p2299}.  The high resolution (FWHM = 0.09\arcsec) is needed for the morphological analysis presented in the next section. Where available ($\sim 5\%$ of the 70\ts$\mu$m sample), the NICMOS images \citep{Scoville:2007p1769} are also used to discuss the effect of band-shifting on the morphology of the sample. Details on these and all of the rest of the multiwavelength data sets are given in Paper I and references therein. 

The full UV--NIR multiwavelength data set was used to derive photometric redshifts, stellar masses, and rest-frame colors using the template fitting code {\it Le Phare}.\footnote{http$://$www.cfht.hawaii.edu/$\sim$arnouts/LEPHARE/cfht$\_$lephare/lephare.html}  See \cite{Ilbert:2009p2146,Ilbert:2010p6983} and \cite{Salvato:2009p2142} for a complete description of these quantities. Spectroscopic redshifts were used where available from z-COSMOS \citep{Lilly:2007p2297}, IMACS (\citealt{Trump:2007p2229}, \citeyear{Trump:2009p5204}), and DEIMOS on {\it Keck II} (Kartaltepe \etal, in prep.).  Photometric redshifts were derived for all galaxies \citep{Ilbert:2009p2146} and X-ray detected AGN \citep{Salvato:2009p2142} in the COSMOS field using fluxes in 30 bands and the median of the probability distribution function determined by {\it Le Phare} using the templates of \cite{Polletta:2007p590} and \cite{Bruzual:2003p6853}. These redshifts have an unprecedented accuracy $dz/(1+z)=0.007$ at $i^{+}<22.5$ (0.02 for AGN) and 0.012 (0.03 for AGN) at $i^{+}<24$ out to $z<1.25$. Spectroscopic redshifts are available for $\sim 40\%$ of the 70\ts$\mu$m selected sample used in this paper. A detailed comparison of the photometric and spectroscopic redshifts for this sample is included in Paper I. The stellar masses used in this paper were derived from the absolute K band magnitude using the analytical relation from \cite{Arnouts:2007p7006} with a \cite{Chabrier:2003p7007} IMF.

\subsection{Luminosities}

In Paper I we presented the full spectral energy distributions from the UV through the far-infrared (FIR) of a large sample of 1503 70\ts${\mu}$m  selected sources from the ACS area of the COSMOS field. We estimated the total infrared luminosity ($L_{\rm IR}$) by fitting a set of infrared templates to the  8, 24, 70 and 160\ts${\mu}$m (where available) data points. These luminosities are presented in Figure 12 and Table 3 of Paper 1 and are included in Table~\ref{catalog} here (column 3) along with their uncertainties. The mean uncertainty on $L_{\rm IR}$ is 0.05 dex for sources with 160\ts${\mu}$m  detections and 0.2 dex for those without. All together, we identified 686 LIRGs, 305 ULIRGs, and 31 HyLIRGs (Hyperluminous Infrared Galaxies, $L_{\rm IR} >10^{13}\ts L_{\odot}$). For details on the sample selection, SED template fitting, and general sample properties, see Paper I.

\section{Morphological Analysis}

In order to understand the structural properties of our 70\ts${\mu}$m selected sample and identify merger signatures we classified each of the 1503 galaxies into several morphological categories described below. Each of the galaxies was classified visually by JSK. The main motivation for choosing the visual inspection method is the advantage of pattern recognition provided by the human eye, particularly for identifying low surface brightness features. Of course visual classification is time consuming and is subject to biases dependent on the individual doing the identification. For this reason, we have also carefully selected a comparison sample of ``normal" galaxies and classified them in the same way. We also compare the results of our visual classifications to the results of several automated classification techniques.

\subsection{Classification}

\begin{figure*}
\epsscale{1}
\plotone{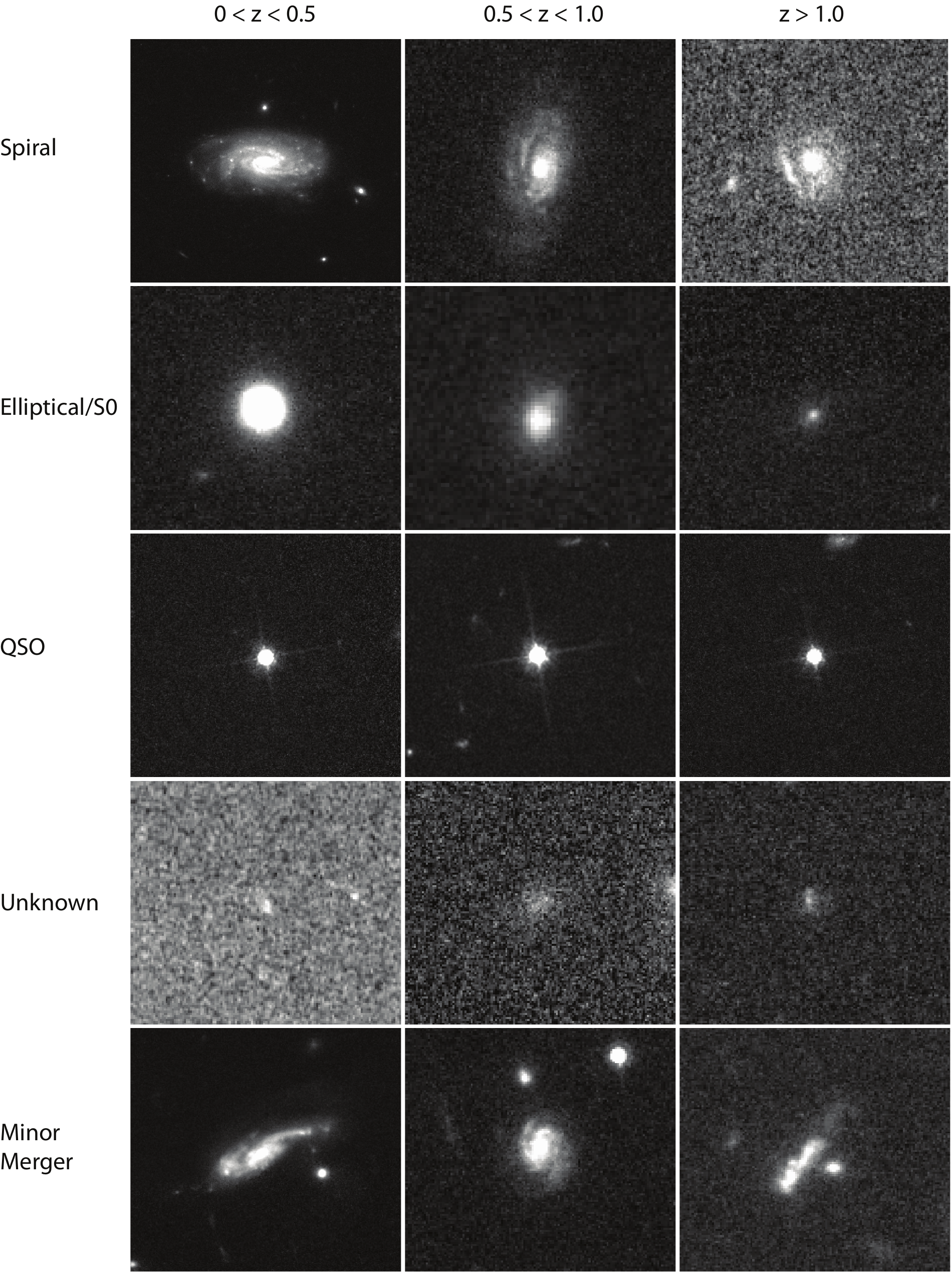}
\caption{Postage stamp images from the HST/ACS mosaic \citep{Koekemoer:2007p2299} of sample galaxies in each of the morphological categories: spiral, elliptical/S0, QSO, unknown, and minor merger. Samples are shown in three different redshift bins to illustrate the changes in resolution and surface brightness with redshift.}
\label{sample1}
\end{figure*}

\begin{figure*}
\epsscale{1.1}
\plotone{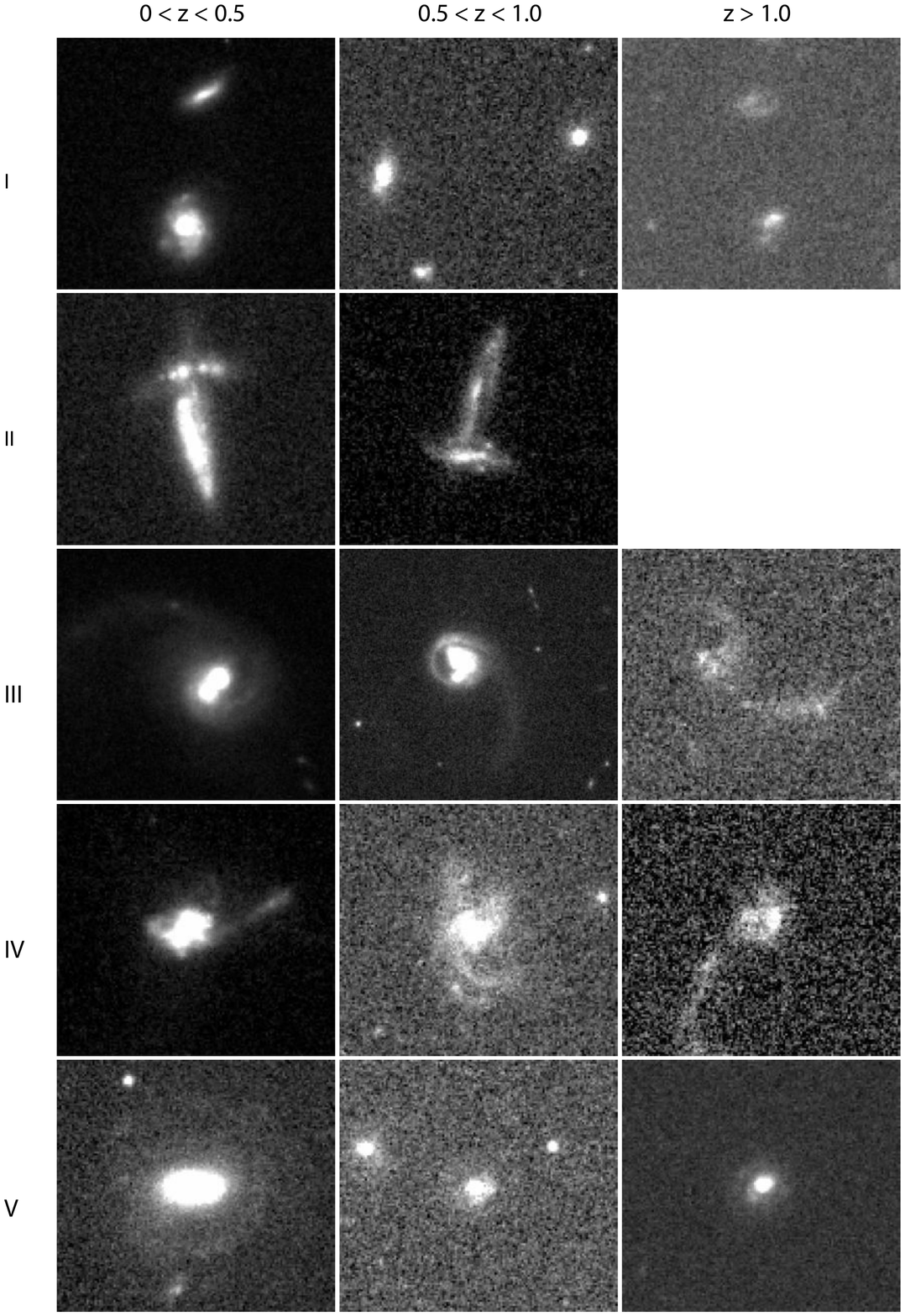}
\caption{Postage stamp images from the HST/ACS mosaic \citep{Koekemoer:2007p2299} of sample major mergers in each different interaction classes, I--V shown in three different redshift bins to illustrate the changes in resolution and surface brightness with redshift.}
\label{sample2}
\end{figure*}

Each of the 1503 70\ts${\mu}$m selected galaxies in our sample were classified into the categories described below. Sample objects in each of these categories (in three different redshift bins) are shown in Figures~\ref{sample1} and \ref{sample2}. Note that some objects can fall into multiple categories. For example, a galaxy can be a spiral and be part of a minor merger. 

{\it Spiral (S)}: Galaxies with a clear spiral/disk structure. This category includes barred and edge on spirals.

{\it Elliptical/S0 (E)}: Galaxies that are spheroidal in appearance and have no obvious sign of a spiral disk. S0 galaxies are included in this category with the usual caveat that it can be difficult to distinguish between spirals and S0s, especially at large inclinations. 

{\it QSO (Q)}: Galaxy with a central point source. In most cases, the QSO overwhelms the host galaxy such that no other information can be discerned about its structure.

{\it Unknown (U)}: Galaxies in this category are too faint or small to classify into any of the other categories or are irregular but not clearly mergers.

{\it Minor Merger (m)}: Galaxies with a slightly disturbed morphology (e.g., warped disks, asymmetric spiral arms, etc.) and no evidence for a large companion (and therefore not likely to be early stage major mergers) or a small companion at the same redshift. 

{\it Major Merger (M)}: Galaxies with a signature of an ongoing or past major merger event. These objects are further grouped into one of several interaction classes based on the classification scheme first presented by \cite{Surace:1998p3674} and \cite{Veilleux:2002p920}. This scheme has its foundation in the interaction stages seen in numerical simulations \citep{Barnes:1992p4505,Mihos:1996p4245} and thus these interaction classes are tied to the merger phase that the galaxy is currently in. However, we note that classifying galaxies at high redshift into these categories can be difficult, particularly due to the difficulty of identifying low surface brightness features such as tidal tails and surrounding debris and thus we expect there to be considerable overlap between these classes, particularly classes II, III, and IV. 

\begin{itemize}
\item {\it First approach (I)}: In this stage the galaxy has a companion at the same redshift ($\Delta z < 0.05$ for sources with photometric redshifts only) but no signs are present to indicate that they are beyond the first passage stage (i.e., no tidal features or disturbed morphologies).

\item {\it First contact (II)}: At this stage, the two galaxies overlap but have not yet formed strong tidal features.

\item {\it Pre-merger (III)}: In these systems, the galaxy is either a member of a pair or a single system with two nuclei. These galaxies have well-developed tidal tails and/or a disturbed morphology. This group is further divided in two based on the nuclear separation of the galaxies, (a) for separations $>10$\ts kpc and (b) for separations $<10$\ts kpc. Some of these pairs are two separate objects in our catalogs and some are a single blended object. Due to lower resolution at high redshift, it is likely that some sources that would be classified as IIIb locally would be IV at high redshift if the two nuclei appear as one.  

\item {\it Merger (IV)}: This stage takes place after the coalescence of the two objects into a single nucleus. These objects still have clear merger signatures such as tidal features and disturbed morphologies. These objects are further divided into two groups depending on the compactness of the galaxy.  \cite{Veilleux:2002p920} use the ratio of the luminosity within 4\ts kpc to the total luminosity to define diffuse mergers (a, ratio $<1/3$) and compact mergers (b, ratio $>1/3$). Here, we use the concentration parameter (the logarithmic ratio of the radii enclosing 80\% and 20\% of the total galaxy flux; see \S 3.4) to separate our sample into diffuse (a, $C<2.7$) and compact (b, $C>2.7$) merger classes.

\item {\it Old merger / merger remnant (V)}: These objects have no tidal features but do still have disturbed central morphologies and possibly knots of star formation and surrounding debris. The lack of tidal features is the primary way to distinguish between class IV and V objects. This category also includes objects with a relaxed core and surrounding debris, indicative of a merger remnant.

\end{itemize}

{\it Groups (G)}: Finally, a small fraction of the galaxies in our sample appear to be a member of a close group of three or more objects. All of these objects are also classified into one of the above interaction classes, but we include this category to note which are part of a larger group.

\begin{figure}
\epsscale{1.1}
\plotone{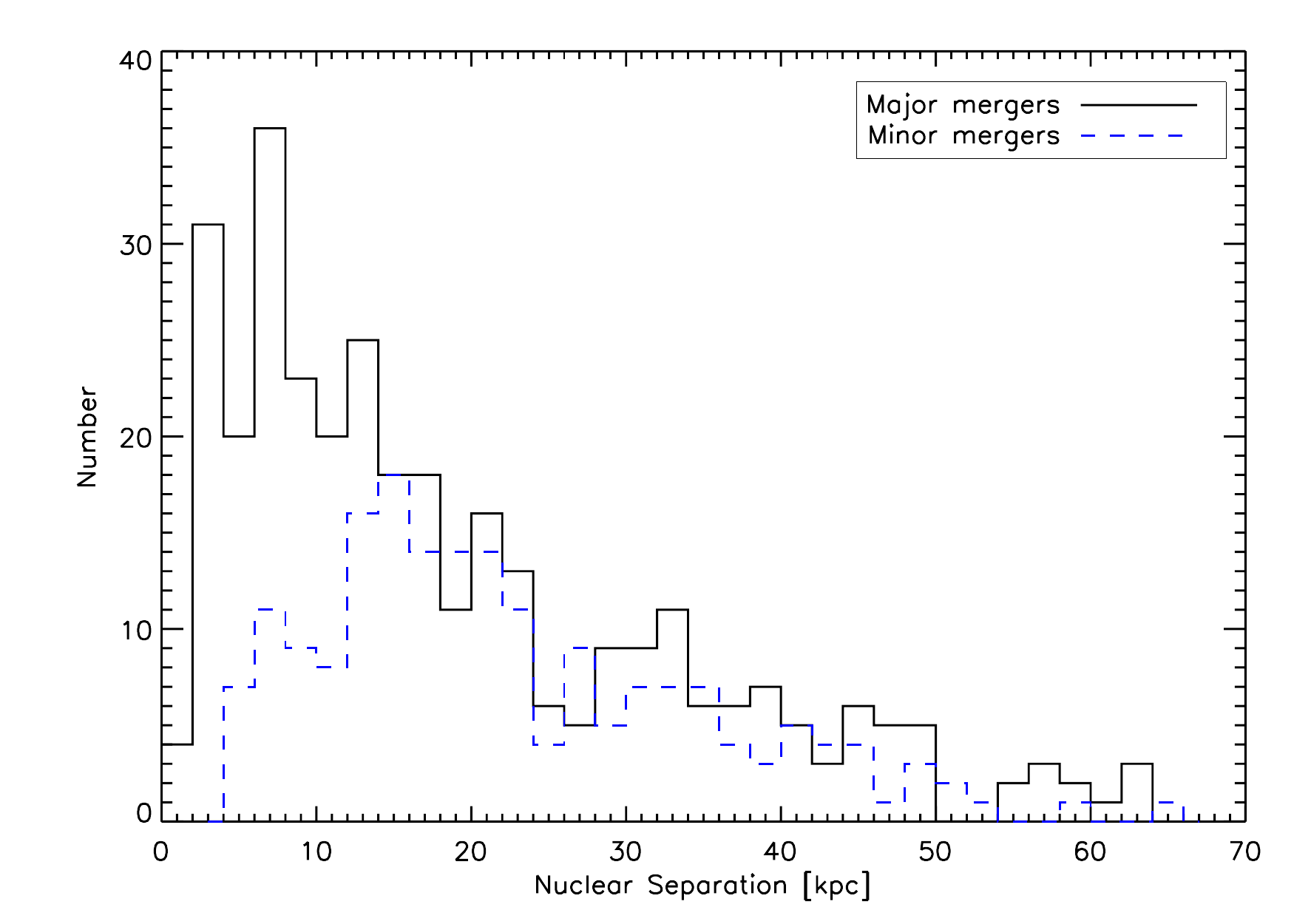}
\plotone{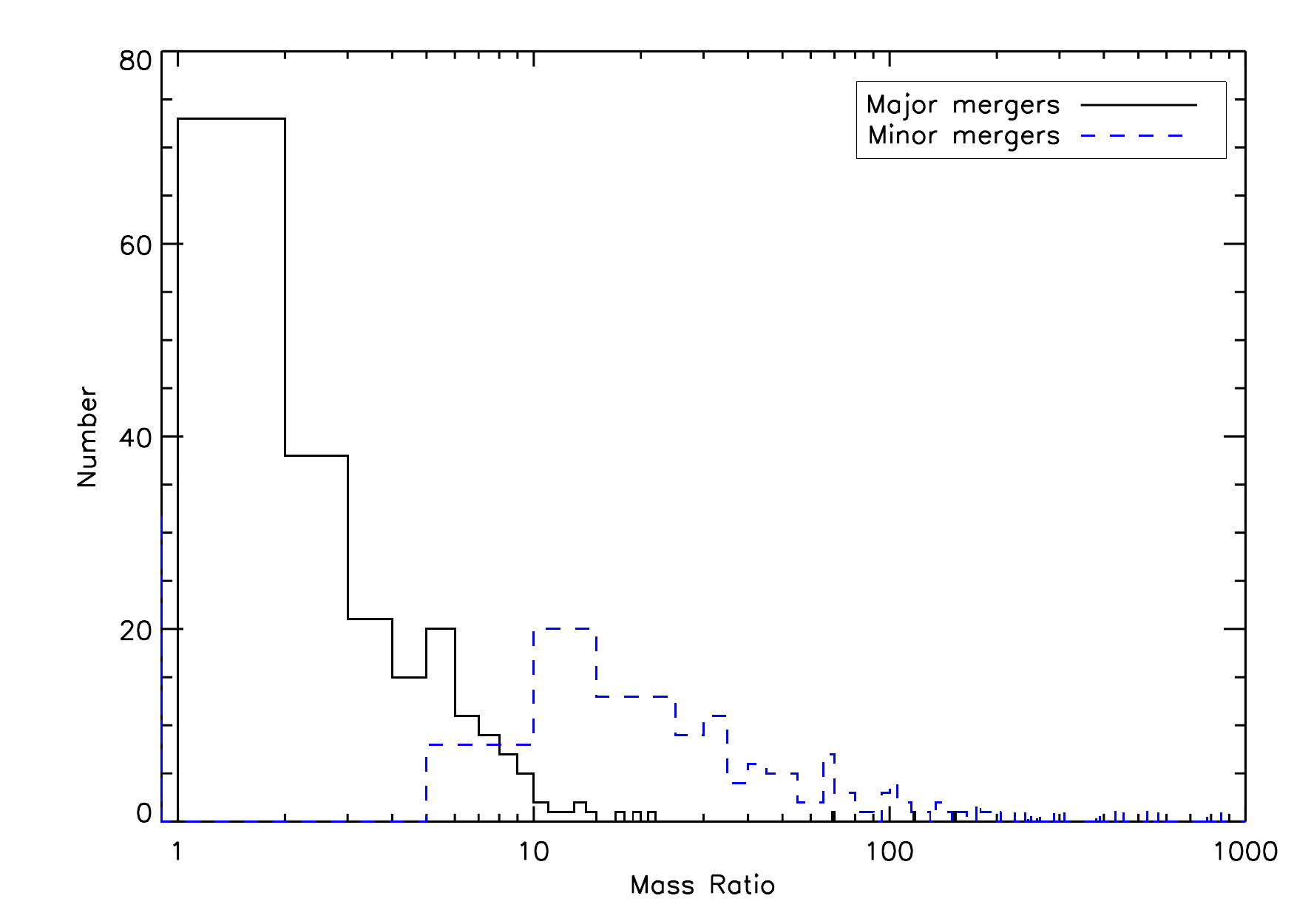}
\caption{Nuclear separation between pairs of galaxies and double nuclei for all major (black solid) and minor (blue dashed) mergers in the 70\ts$\mu$m sample (top) and their stellar mass ratios (bottom). Note that most of the mergers have nuclear separations $< 30$\ts kpc with a tail that goes out to $\sim 70$\ts kpc. Most of the major mergers ($\sim 90\%$) have mass ratios $<5:1$ while the minor mergers have ratios from 5:1 to beyond 100:1. For consistency with the definition typically used in the literature for major and minor mergers, we adopted a strict stellar mass ratio dividing line of $3:1$ and reclassified the 70 major mergers with mass ratio $>3:1$ as minor mergers.}
\label{ns}
\end{figure}

The results of the classification are presented in Table~\ref{catalog}. Column 6 lists the classification(s) assigned to each object in our sample along with their redshifts (column 4) and total infrared luminosity (column 3). The nuclear separations between major and minor merger galaxy pairs and double nuclei are presented in column 7 and plotted in Figure~\ref{ns}. The majority of these (74\%) have nuclear separations less than 30\ts kpc, though there is a significant tail at separations out to $\sim 65$\ts kpc. The separation into the major and minor merger classes was initially done visually, based on the relative appearance of the two objects. The bottom panel of Figure~\ref{ns} shows the stellar mass ratios for each of the pairs for which photometry is available for the galaxies individually. Most of the objects classified as major mergers have small ratios (84\% with a ratio $< 3:1$, 90\% with a ratio $<5:1$). The minor mergers span a larger range, from 5:1 to over 100:1. For consistency with the definition of major/minor mergers commonly used in the literature, we adopt a strict division of mass ratio $3:1$ and reclassify the major mergers with larger mass ratios as minor mergers. This only affects 70 galaxies in our sample.  Note that the stellar masses used in this paper and listed in Table~\ref{catalog} are for the optical counterpart to the 70\ts$\mu$m source. For pairs that are blended in the optical photometry, the mass given is for both objects. For pairs with photometry for both objects, only the mass of the one chosen as the most likely counterpart is given. However, the total mass of the system can be obtained using the given mass ratio.

In order to test the robustness of this classification scheme at high redshift, we have artificially redshifted the 88 (U)LIRGs in the GOALS sample (Great Observatories All-sky LIRG Survey: \citealt{Armus:2009p5172}) with HST/ACS images out to redshifts 0.5, 1, 2, and 3 using the code FERENGI \citep{Barden:2008p2203}. We then classified each of the GOALS sources using the same classification scheme at each redshift and compared the results. We find that 46 of the 88 sources (52\%) retain their classification at all redshifts. For the remaining 42, the fraction of sources that change classification is dependent on redshift. Ten sources (11\%) change at $z=0.5$, 13 sources (15\%) change at $z=1$, and 19 sources (22\%) change at $z=2$.  The sources that change category typically go from one merger classification to another, e.g., from a class IIIb to a class IV. In a few cases, an object classified as a class III becomes a class II as surface brightness dimming causes the tidal features to become too faint to detect. It is worth noting that there are very few objects in our 70\ts$\mu$m sample at $z>2$ that we were able to classify (18 objects). Among these, objects that are pairs or double nuclei are easy to distinguish from the rest, but we note that some of the objects classified as more advanced mergers (i.e., class IV or V) may also have double nuclei below the resolution limit at high redshift. The artificial redshifting test indicates that the classification (including the interaction class) does not change for 74\% of the sources out to $z=1$. 

As an additional test, we chose a randomly selected subset of objects (600 in total) spanning the full range of redshift, luminosity, and morphology to be classified by at least 3 people. These objects were individually classified by JSK, AMK, MS, HA, DJT, JT, VFB, \& SLF in the same manner and using the same classification scheme as the full sample. For galaxies classified as mergers, all three classifiers agree for 51\% of the objects and 2 out of 3 agree for another 33\%. Within mergers, the most common disagreement is whether to classify the merger as class IV or V, as expected due to the difficulty of identifying tidal features. This general agreement is encouraging and the points of disagreement are as expected. For the sample as a whole, all three classifiers agree for 34\% of the objects and 2 out of 3 agree for another 42\%. Most of the discrepancies are between whether or not an object is a spiral or a minor mergers since the slight asymmetries/warps can be difficult to recognize. In some cases, a pair of galaxies classified as a merger by other classifiers could be rejected based on the redshifts of the objects. 

\begin{figure}
\hspace*{-0.2in}
\epsscale{1.25}
\plotone{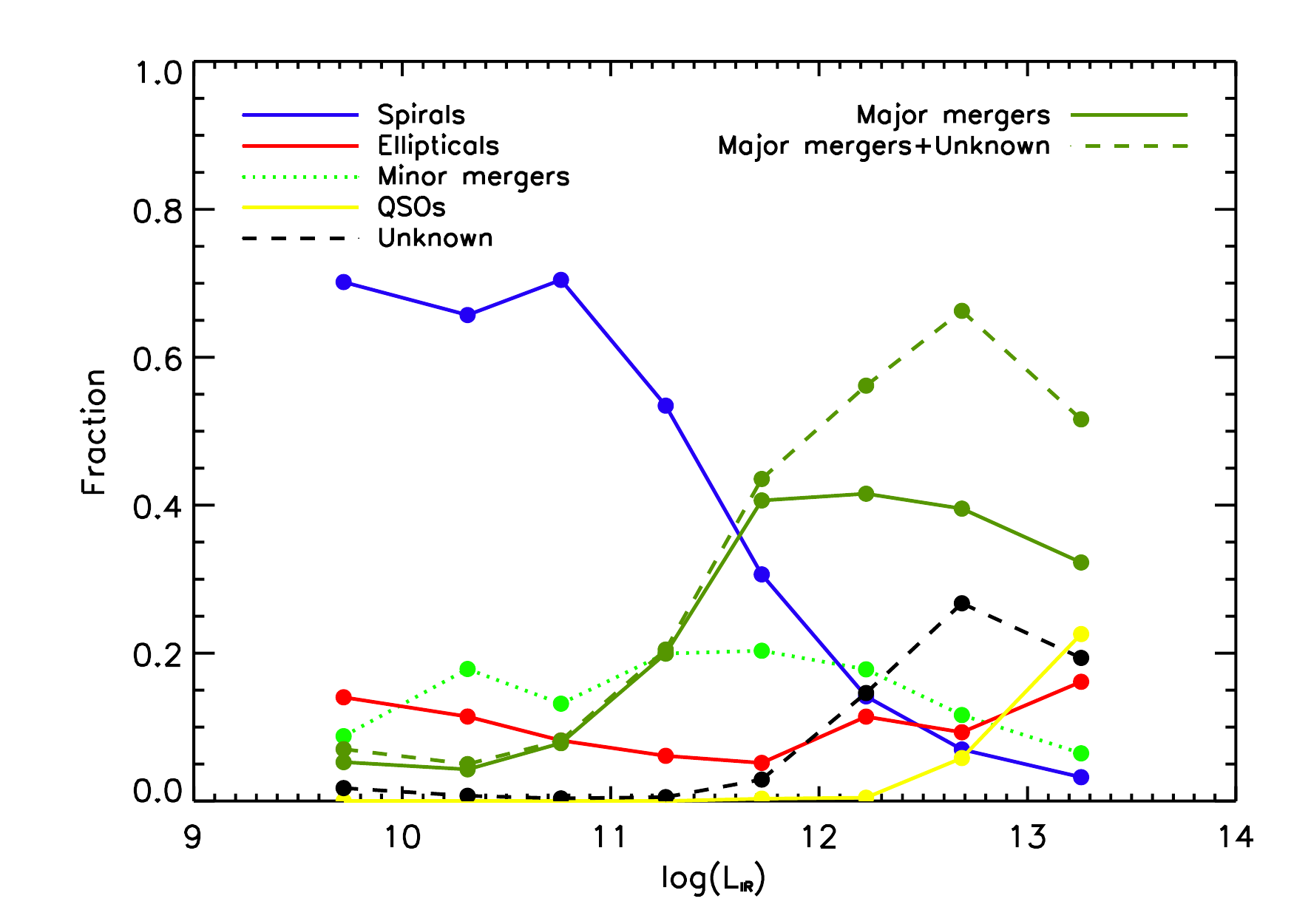}
\caption{Fraction of 70\ts$\mu$m sources that are morphologically classified as spirals (blue solid line), ellipticals (red solid line), minor mergers (green dotted line), major mergers (dark green solid line), QSOs (yellow solid line), unknown (black dashed line), or major mergers+unknown (green dashed line) as a function of $L_{\rm IR}$. Note that the fraction of galaxies classified as spirals drops dramatically with $L_{\rm IR}$ while the fraction of major mergers increases. There is also a larger fraction of galaxies classified as unknown at high $L_{\rm IR}$ which corresponds to high redshift.}
\label{fraction}
\end{figure}

The relationship between galaxy morphology and luminosity is presented in Table~\ref{morphology} for all of the visual classifications and in Table~\ref{mergers} for major mergers broken down into interaction class. For these tables we only include each object in one category based on the dominating classification. For example, if a spiral has a minor companion but no signs of interaction then the object is placed in the spiral category. On the other hand, if the spiral has a disturbed morphology such as a warped disk or asymmetric arm, then it is placed in the minor merger category. For this reason, pairs of galaxies in interaction class I are not included in the major merger category in either table. Instead, objects in this class are included in either the spiral or elliptical category since the interaction has not yet taken place and there is no evidence of disturbed morphology or tidal features. This is an important consideration because some fraction of the objects in class I are chance projections while the objects in the other interaction classes are bona fide mergers as evidenced by the presence of merger signatures such as tidal tails.

The merger fraction in Table~\ref{morphology} for each morphological class are also plotted in Figure~\ref{fraction} as a function of $L_{\rm IR}$. This  clearly shows that the fraction of spirals drops dramatically as $L_{\rm IR}$ increases. Approximately 70\% of sources with $L_{\rm IR} < 10^{10}\ts L_{\odot}$ are spirals while less than 10\% of objects above $10^{12.5}\ts L_{\odot}$ are.  The fraction of E/S0 galaxies among 70\ts$\mu$m sources does not appear to change very much. Approximately 15\% of low luminosity sources are E/S0 while $\sim 10\%$ of sources above $10^{10}\ts L_{\odot}$ are. This fraction remains roughly constant with possibly a slight increase for HyLIRGs (16\%). The fraction of minor mergers also decreases slightly with luminosity but this is partly due to the fact that minor companions and small distortions become difficult to identify at higher redshifts.

Consistent with results in the local universe, the fraction of objects that are major mergers increases with $L_{\rm IR}$. Only 5\% of sources with $L_{\rm IR} < 10^{10}\ts L_{\odot}$ are major mergers. This increases slightly (to $\sim 8\%$) for sources with $10^{10.5} < L_{\rm IR} < 10^{11}\ts L_{\odot}$ to 20\% for low luminosity LIRGs ($10^{11} < L_{\rm IR} < 10^{11.5}\ts L_{\odot}$) to $\sim 40\%$ for high luminosity LIRGs and ULIRGs. The fraction of major mergers drops to 32\% for HyLIRGs. However, these fractions can be seen as lower limits since the fraction of galaxies classified as ``unknown" also increases strongly with luminosity. This is most likely due to the fact that the high luminosity sources are also at higher redshift making them more difficult to classify morphologically. Nearly 30\% of sources fall into this category at the high luminosity end. Additionally, the fraction of objects classified as major mergers in this study should be considered a lower limit since not all mergers are expected to produce prominent merger signatures, such as tidal tails, that would be visible at high redshift (i.e., prograde versus retrograde mergers). It is possible that deeper imaging of objects classified as spirals, ellipticals, or unknown would reveal such signatures. The fraction of QSOs also increases with $L_{\rm IR}$ -- there are none with $L_{\rm IR} < 10^{11.5}\ts L_{\odot}$ and  the fraction rises in the three subsequent bins with $\sim$20\% of HyLIRGs falling into this category. However, this category only includes objects morphologically classified as a point source and is not indicative of all AGN in the sample as will be further discussed in \S5.

The fraction of major mergers in each interaction class also correlates with $L_{\rm IR}$. Most notable is that the majority of class I and II objects have low luminosities ($<10^{11}\ts L_{\odot}$) while most objects at the high luminosity end are class III and IV. The fraction of class V objects remains fairly constant for all luminosities, though it is difficult to determine if this is a real effect or simply due to class V objects being difficult to identify at high redshift.

\begin{figure*}[t]
\plotone{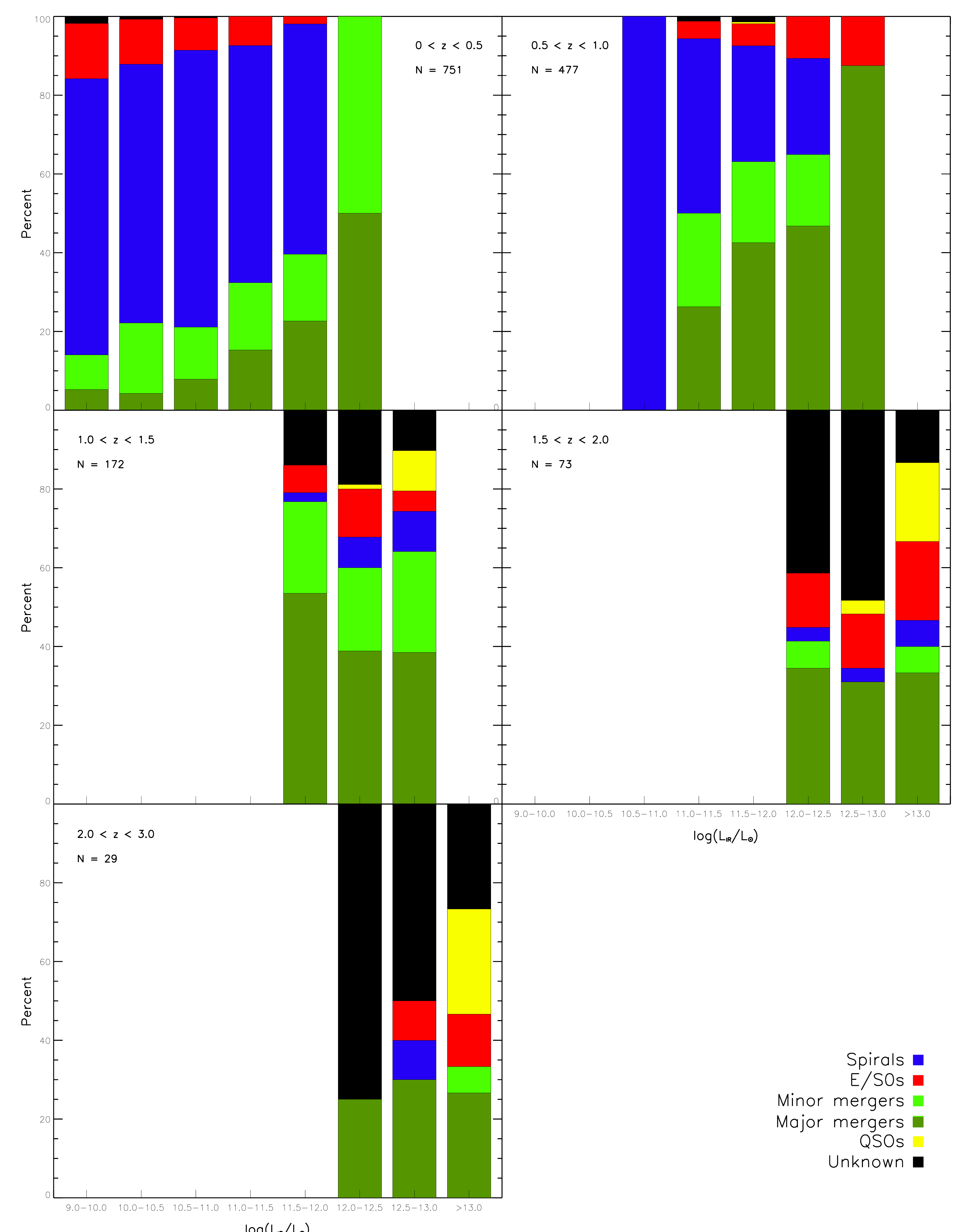}
\caption{Histograms color coded by morphology for the 70\ts$\mu$m selected sample as a function of $L_{\rm IR}$ divided into several different redshift bins. The detailed numbers for each bin are shown in Table 4.}
\label{bars}
\end{figure*}

In order to disentangle trends with redshift from trends with luminosity, we have divided up the sample into several redshift and luminosity bins, as shown in Table 4 and illustrated in Figure 5. We find that out to $z\sim 1$, where our morphological classifications are most robust, the trend of morphology with luminosity is quite strong. In the $0.5<z<1.0$ redshift bin, 26\% of sources with $L_{\rm IR} = 10^{11.0}-10^{11.5}\ts L_{\odot}$ have signatures of major mergers while 44\% are spirals and 24\% are minor mergers. For sources with  $L_{\rm IR} = 10^{11.5}-10^{12.0}\ts L_{\odot}$, 43\% are major mergers, 21\% are minor mergers, and 29\% are spirals. At ULIRG luminosities of  $L_{\rm IR} = 10^{12.0}-10^{12.5}\ts L_{\odot}$, 47\% are major mergers, 21\% are minor mergers, and 24\% are spirals. For sources with extreme luminosities of  $L_{\rm IR} = 10^{12.5}-10^{13.0}\ts L_{\odot}$, {\it 88\% are major mergers, while none are minor mergers or spirals}. At higher redshifts ($z>1$) the fraction of sources classified as unknown increases dramatically, and the total number of sources at these redshifts is small, so it is unclear from this dataset whether these fractions would remain the same out to $z\sim 2$ or 3. However, even though these numbers are lower limits, $\sim 30\%$ to 40\% of ULIRGs and HyLIRGs out to $z\sim 3$ have clear signatures of major mergers. Additionally, $>20\%$ of HyLIRGs are QSOs.

\subsection{Comparison Sample}

\begin{figure}
\hspace*{-0.2in}
\epsscale{1.2}
\plotone{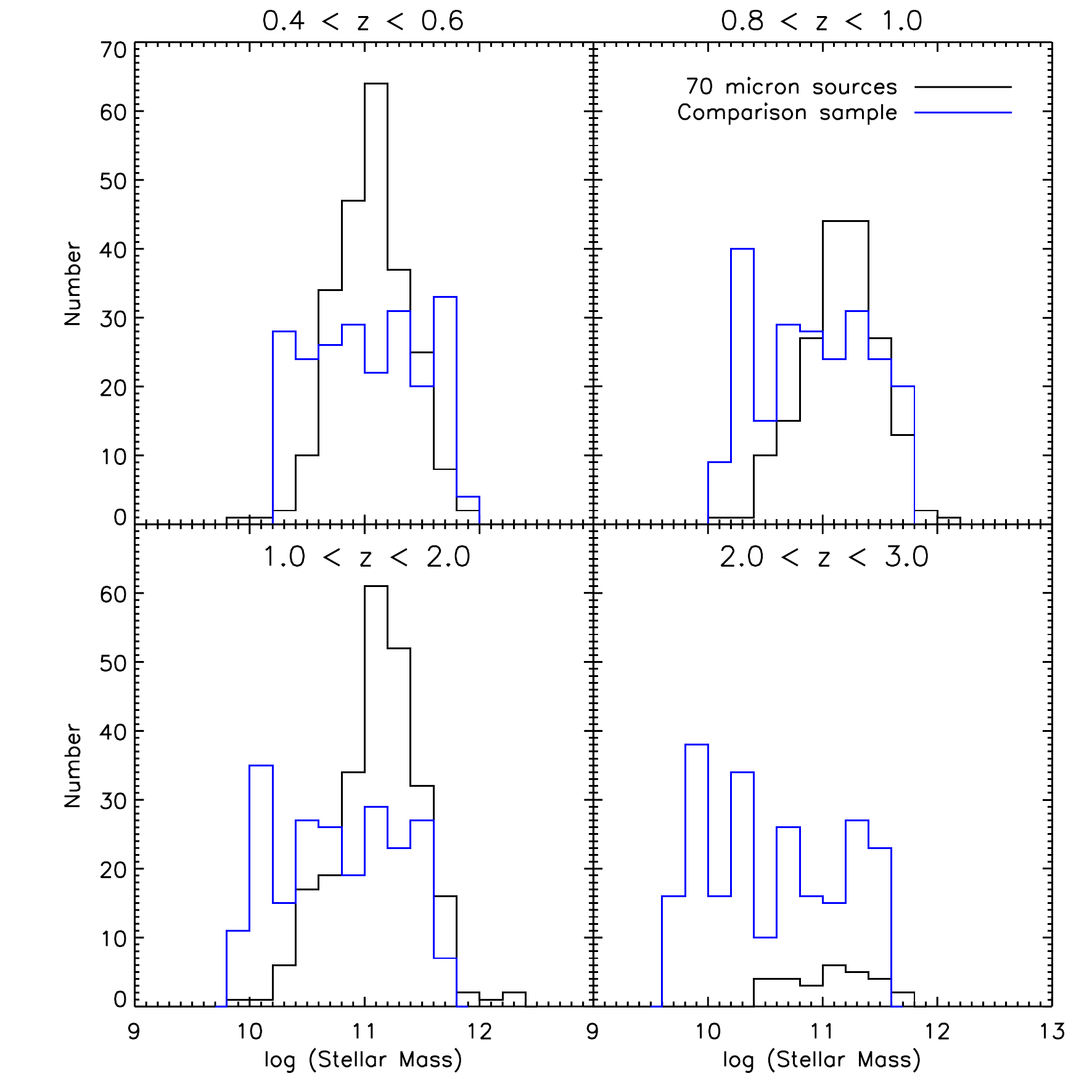}
\caption{Distribution of stellar masses for the 70\ts$\mu$m selected sample (black) and the optically selected comparison sample (blue). The comparison sample is limited to the same mass range as the 70\ts$\mu$m sample to allow for a more robust comparison of similar mass galaxies.}
\label{compare_mass}
\end{figure}

Given the subjective nature of the visual classifications applied to our 70\ts$\mu$m selected sample we have also classified a sample of optically selected galaxies for comparison. We chose four different redshift bins for the comparison ($0.4<z<0.6$, $0.8<z<1.0$, $1.0<z<2.0$, and $2.0<z<3.0$) and randomly selected $\sim 200$ (approximately the same number of 70\ts$\mu$m selected galaxies in each of these bins) within the ACS coverage of the COSMOS field. We also limited the optical sample to the same range of stellar masses as the 70\ts$\mu$m sample ($10< {\rm log}(M/M_{\odot})<12$) in order to avoid any systematic differences between the two samples. The stellar mass distribution of both the 70\ts$\mu$m and optically selected comparison samples is plotted in Figure~\ref{compare_mass} for each of these four bins.

\begin{figure}
\hspace*{-0.2in}
\epsscale{1.2}
\plotone{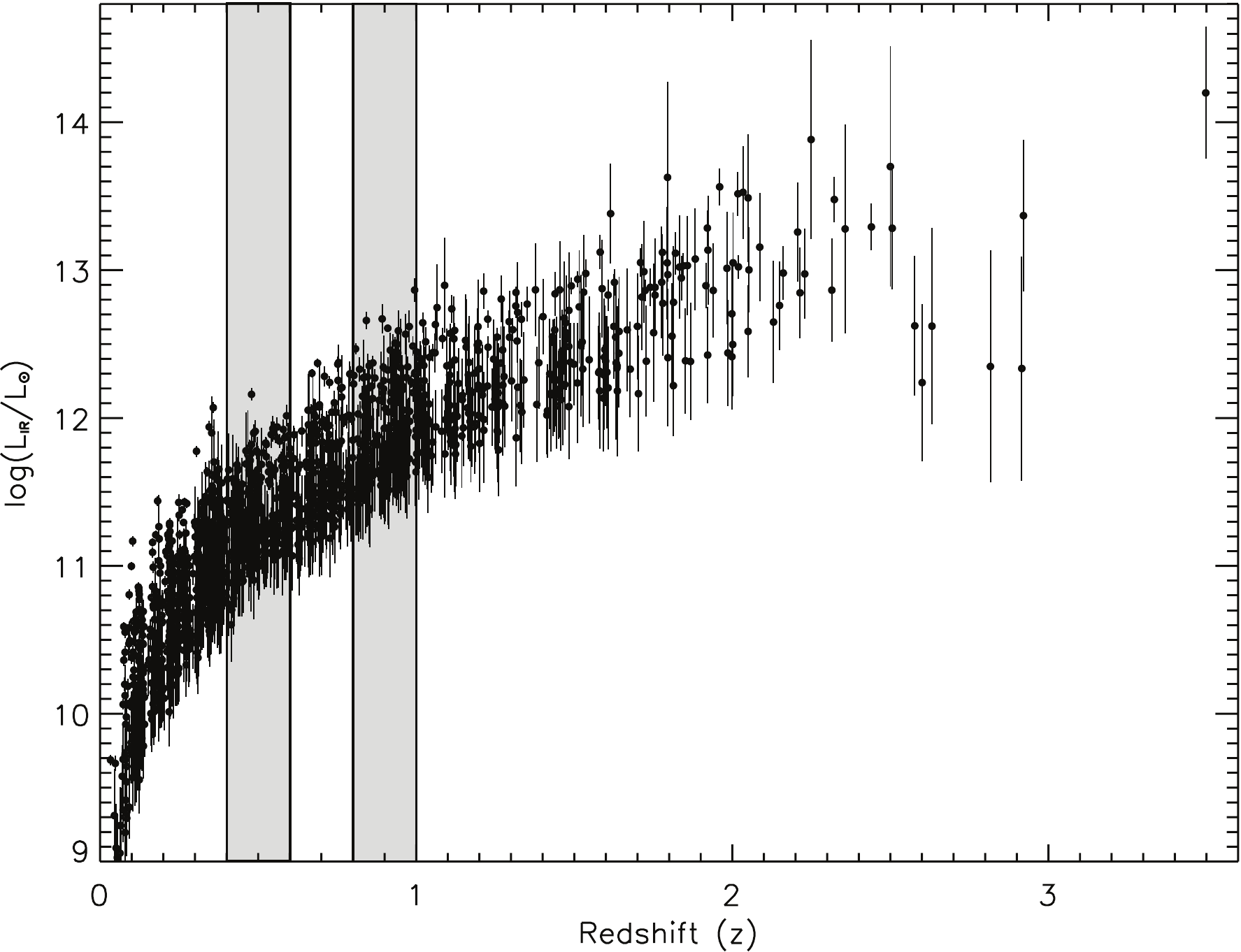}
\epsscale{1.25}
\hspace*{-0.2in}
\plotone{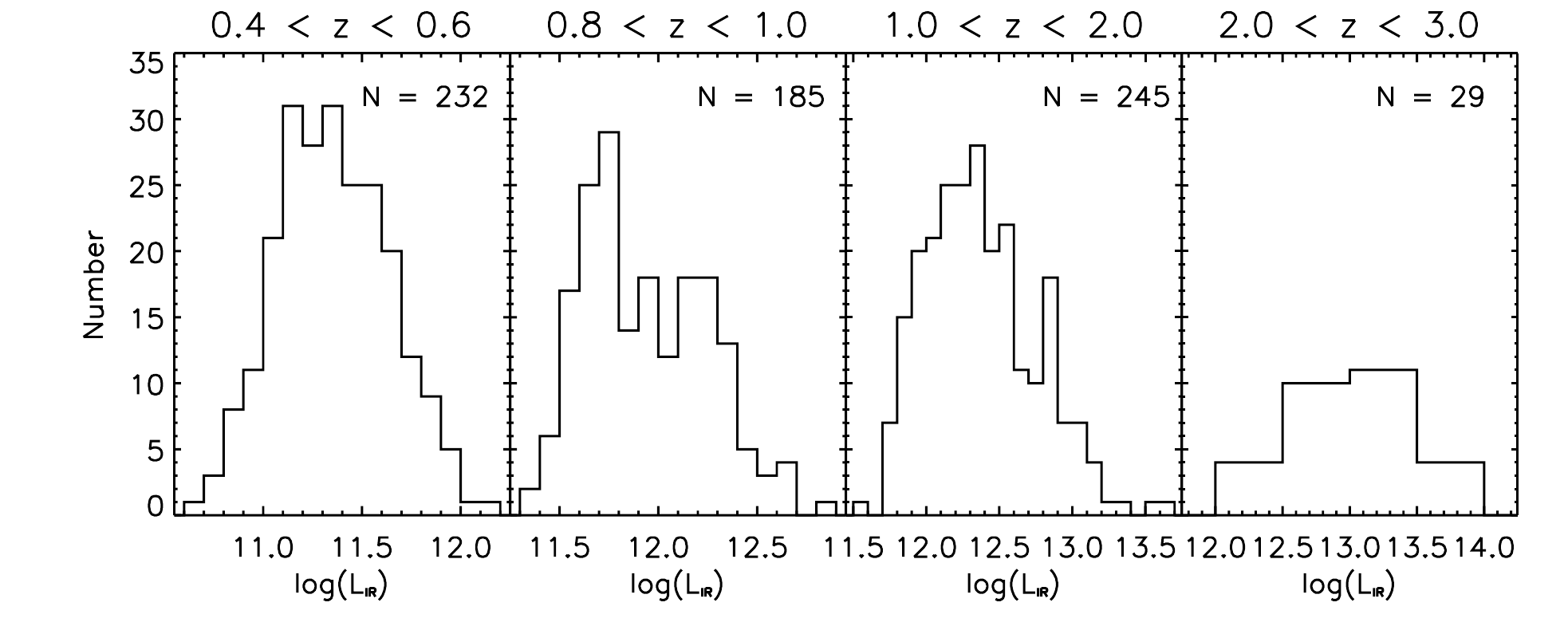}
\caption{Top: $L_{\rm IR}$ as a function of redshift for the entire 70\ts${\mu}$m selected sample with the first two redshift bins used for the comparison sample highlighted. Bottom: $L_{\rm IR}$ distribution of the 70\ts$\mu$m selected sample in all four redshift bins used for the comparison sample. Note that the four redshift bins probe slightly different ranges in $L_{\rm IR}$.}
\label{lir}
\end{figure}

We then classified each of these galaxies into the same morphological categories as described in the previous section. The results of the comparison are summarized in Table~\ref{comparison}. In each of the redshift bins, the 70\ts$\mu$m selected sample contains significantly more mergers and fewer spiral and elliptical galaxies than the optically selected sample. This indicates that the high fraction of galaxy mergers seen in the 70\ts$\mu$m sample is not simply a result of a biased classification (i.e., looking for merger signatures in the first place). It also helps to disentangle trends with redshift from trends with luminosity. For example, the fraction of minor mergers decreases with redshift between each of the two comparison bins. This is likely due to small companions and slightly disturbed morphologies being more difficult to see at high redshift.  Note that as expected, the fraction of objects classified as unknown increases with redshift for both the 70\ts$\mu$m sample and the comparison sample.

As a test of the classification of our comparison sample, we compared our results in the $0.4<z<0.6$ redshift bin to the results of \cite{Lotz:2008p347} and \cite{Jogee:2009p7014} over similar mass and redshift ranges. \cite{Lotz:2008p347} find that $13\pm3\%$ of their optically selected luminous ($L_B>0.4\ts L_B^*$) galaxies are classified as merger candidates using the Gini-M$_{20}$ technique (described in \S 3.5). Similarly, \cite{Jogee:2009p7014} find that $8-9\%$ of galaxies with stellar masses $M>2.5\times 10^{10}\ts M_{\odot}$ are visually classified as mergers at $0.24<z<0.8$. In our optical comparison sample, we find that $13\%$ of the 220 objects at $0.4<z<0.6$ are classified as ongoing major mergers (interaction classes II--V). Given the small numbers of galaxies in the comparison sample, our results are generally consistent with both of these two studies at low redshift.

\subsection{Effects of Band-Shifting}

The ACS images used for our morphological analysis were all taken using the F814W (I-band) broad-band filter. At higher redshifts, this filter traces shorter rest-wavelengths and by $z>1$ the images are really probing the rest-frame UV emission. The necessity of the high-resolution HST images makes it difficult to study and account for the morphological $k$-corrections necessary as a result. In order to test for the effect this band-shifting has on our morphological classifications, we compared our classifications obtained in the optical with those obtained in the near-infrared using the small sample of objects covered by the NICMOS parallel observations. 

\begin{figure}
\hspace*{-0.2in}
\epsscale{1.2}
\plotone{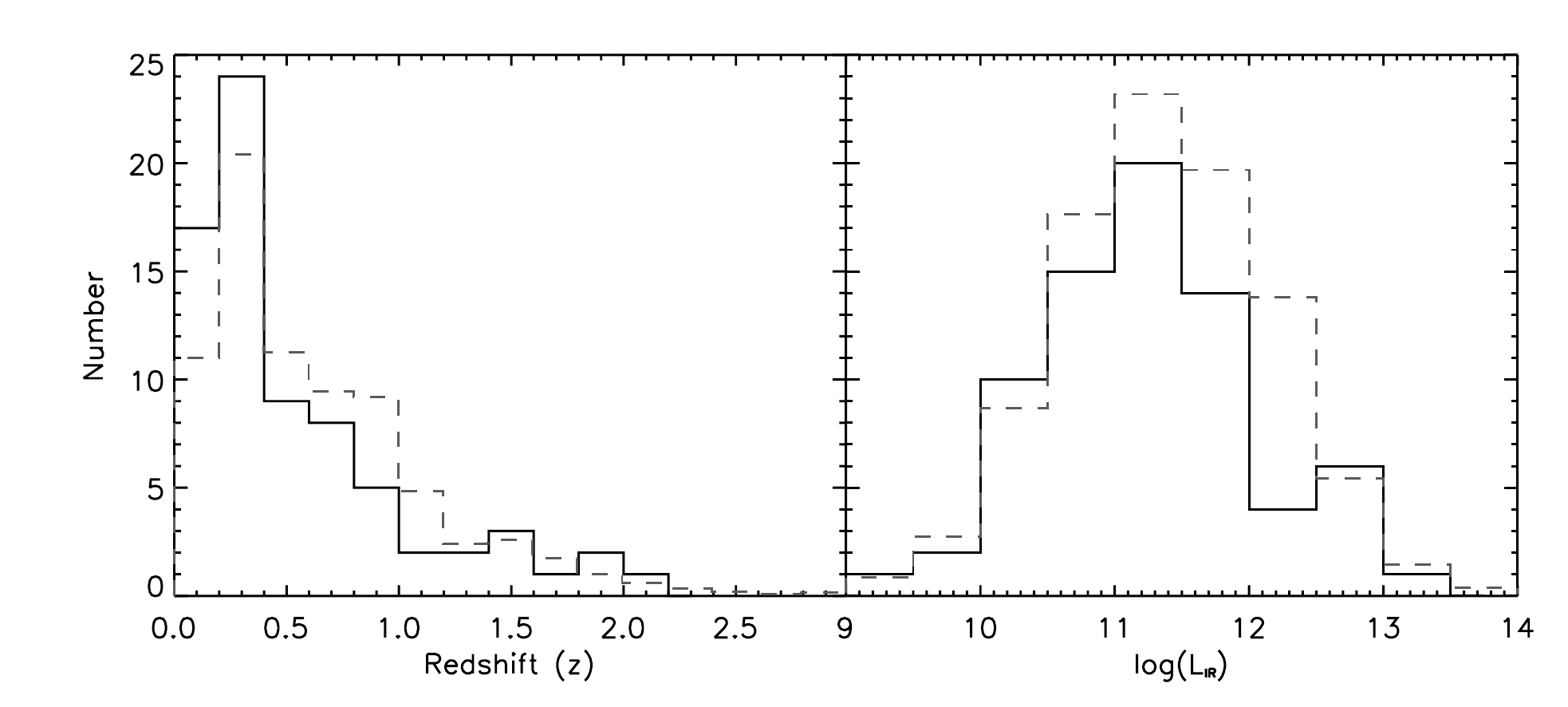}
\caption{Redshift and infrared luminosity distributions for the 74 70\ts$\mu$m sources with HST-NICMOS imaging in the H band (black solid line, representing $\sim 5\%$ of the full sample) with the normalized distributions of the full 70\ts$\mu$m sample over plotted (gray dashed line). The sources span the full redshift and luminosity range of the full sample, but there are very few beyond $z=0.8$ (16 sources) where the near-infrared imaging covers the rest-frame optical.}
\label{nicmos}
\end{figure}

In total 74 of our 70\ts$\mu$m sources have H-band imaging from NICMOS available. The redshift and luminosity distributions for this sub-sample is shown in Figure~\ref{nicmos}. While this sub-sample spans the full range of redshifts and luminosities, very few (11) are at high redshift ($z>1$). We then classified these 74 objects into the same morphological categories used for the ACS images. We find that the classifications agree for all but 14 (19\%) of our sample. For those that disagree, 9 ($<z>=1.04$) are simply too faint or small in the NICMOS images to classify, one object ($z=0.24$) classified as a spiral in the optical was classified as an S0 in the near-infrared, one object ($z=0.22$) classified as an old merger in the optical was classified as an elliptical in the near-infrared, two objects ($z=1.44$ and $z=2.00$) that were were classified as unknown in the optical due to their faintness were classified as a merger and an old merger in the near-infrared, and one object ($z=0.66$) that was classified as a pre-merger (class III) in the optical was classified as a pair on first approach (class I) in the near-infrared due to the low resolution of the NICMOS image. Though there are too few objects in this sub-sample to draw statistical conclusions, the overall agreement is encouraging, particularly since those objects that changed classes when looking at the near-infrared tend to change from one merger class to another or were simply too faint or of insufficient resolution. 

It is also worth noting that any bias introduced in the morphological classifications due to band-shifting will affect the comparison sample as well so we would expect to see any resulting trends in both samples.

\subsection{Automated Morphological Parameters}

One of the difficulties of studying the morphologies of objects from large surveys is that it can be time consuming and cumbersome to classify large samples by eye. Automated classification techniques have many benefits over visual classification; they are faster, less subjective, and more likely to provide consistent results. They also offer a way of studying information about galaxy structure beyond classification into the simple Hubble sequence. This is particularly important at high redshift where galaxies tend not to conform to Hubble type morphologies.

\begin{figure}
\epsscale{1.2}
\plotone{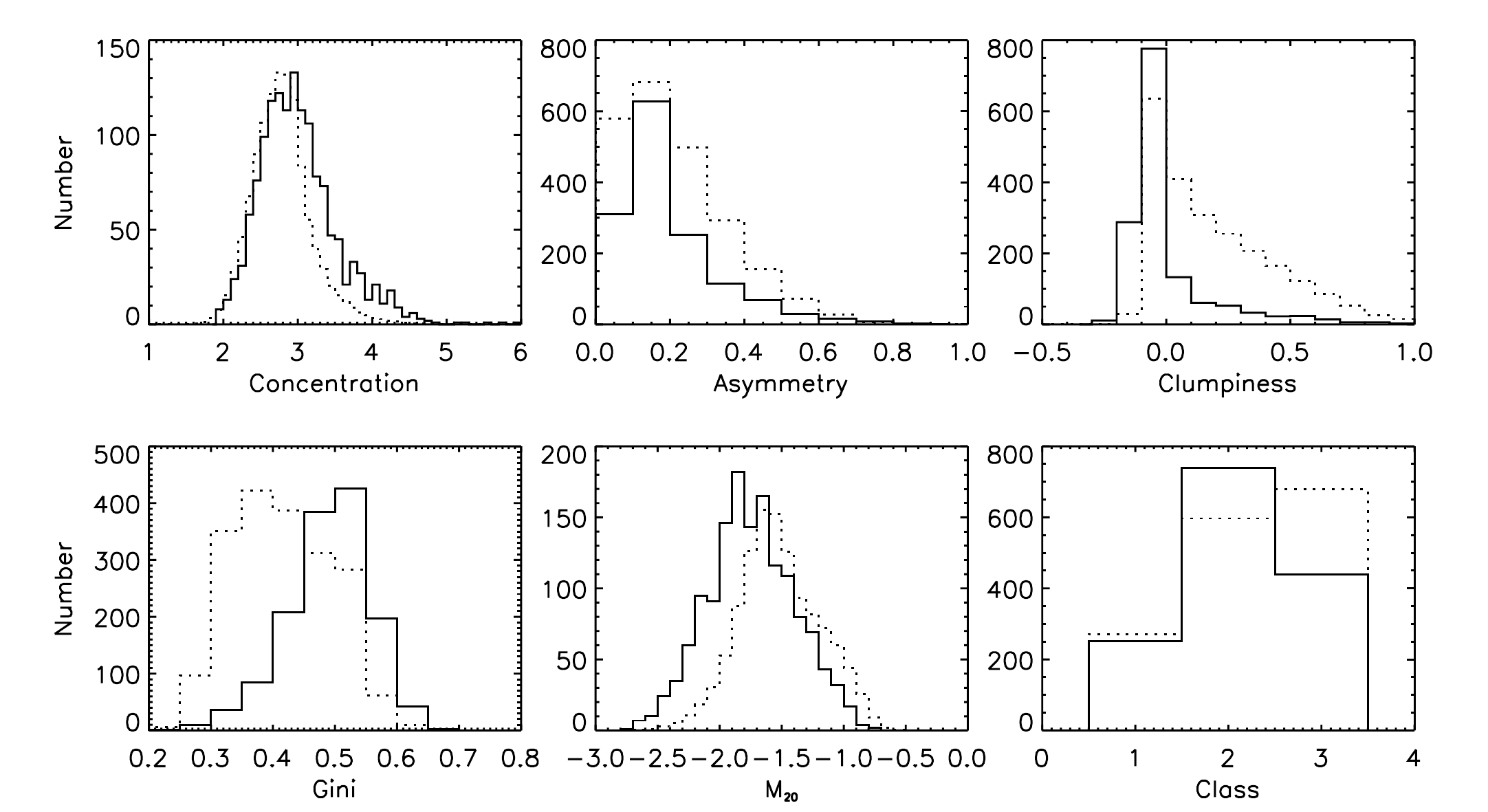}
\caption{Distribution of the five automated morphological parameters (concentration, asymmetry, clumpiness, gini, and M$_{20}$) as well as the Cassata \etal (2010) classification for the 70\ts$\mu$m sample. Over plotted (dotted line) is the normalized distribution for the entire optically selected ($I<23.8$) morphological sample. Note the differences in clumpiness, Gini, M$_{20}$, and automated classification between the full optically selected sample and the 70\ts$\mu$m selected sample. }
\label{auto}
\end{figure}

We use several quantitative morphological measurements of our sample of 70\ts$\mu$m selected galaxies and compare the results with our visual classifications. We use the catalog of Cassata \etal (2010, in preparation) which includes the five parameters discussed here (concentration, asymmetry, clumpiness, Gini, and M$_{20}$) measured using a ``quasi-Petrosian" technique. Cassata \etal also used the entire multi-dimensional parameter space to automatically assign each galaxy to elliptical/S0, spiral, or irregular classes. 1433 objects out of our sample of 1503 (95\%) 70\ts$\mu$m sources have a high enough $S/N$ (i.e., $I<23.8$) to reliably measure each of these parameters.  We show the distribution of all six of these quantities in Figure~\ref{auto} along with the normalized distribution for the entire optically selected galaxy population (with $I<23.8$). 

The concentration index \citep[$C$:][]{Kent:1985p4522,Abraham:1994p4473,Abraham:1996p4462,Bershady:2000p4579} is a measure of the logarithmic ratio of the radii enclosing 80\% and 20\% of the total galaxy flux. The values range from $\sim 2-5$ where a galaxy with a value of $ C=2.7$ has an exponential disk and a galaxy with $C=5.2$ has an $r^{1/4}$ law profile \citep{Bershady:2000p4579}. Our 70\ts$\mu$m sample has $<C>=2.95$, slightly higher than the full optically selected sample ($<C>=2.77$). The median value and standard deviation for all of the morphological parameters are listed in Table~\ref{auto_table} as a function of $L_{\rm IR}$. Surprisingly, the median concentration parameter is constant for all ranges in $L_{\rm IR}$. 

The rotational asymmetry parameter \citep[$A$:][]{Abraham:1996p4462,Conselice:2000p4573} is measured by subtracting a galaxy's image rotated by 180 degrees about its center from itself. \cite{Conselice:2003p4555} defines galaxies with $A>0.35$ as being major mergers but notes that in the local universe mergers span a wide range in asymmetry and thus this parameter is only sensitive to certain phases of the merging process. Our sample has $<A>=0.15$, similar to that of the full optical sample, and 172 galaxies (12\%) have $A>0.35$.  The median value of $A$ increases with $L_{\rm IR}$ from 0.11 at $<10^{10} L_{\odot}$ to 0.25 for ULIRGs. It should also be noted that at high redshifts the value of $A$ can decrease by $0.05-0.15$ due to the decreased resolution and the effects of surface brightness dimming \citep{Conselice:2000p4573,Conselice:2003p4567,Lotz:2008p4477} though some of this is countered by morphological $k$-corrections \citep{Conselice:2008p4097}.  This has a small effect on our sample, i.e., 230 (16.5\%) of the galaxies have $A>0.3$.

The clumpiness parameter \citep{Conselice:2003p4555} measures the patchiness of the light distribution within a galaxy. This patchiness can be related to a number of physical aspects of a galaxy including its inclination, the presence of dust lanes, and star formation. For ellipticals, the smooth light distribution results in a clumpiness value near 0. Our sample has a median clumpiness of $-0.05$ and is on average smoother than the full optical sample (0.12). Table~\ref{auto_table} indicates that there is a trend for galaxies to become clumpier at higher $L_{\rm IR}$ and the overall spread in values increases as well.

The Gini coefficient \citep[$G$:][]{Abraham:2003p4427,Lotz:2004p4487} measures the distribution of flux values among the pixels of the image of a galaxy. Gini is highly correlated with the concentration parameter but is different in that it does not depend on the central position of the galaxy \citep{Lotz:2004p4487}. Galaxies with bright nuclei and compact galaxies have high values of Gini whereas diffuse galaxies such as late type spirals have low values. The Gini coefficient shows the greatest difference between the 70\ts$\mu$m sample ($<G>=0.50$) and the full optical sample ($<G>=0.41$) and there is no clear trend between Gini and $L_{\rm IR}$. However, the standard deviation of the Gini values becomes very small at $>10^{11}\ts L_{\odot}$, indicating that there is very little spread in Gini at high $L_{\rm IR}$. 

$M_{20}$ is a measure of the second-order moment of the brightest 20\% of pixels in a galaxy's image \citep{Lotz:2004p4487}. This parameter is particularly useful because it is sensitive to merger signatures, especially multiple nuclei in a galaxy. The 70\ts$\mu$m selected sample has on average higher $M_{20}$ values ($<M_{20}>= -1.74$) than the full optical sample ($<M_{20}> = -1.54$). There is also a clear trend of higher $M_{20}$ values with $L_{\rm IR}$.

It is worth noting that the values presented in Table 6 have not been $k$-corrected. It is expected that these values will change somewhat as a function of rest-frame wavelength  \citep[e.g.,][]{TaylorMager:2007p7021}. For example, galaxies have been observed to be clumpier at shorter wavelengths due to regions of prominent star formation. Therefore, trends seen here with $L_{\rm IR}$ might also be due to observing the rest-frame UV at $z>1$.  In order to study these automated classifications systematically, high-resolution images probing the same rest-frame wavelength across the full redshift range is needed. This will be possible with future work using near-infrared imaging from WFC3 on HST for the objects with $z>1$.

\subsection{Comparison of Visual and Automated Classifications}

\begin{figure}
\epsscale{1.15}
\plotone{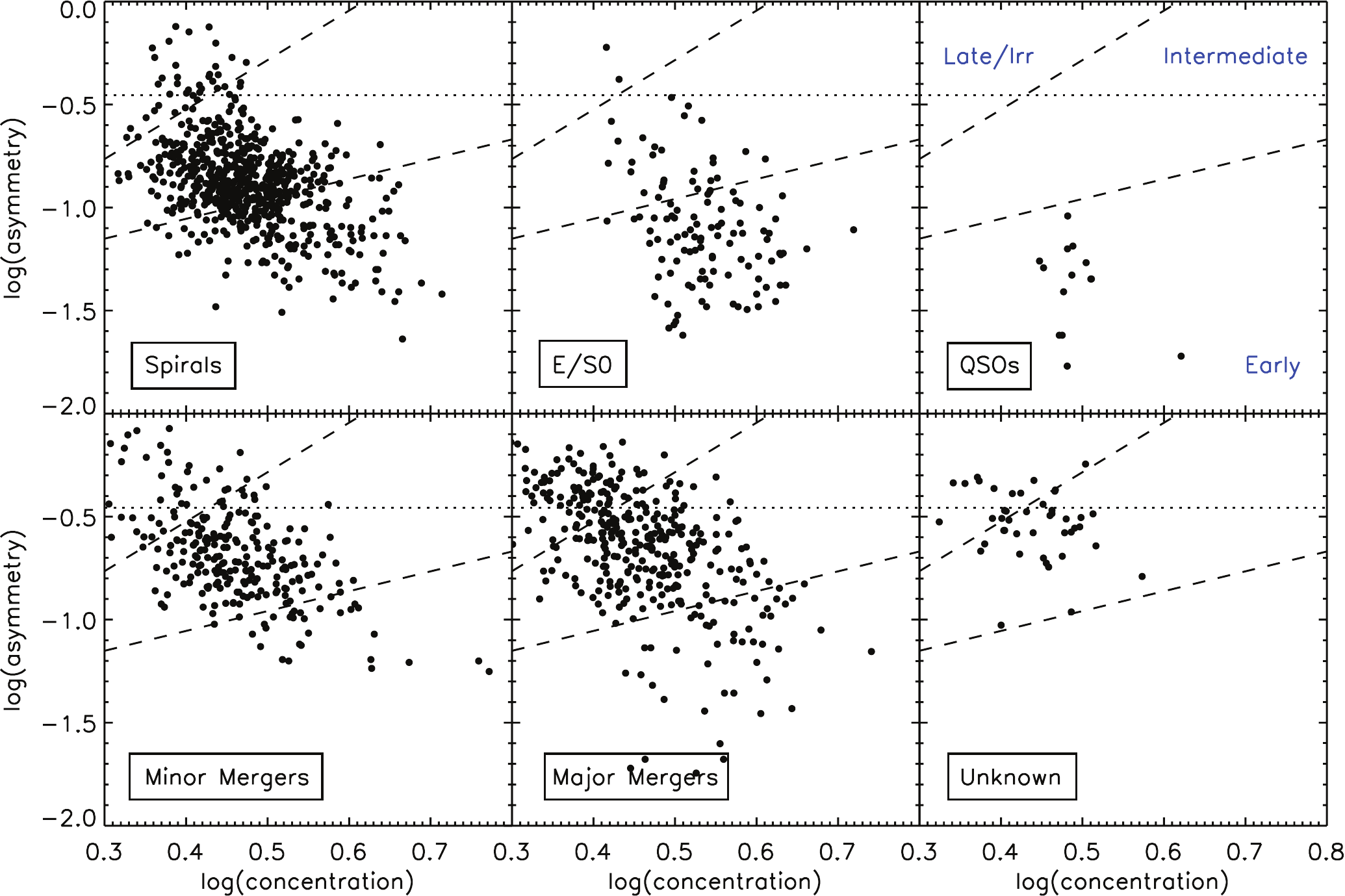}
\caption{Asymmetry versus concentration for the 70\ts$\mu$m selected sample divided by their visual classifications. The dotted line is at $A=0.35$ (log$(A)=-0.46$), the dividing line above which objects are expected to be major mergers \citep{Conselice:2003p4555}. The dashed lines represent the boundaries between early, intermediate, and late/irregular galaxies from \cite{Bershady:2000p4579}. Note that there is general agreement between the visual classification and the location of the galaxy on the concentration-asymmetry plane, though with considerable overlap.}
\label{con_asym}
\end{figure}

The last three columns of Table~\ref{auto_table} lists the percentage of objects in each Cassata \etal classification (ellipticals/S0, spirals, or irregulars) as a function of $L_{\rm IR}$. As for the visual classifications in Table~\ref{morphology}, a clear trend can be seen. The fraction of spirals decreases from $\sim 50\%$ at the low luminosity end to 25\% for ULIRGs. Likewise, the fraction of irregulars increases strongly from $\sim 20\%$ at low luminosities to $>50\%$ for ULIRGs. While this fraction is not as high as that obtained with the visual classification, the relative agreement is an indication that both classification methods are robust.

In Figure~\ref{con_asym} we plot the asymmetry and concentration parameters against one another separated by the visual classification assigned to each galaxy. In the case where a galaxy has multiple classifications we use the one that appears to be dominant (i.e., a spiral-spiral pair in interaction class I would be a spiral and an unperturbed spiral or elliptical with a small minor companion would be considered a spiral or elliptical, respectively). The location of galaxies on this plane has traditionally been used as a method for separating irregular/peculiar galaxies from spirals and ellipticals \citep[e.g.,][]{Abraham:1996p4462,Bershady:2000p4579,Conselice:2000p4573,Cassata:2005p4586,Shi:2006p4394}.  The horizontal dotted line marks the asymmetry value of 0.35 above which objects can be considered to be mergers \citep{Conselice:2003p4555}. The two dashed lines represent the boundaries of \cite{Bershady:2000p4579} separating early type galaxies (bottom of plot), intermediate spirals (between the lines), and late spirals and irregular galaxies (top). On average, this separation appears to work well. The majority of the galaxies visually classified as spirals fall between the two lines, though many also fall into the early and irregular areas as well. The same is true for the E/S0 galaxies, the majority fall below the line into the area that should be dominated by early type galaxies and a few scatter up into the spiral area. All of the objects classified as QSOs fall into the early-type area of the diagram. Most of the minor mergers fall into the spiral area with a few scattering into the early and irregular type areas. This is not surprising since the morphologies of the minor mergers are only slightly disturbed. The major merger galaxies have the largest spread in the concentration-asymmetry plane. Some ($\sim 20\%$) of them fall above the \citeauthor{Bershady:2000p4579} dividing line indicating that they have irregular morphologies and some (25\%) fall above the \cite{Conselice:2003p4567} asymmetry value of 0.35. However, most fall in the area occupied by spirals and a few into the early-type area. Interestingly, most of the galaxies that were too faint or small to classify visually have high asymmetry values at the upper end of the diagram. These likely have irregular morphologies and are possibly mergers as well. While the concentration/asymmetry plane does a good job on average of separating galaxies of the different classes there is considerable overlap. Using this plane to select major mergers would result in missing the majority of mergers.

\begin{figure}
\epsscale{1.15}
\plotone{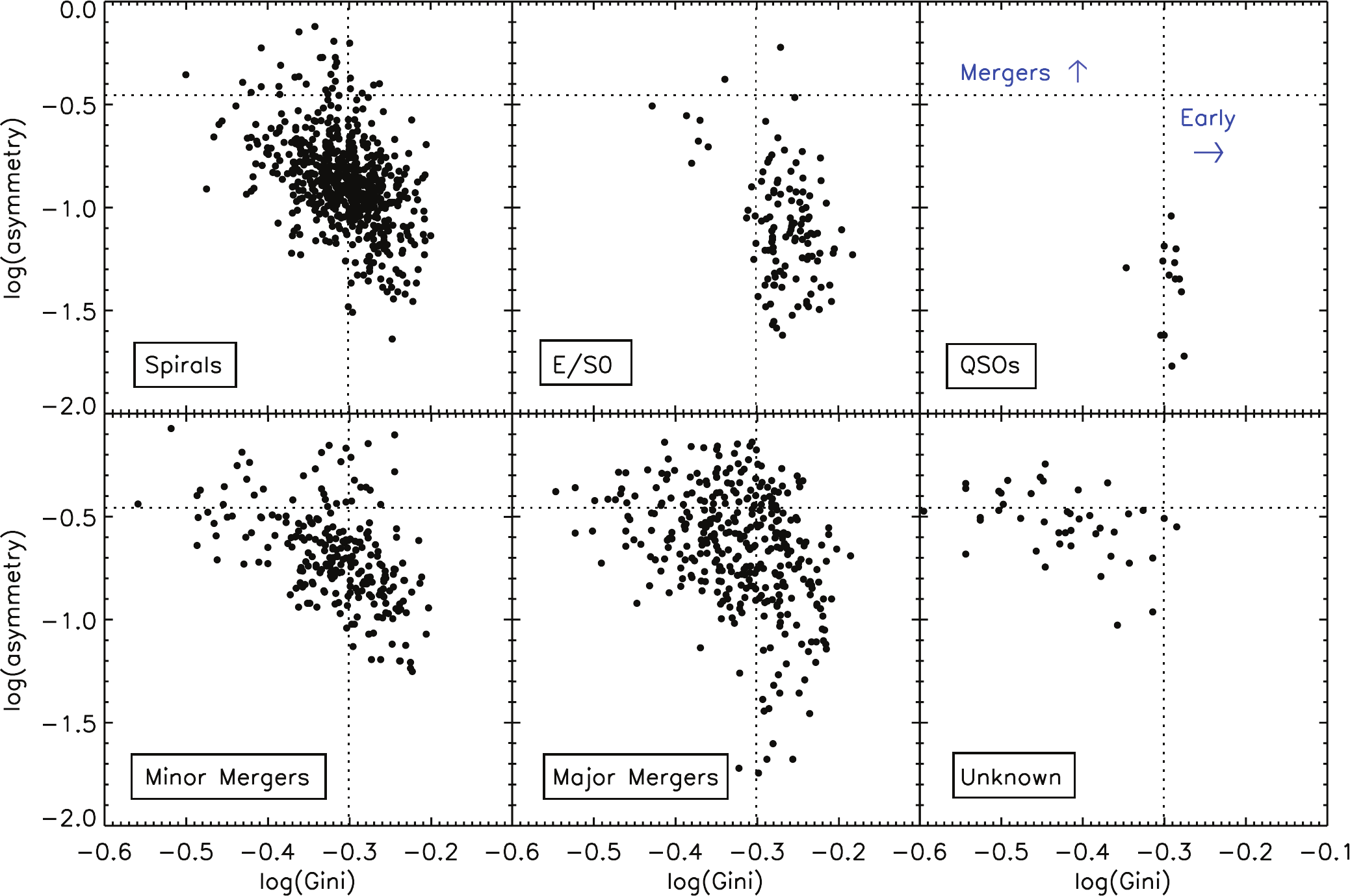}
\caption{Asymmetry versus Gini for the 70\ts$\mu$m selected sample divided by their visual classifications. The two dotted lines represent $A=0.35$ (log$(A)=-0.46$), the dividing line above which objects are expected to be major mergers (Conselice 2003) and $G=0.5$ (log$(G)=-0.30$), above which galaxies are expected to be elliptical. Once again, there is general agreement between the visual classification and the placement of the galaxies on the Gini-Asymmetry plane, particularly for the E/S0 galaxies.}
\label{gini_asym}
\end{figure}

Since the Gini coefficient and the concentration parameter are closely related, the results in the Gini-asymmetry plane are similar (Fig.~\ref{gini_asym}).  \cite{Abraham:2007p4444} and \cite{Capak:2007p2294} use the Gini coefficient to separate early type galaxies as objects with $G>0.5$, over plotted as the dotted vertical line. While this division selects nearly all galaxies visually classified as ellipticals/S0, it also selects many spirals and merging galaxies.

\begin{figure}
\epsscale{1.15}
\plotone{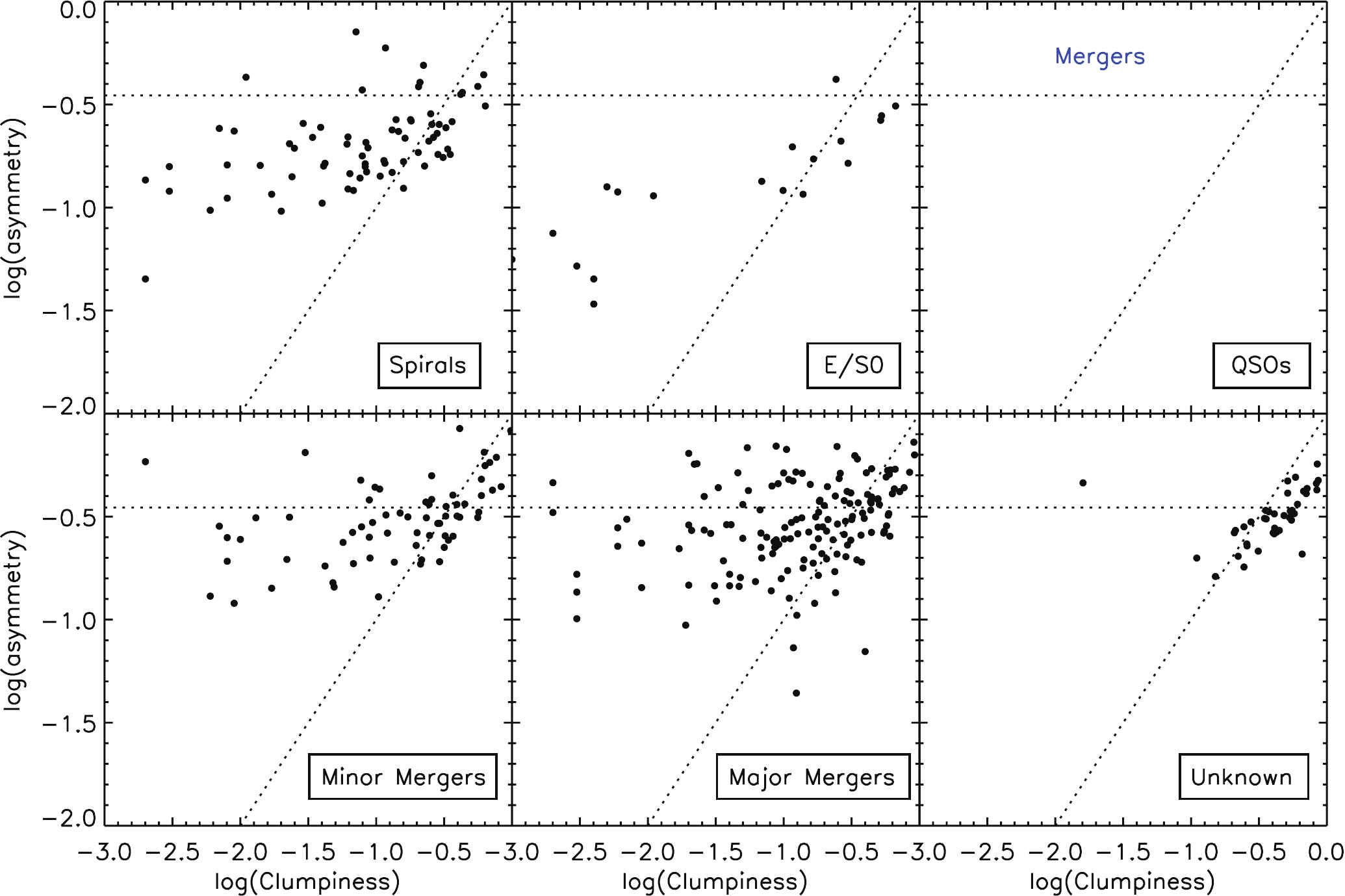}
\caption{Asymmetry versus Clumpiness for the 70\ts$\mu$m selected sample divided by their visual classifications. The two dotted lines represent $A=0.35$ (log$(A)=-0.46$) and $A>Clumpiness$, above which objects are expected to be major mergers \citep{Conselice:2008p4097}. Though most of the objects that meet these criteria are visually classified as major mergers, some galaxies of each morphological type do as well. Very few of the unknown objects or ellipticals meet these criteria.}
\label{clump_asym}
\end{figure}

\begin{figure}[b]
\epsscale{1.15}
\plotone{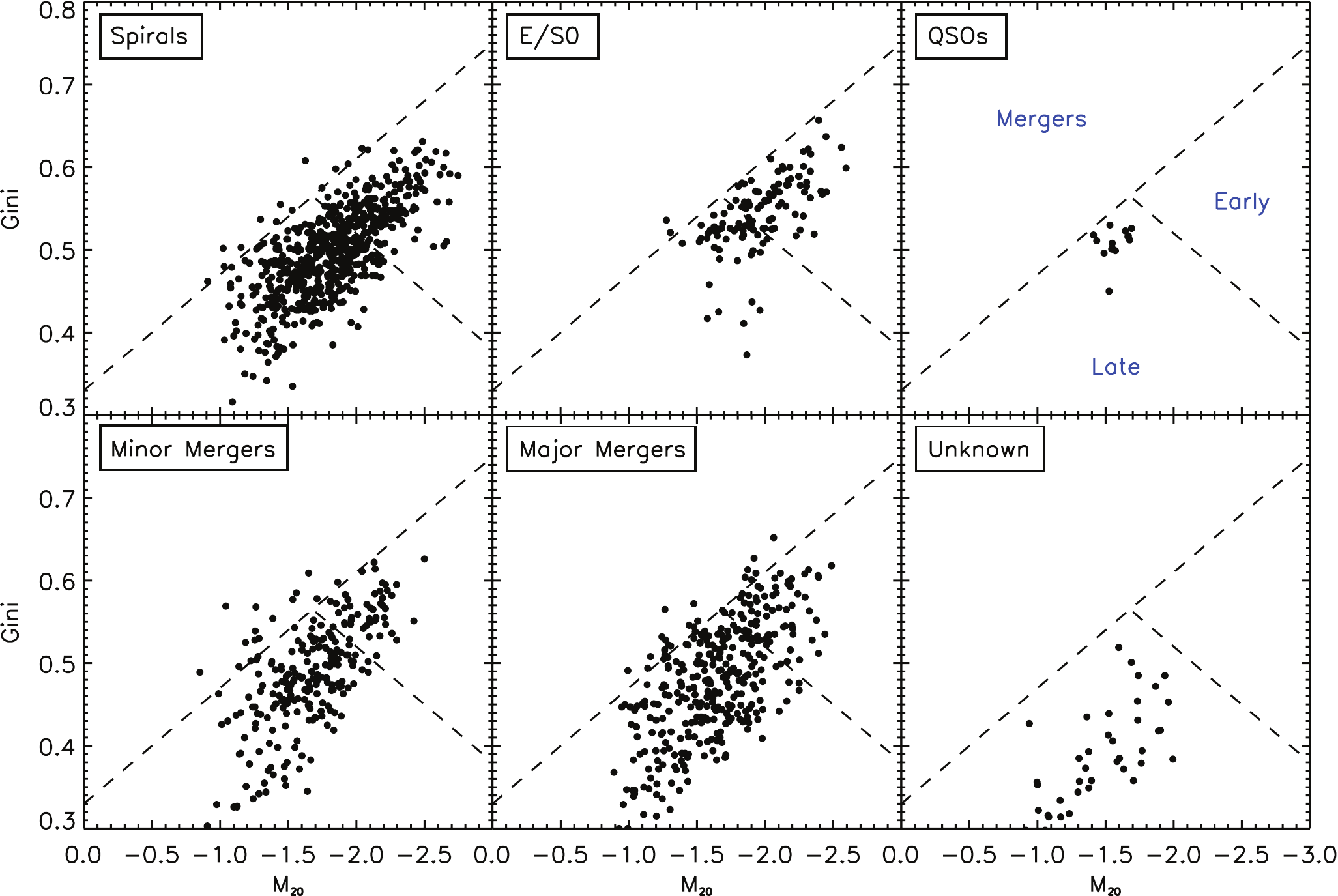}
\caption{Gini versus M$_{20}$, as plotted in \cite{Lotz:2004p4487}, for the 70\ts$\mu$m selected sample divided by their visual classifications. The dashed lines represent the boundaries between mergers, early, and late type morphologies as derived from EGS galaxies at $0.2<z<1.2$ by \cite{Lotz:2008p347}. Although there is general agreement between the spiral/elliptical visual classifications and the location on the Gini-M$_{20}$ plane, though with considerable overlap, very few of the major mergers occupy the major merger portion of the diagram though some spirals and minor mergers do. }
\label{gini_M20}
\end{figure}

\begin{figure*}
\epsscale{1.1}
\vspace{-0.1in}
\plottwo{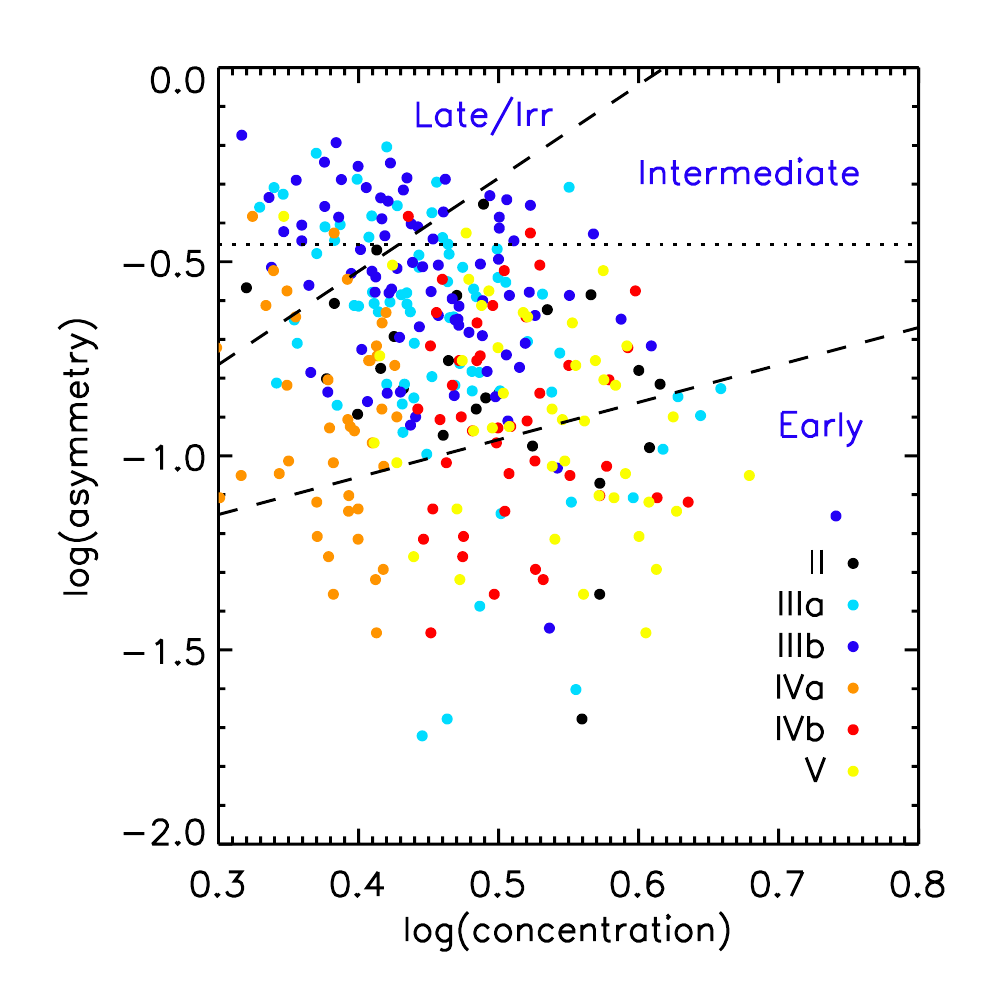}{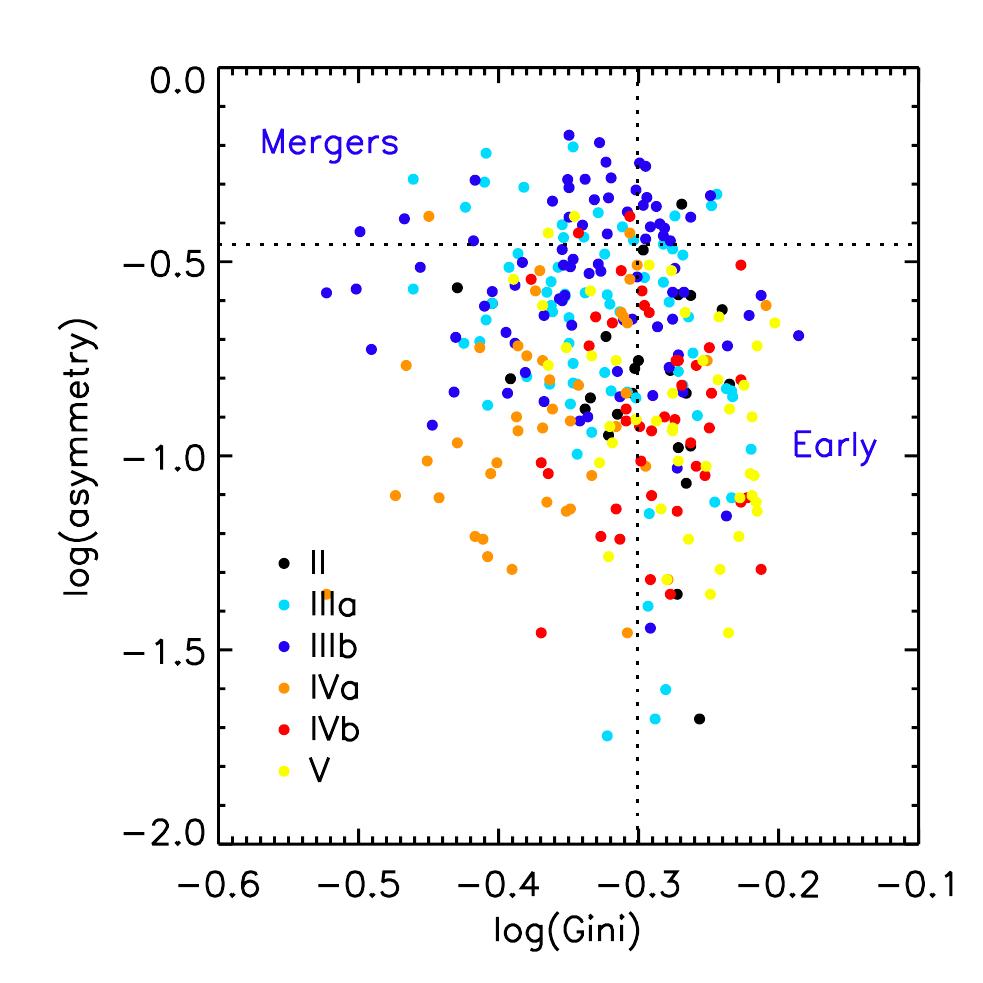}\\
\vspace{-0.2in}
\epsscale{1.1}
\plottwo{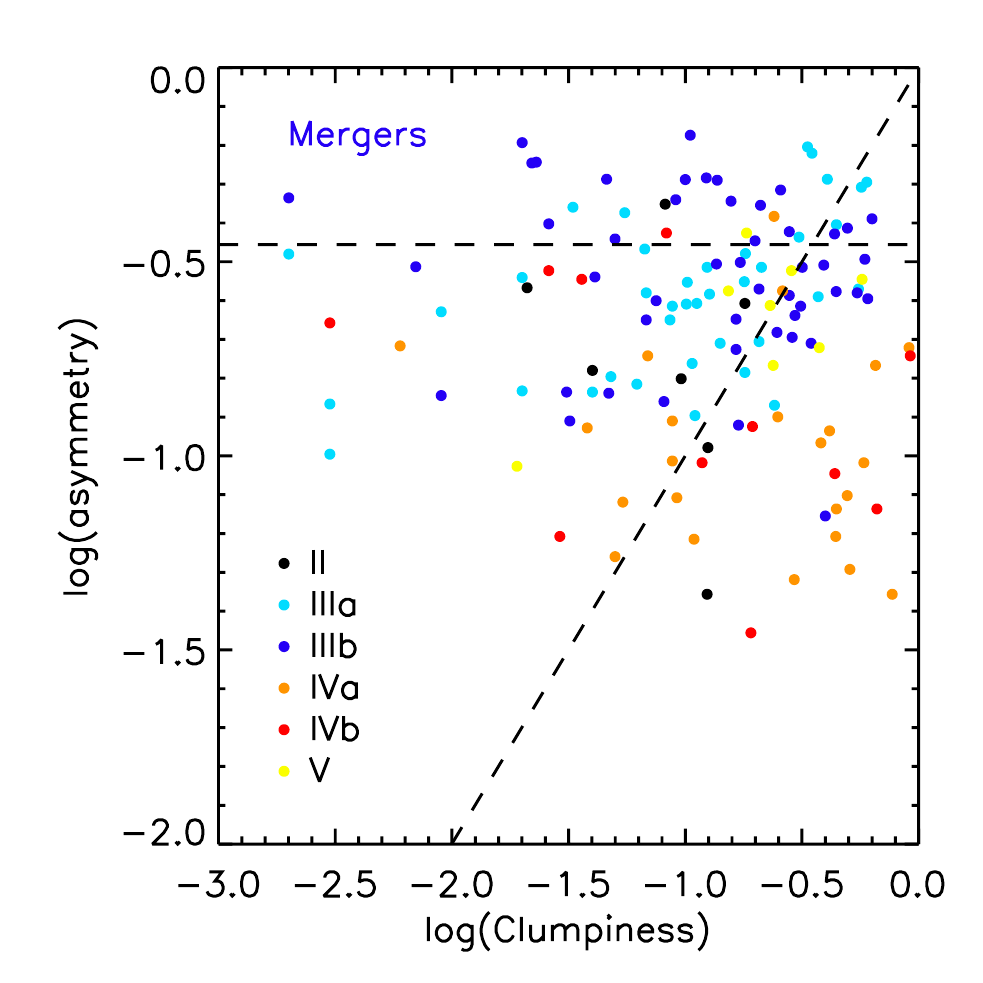}{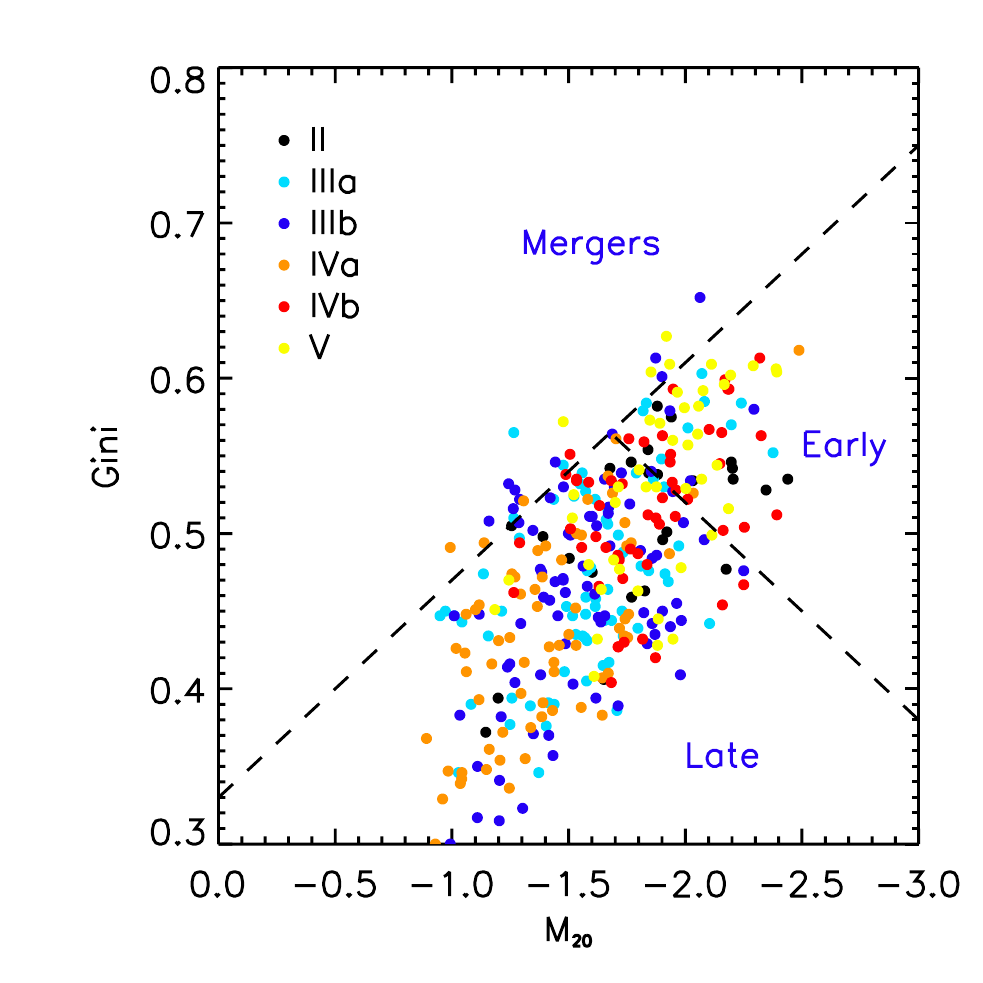}
\caption{Asymmetry versus concentration (top left), asymmetry versus Gini (top right), asymmetry versus clumpiness (bottom left), and Gini versus M$_{20}$ (bottom right) as in Figs.~\ref{con_asym}, \ref{gini_asym}, \ref{clump_asym}, and \ref{gini_M20}, respectively, for all of the 70\ts$\mu$m sources visually classified as major mergers color coded by their interaction classes. The dividing lines between classes are the same as shown in Figs.~\ref{con_asym}-\ref{gini_M20} and the morphological classes based on these divisions are labeled in blue. Interaction Class III and IV objects separate out neatly in the concentration-asymmetry and Gini-asymmetry planes while the class V objects separate out in the Gini-M$_{20}$ and Gini-asymmetry planes. }
\label{auto_IC}
\end{figure*}

\cite{Conselice:2008p4097} proposed an additional criterion of asymmetry values greater than the values for clumpiness (their Eq. 5) to identify mergers. This is expected to exclude galaxies that are asymmetric only due to clumpy star formation and not an ongoing merger. Figure 12 shows the 70\ts$\mu$m selected galaxies plotted in the asymmetry-clumpiness plane with these two selection criteria over plotted as dashed lines. Although most of the galaxies that meet these criteria (131 objects) are major mergers by their visual classification, some galaxies with other classifications do as well, particularly minor mergers and spirals. It is encouraging to note that very few, if any, of the objects classified as unknown, elliptical, or QSOs would be considered major mergers by this classification. \cite{Jogee:2009p7014} find that the fraction of visually classified mergers that match these criteria ranges from $14-73\%$, depending on the mass and redshift range studied. For the typical mass and redshift range of our sample, a recovery fraction of $\sim 50\%$ represents the best comparison. Our recovery fraction ($21\%$, $74/351$) is substantially lower, though this is likely because we are including galaxies at a higher redshift than the \cite{Jogee:2009p7014} study. \cite{TaylorMager:2007p7021} find that a galaxy's morphological parameters change as a function of rest-frame wavelength, such that their asymmetry and clumpiness values increase at shorter wavelengths. They suggest modifying the major merger selection criteria to $A>0.5$ and $A>$ clumpiness when looking at the rest-frame UV. Since the values of $A$ can also decrease by $0.05-0.15$ at high redshift \citep{Conselice:2008p4097}, we adopt the original criteria in Figure~\ref{clump_asym} for this discussion.

\cite{Lotz:2004p4487} use the Gini-$M_{20}$ plane to select mergers. We plot our sample in this plane in Figure~\ref{gini_M20}, again separated by their visual classification. The boundaries between mergers (top/left portion of plot), early-type (right side), and late-type (bottom/center) galaxies are over plotted with dashed lines. These boundaries are from \cite{Lotz:2008p347} based on galaxies from the Extended Groth Strip (EGS) over the redshift range $0.2<z<1.2$. These boundaries work well for the visually classified spiral and elliptical galaxies, though with considerable overlap. Almost all of the unknowns and minor mergers fall into the spiral area. It is interesting to note that very few of the objects in our sample satisfy the merger criteria here. Even for the objects visually classified as major mergers, only a handful lie above the line. The ULIRG sample used by \cite{Lotz:2004p4487} to define the criteria showed that mergers with double nuclei had the largest offset while those in pairs had the smallest. ULIRGs with a single nucleus also have a small offset but a slightly higher Gini coefficient.  \cite{Lotz:2008p4477} confirms this result with simulations of galaxy mergers and finds that mergers are the most disturbed in Gini-$M_{20}$ space at the moment of first passage and at final coalescence.

We plot all four of these parameter planes for the objects visually classified as major mergers in Figure~\ref{auto_IC} color coded by their interaction classes. Almost all of the objects that would be classified as irregulars in the concentration-asymmetry plane have interaction classes IIIa and IIIb which means they are all either pairs or single objects with double nuclei. Most of the class IV and V objects lie in the early-type area of the diagram. These classes separate out better in the asymmetry/Gini plane where almost all of the class IV and V objects have $A<0.2$ and are cleanly separated from each other by the $G=0.5$ vertical dividing line. The class IIIa and IIIb objects all have higher asymmetries as expected from their disturbed morphologies and the IIIb objects have higher Gini coefficients on average than the IIIa objects. This result is consistent with the IIIb objects being more compact advanced mergers than the IIIa objects.

The interaction classes do not separate out as well in Gini-$M_{20}$ space. The class IV and V objects are still divided by their Gini coefficient and the class V objects have smaller values of $M_{20}$ (to the right of the diagram). This places them in the early-type region of the diagram. The class IIIa and IIIb objects are spread out over the diagram. This result is not completely unexpected however and this plot looks remarkably similar to Figure 5 in \cite{Lotz:2008p4477} which shows a large spread in Gini-M$_{20}$ space for mergers at various stages from simulations.

\begin{figure}
\hspace*{-0.25in}
\epsscale{1.25}
\plotone{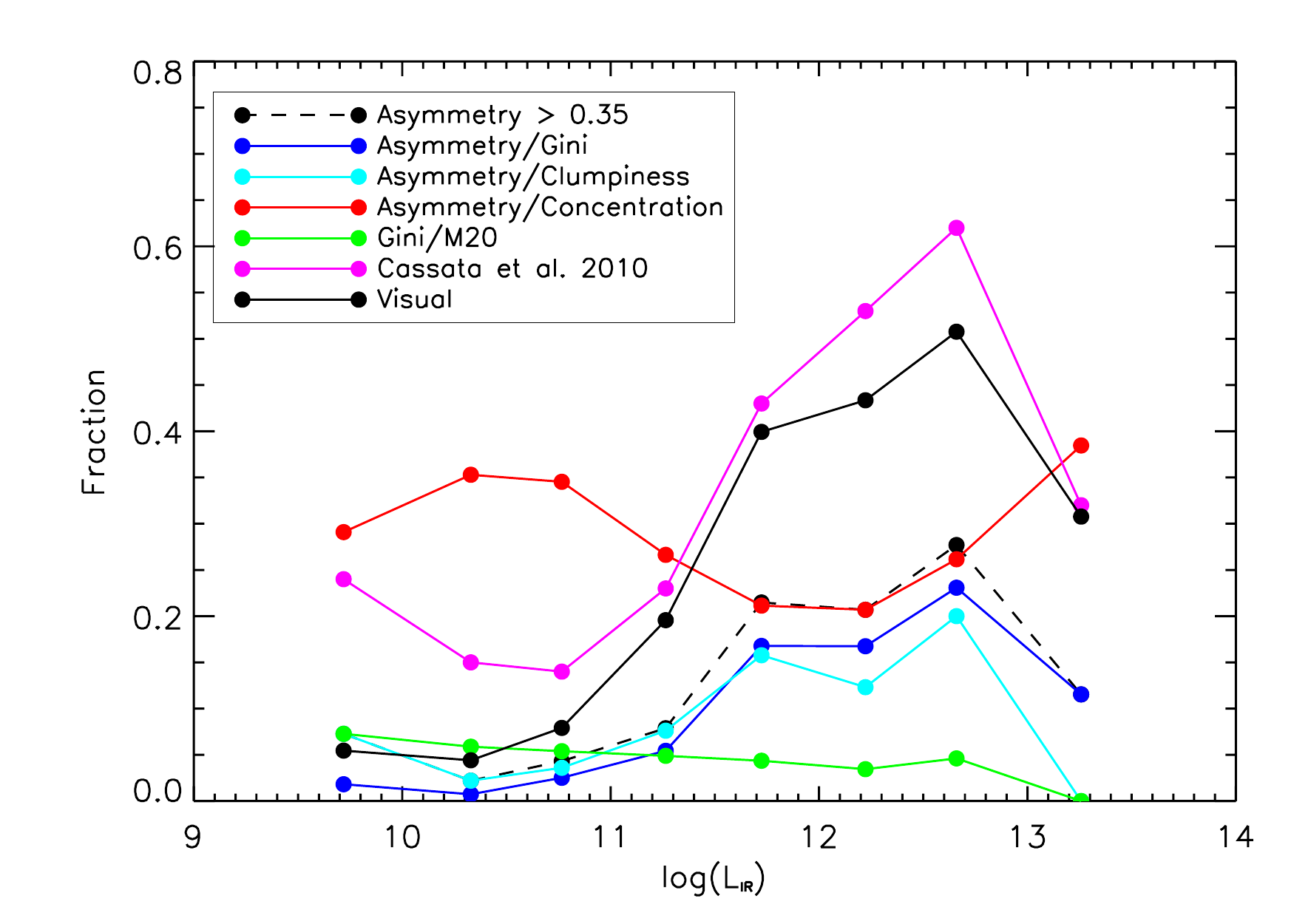}
\caption{Fraction of galaxies in the 70\ts$\mu$m sample identified as mergers using each of the different classification techniques as a function of $L_{\rm IR}$. Note that the fraction of galaxies classified as mergers visually does not quite match the fraction depicted in Figure 4 since the fraction here is limited to objects with reliable automated morphological parameters for a fair comparison.}
\label{frac_auto}
\end{figure}

Using each of above discussed classification methods, we plot the fraction of galaxies in the 70\ts$\mu$m selected sample that would be identified as mergers as a function of $L_{\rm IR}$ in Figure~\ref{frac_auto}.  Each of them shows an increase in the fraction of mergers with $L_{\rm IR}$, though the Cassata \etal (2010, in prep) classification shows the strongest trend. Most of the methods show a decline for HyLIRGs, likely due to the increasing fraction of QSOs in this bin. The \cite{Bershady:2000p4579} selection (red line in the plot) shows a high fraction of mergers at low $L_{\rm IR}$. Most of these objects were classified visually as spirals. Figure~\ref{con_asym} confirms the overlap between spirals/mergers and since this selection is intended to identify late-type galaxies as well as mergers, this is to be expected.

\section{Galaxy Color}
Optically selected galaxy samples have been observed to lie in two distinct regions on the color-magnitude plane, the ``blue cloud" of star forming and typically late type galaxies and the ``red sequence" of quiescent typically early type spheroids \citep[e.g.,][]{Strateva:2001p4590, Hogg:2003p4595,Blanton:2003p4606,Baldry:2004p4620}. This color bimodality has been observed and well-studied over a large redshift range \citep[e.g.,][]{Bell:2004p4626,Weiner:2005p4632}. Understanding the location of a galaxy on the color-magnitude plane and how it got there is key to understanding galaxy evolution and the formation of ellipticals. Recently, sub-samples of objects (e.g., AGN) have been found to preferentially inhabit the region between these two color peaks -- the so-called ``green valley." Much work has gone into determining whether these objects represent a transition stage between blue cloud galaxies and the red sequence. In this section, we look at the location of our 70\ts$\mu$m selected sample of objects in color-magnitude space and correlate this with $L_{\rm IR}$ and morphology.

\subsection{Color-Magnitude Diagrams}

\begin{figure*}
\epsscale{1}
\plotone{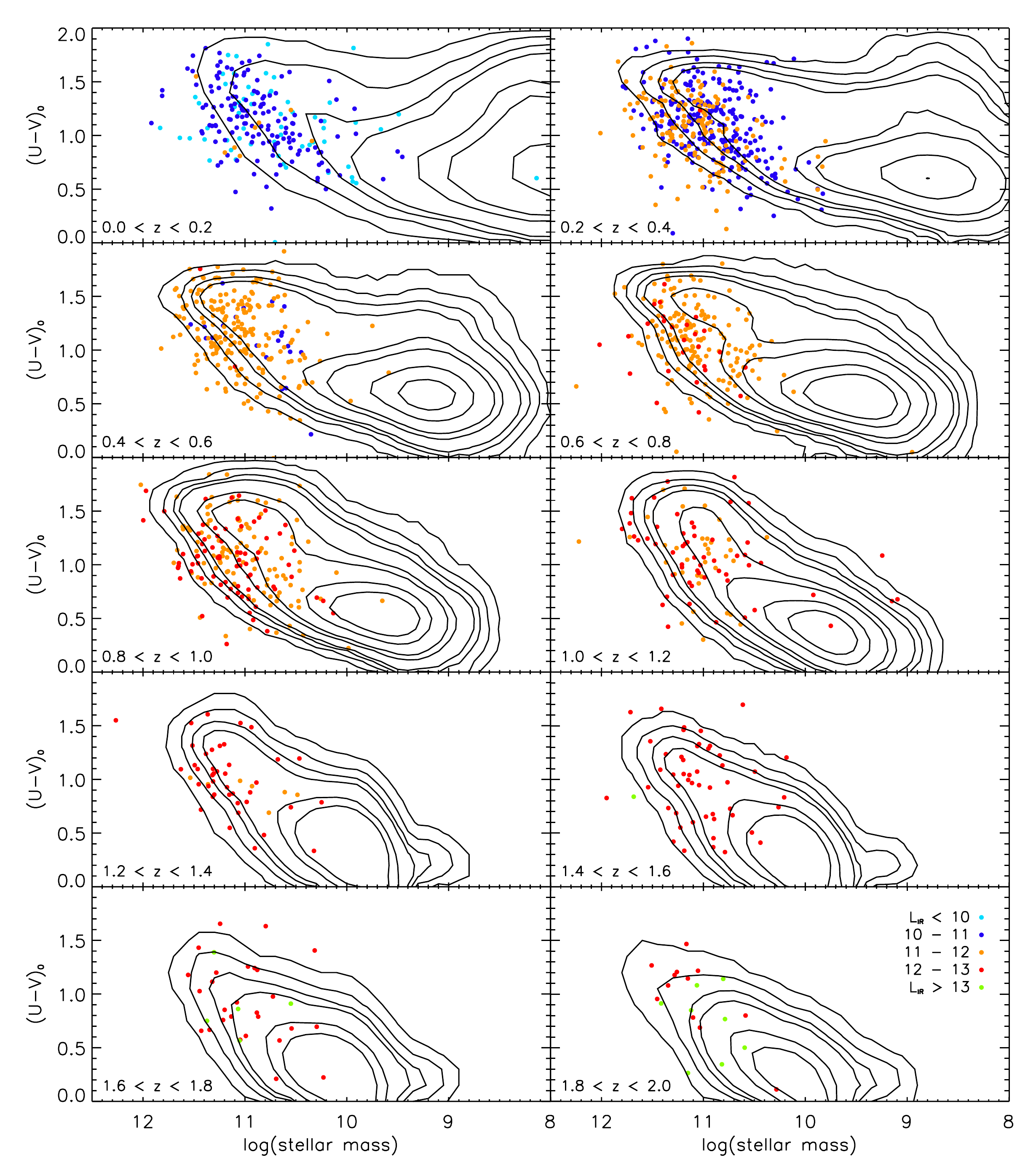}
\caption{$U-V$ color as a function of stellar mass for several redshift bins. The contours represent the distribution for the entire optically selected sample ($I<25$) and the dots represent the 70\ts$\mu$m selected sample colored by luminosity. The contour levels represent the number of objects in $U-V$ and log(stellar mass) bins of 0.1. The levels are, from the inside out, 300, 200, 100, 70, 30, 20, 10, and 3 objects. Note that the red sequence is clear at low redshifts ($z<1$) but is less well-defined at higher redshifts. The 70\ts$\mu$m selected sources are generally at the higher mass end of the diagram and have colors that span the full range from the blue cloud to the red sequence, though they predominantly lie in between in the green valley.}
\label{UV_mass}
\end{figure*}

\begin{figure*}
\plotone{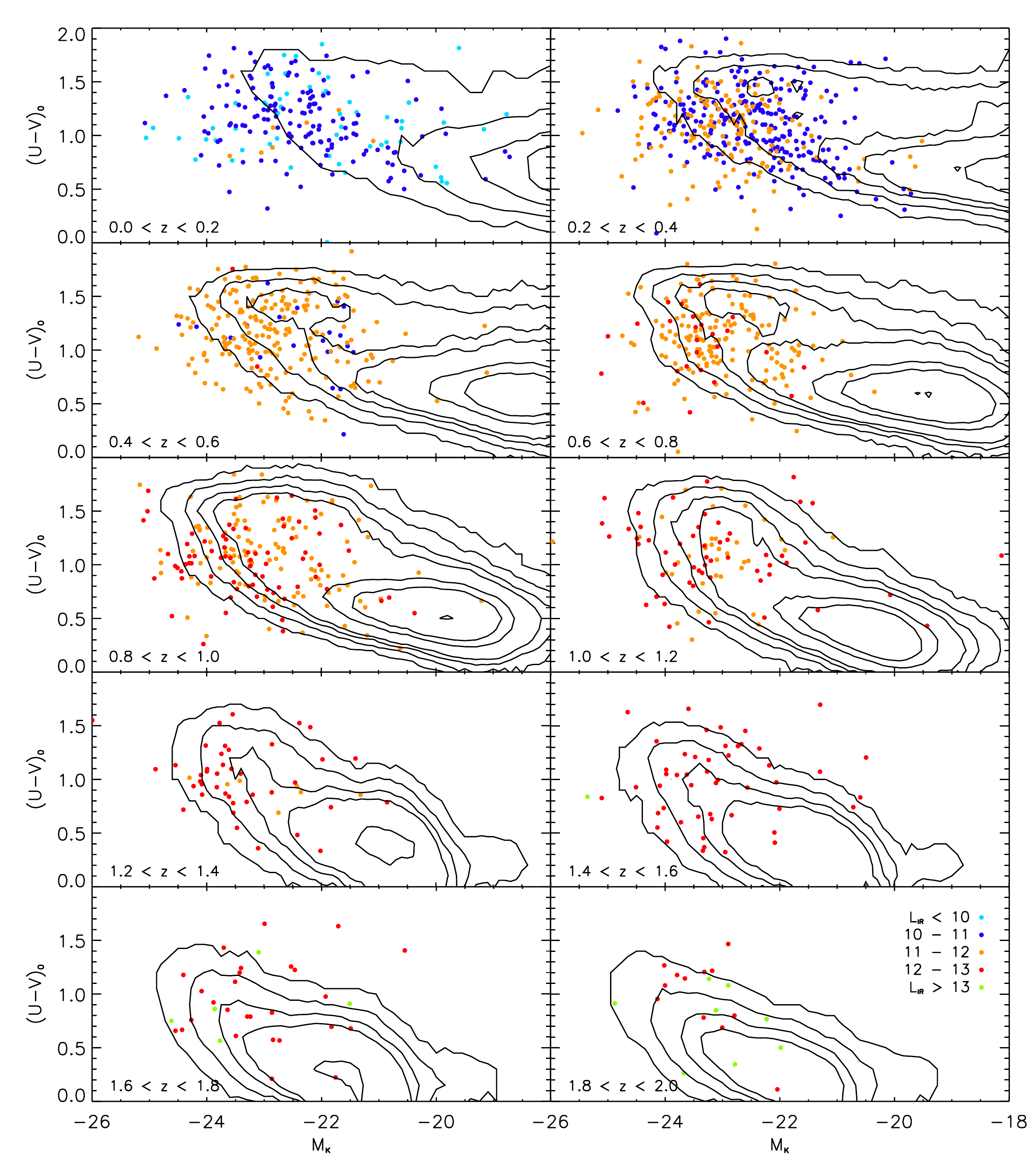}
\caption{$U-V$ color as a function of absolute K-band magnitude ($M_K$) for several redshift bins. The contours represent the distribution for the entire optically selected sample ($I<25$) and the dots represent the 70\ts$\mu$m selected sample colored by luminosity. The contour levels represent the number of objects in $U-V$ and $M_{K}$ bins of 0.1. The levels are, from the inside out, 300, 200, 100, 70, 30, 20, 10, and 3 objects. Note that the red sequence is clear at low redshifts ($z<1$) but is less well-defined at higher redshifts. The 70\ts$\mu$m selected sources are generally at the higher mass end of the diagram and have colors that span the full range from the blue cloud to the red sequence, though they predominantly lie in between in the green valley.}
\label{UV_MK}
\end{figure*}

\begin{figure}
\hspace*{-0.2in}
\epsscale{1.2}
\plotone{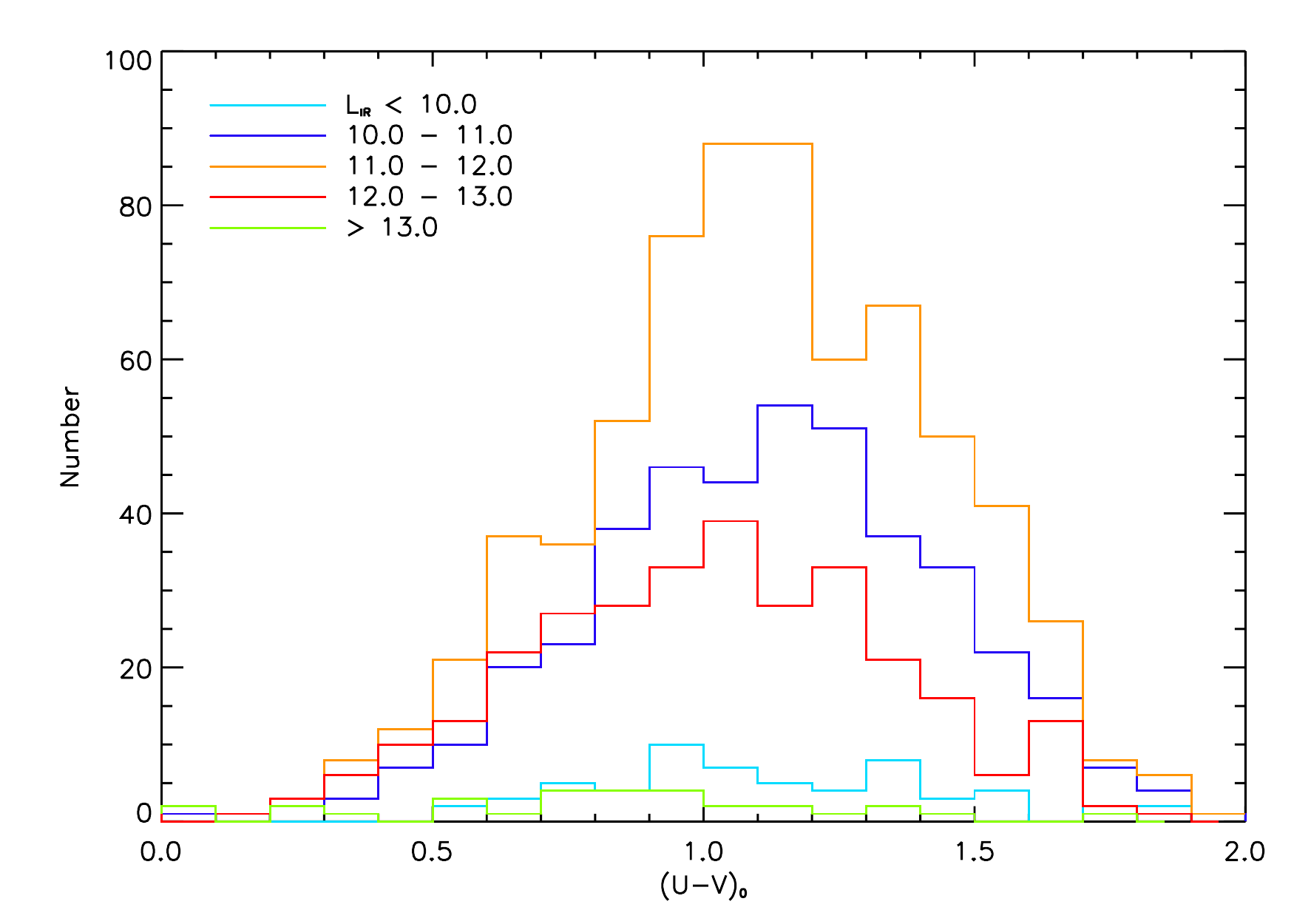}
\caption{Distribution of $U-V$ color in several $L_{\rm IR}$ bins coded by color. Note that LIRGs and ULIRGs have a blue tail. This is due to the very blue colors of the QSOs in those luminosity bins.}
\label{UV_LIR}
\end{figure}

Since the U and V bands straddle the 4000\ts\AA\ break, the $U-V$ rest frame color is particularly sensitive to the age and metallicity of a galaxy's stellar population. For this reason, as well as for direct comparison with other studies in the literature, we use these two bands when discussing a galaxy's optical color. The rest frame $U-V$ color for each of the 70\ts$\mu$m sources in our sample out to $z=2$ is plotted in Figures~\ref{UV_mass}-\ref{UV_MK} as a function of stellar mass and $M_{K}$. The points are color coded by $L_{\rm IR}$ and plotted on top of contours representing the distribution of the entire COSMOS optical sample ($I<25$). The contour levels represent the number of objects in $U-V$ and $M_{K}$ bins of 0.1. The levels are, from the inside out, 300, 200, 100, 70, 30, 20, 10, and 3 objects. The blue cloud and red sequence are well defined at low redshifts ($z <1$).  One thing that is immediately obvious is that the 70\ts$\mu$m sources are more luminous and more massive than typical blue cloud galaxies. In fact, a trend can be seen where sources with increasing $L_{\rm IR}$ are also more luminous in the optical and have higher stellar masses. While the objects seem to span a wide range in color, from typical blue cloud colors ($U-V\sim 0.5$) to well into the red sequence ($U-V>1.5$), most objects lie in between these two extremes in the green valley. The histogram in Figure~\ref{UV_LIR} shows the $U-V$ distribution color coded by $L_{\rm IR}$. Sources with $L_{\rm IR} < 10^{10}\ts L_{\odot}$ are on average bluer than more luminous sources, though ULIRGs and HyLIRGs have a bluer tail than low luminosity sources. HyLIRGs appear to be spread across all colors with no obvious peak while sources with $10^{10} <  L_{\rm IR} < 10^{12}\ts L_{\odot}$ are slightly redder on average than ULIRGs.

\begin{figure*}
\epsscale{0.8}
\plotone{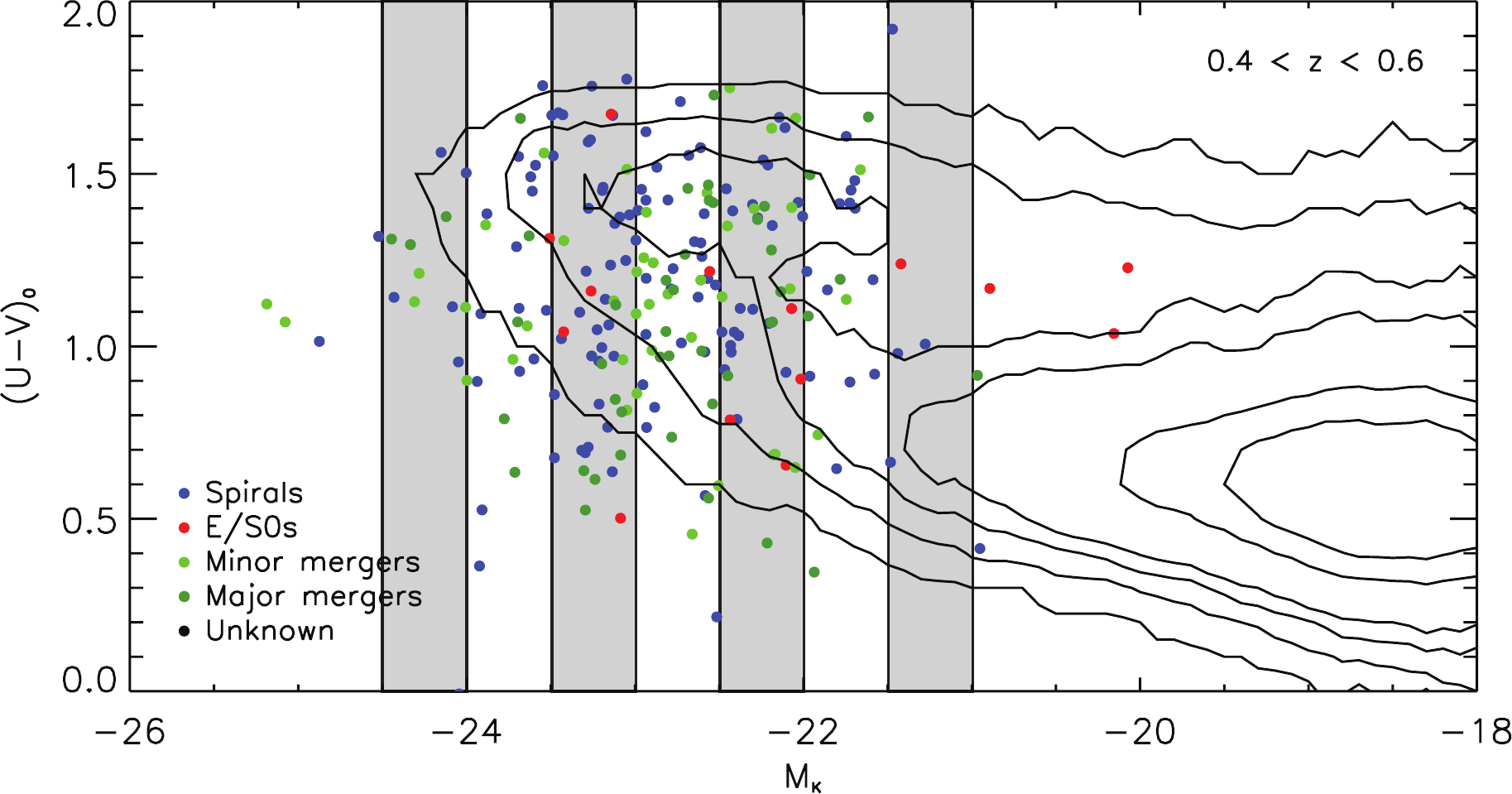}
\epsscale{1.1}
\hspace*{-0.6in}
\plotone{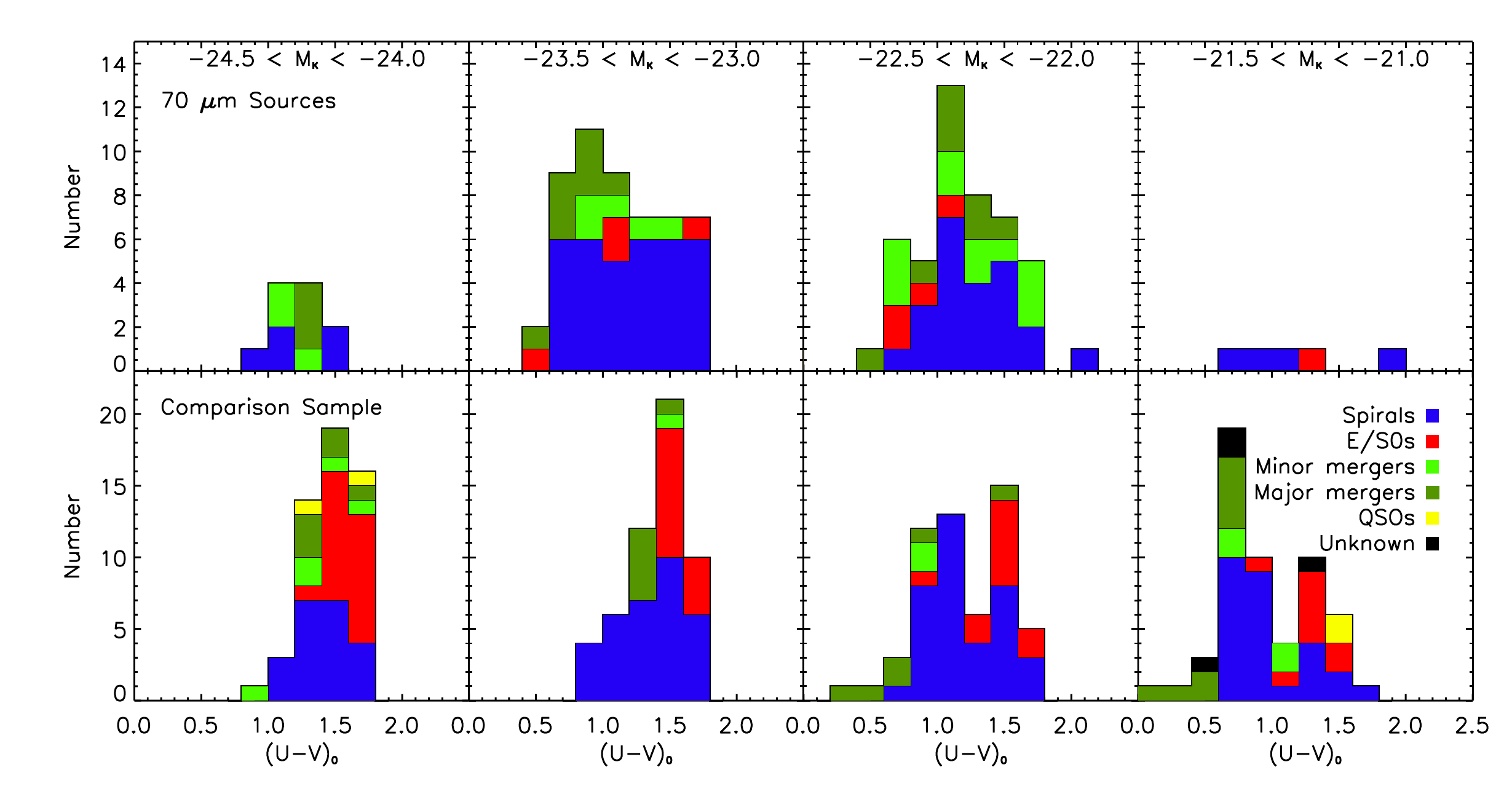}
\caption{Top: Color magnitude diagram for the $0.4<z<0.6$ redshift bin (as in Fig.~\ref{UV_MK}) with the 70\ts$\mu$m selected sources color coded by their visual morphological classification. The shaded regions highlight the four $M_{K}$ slices presented in the bottom panel. Bottom: $U-V$ histogram for the 70\ts$\mu$m sample (top panel) and a comparison sample of $\sim200$ randomly chosen optically selected galaxies, 50 in each $M_{K}$ slice (bottom panel), in the $0.4<z<0.6$ redshift bin in four different $M_{K}$ slices color coded by morphology.}
\label{UV_comp1}
\end{figure*}

\begin{figure*}
\epsscale{0.8}
\plotone{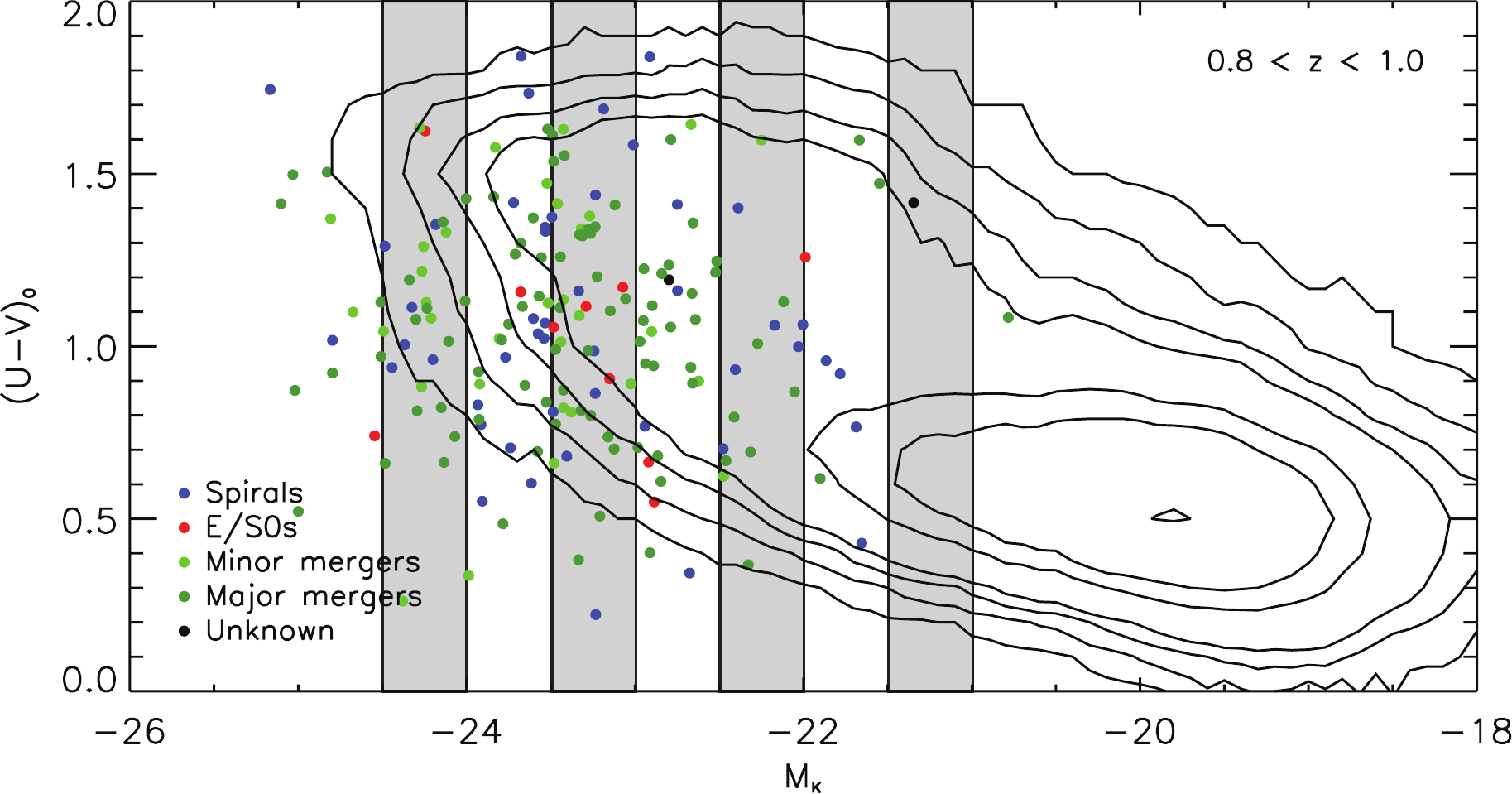}
\epsscale{1.1}
\hspace*{-0.6in}
\plotone{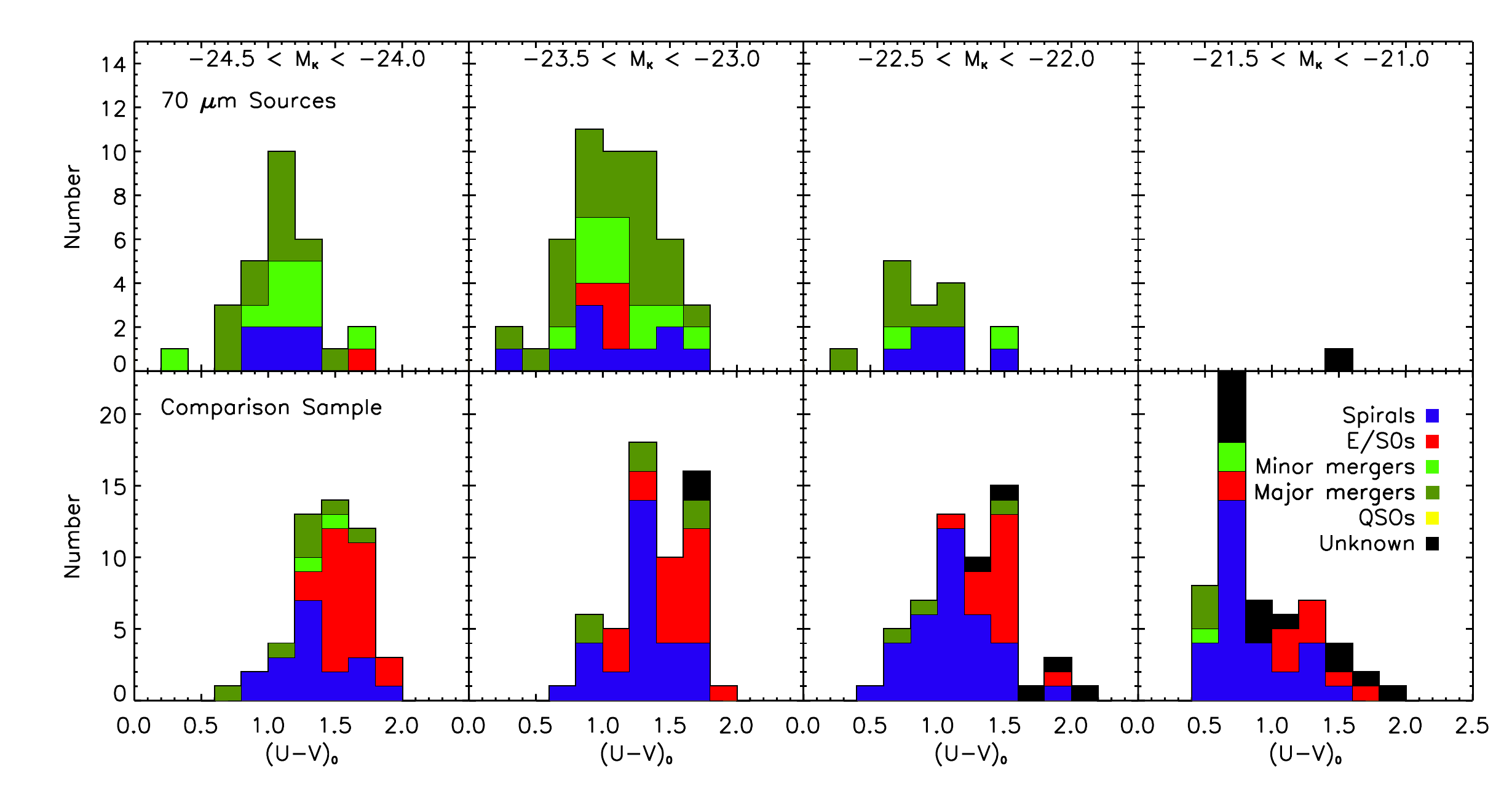}
\caption{Top: Color magnitude diagram for the $0.8<z<1.0$ redshift bin (as in Fig.~\ref{UV_MK}) with the 70\ts$\mu$m selected sources color coded by their visual morphological classification. The shaded regions highlight the four $M_{K}$ slices presented in the bottom panel. Bottom: $U-V$ histogram for the 70\ts$\mu$m sample (top panel) and a comparison sample of $\sim200$ randomly chosen optically selected galaxies, 50 in each $M_{K}$ slice (bottom panel), in the $0.8<z<1.0$ redshift bin in four different $M_{K}$ slices color coded by morphology.}
\label{UV_comp2}
\end{figure*}

\begin{figure*}
\epsscale{0.8}
\plotone{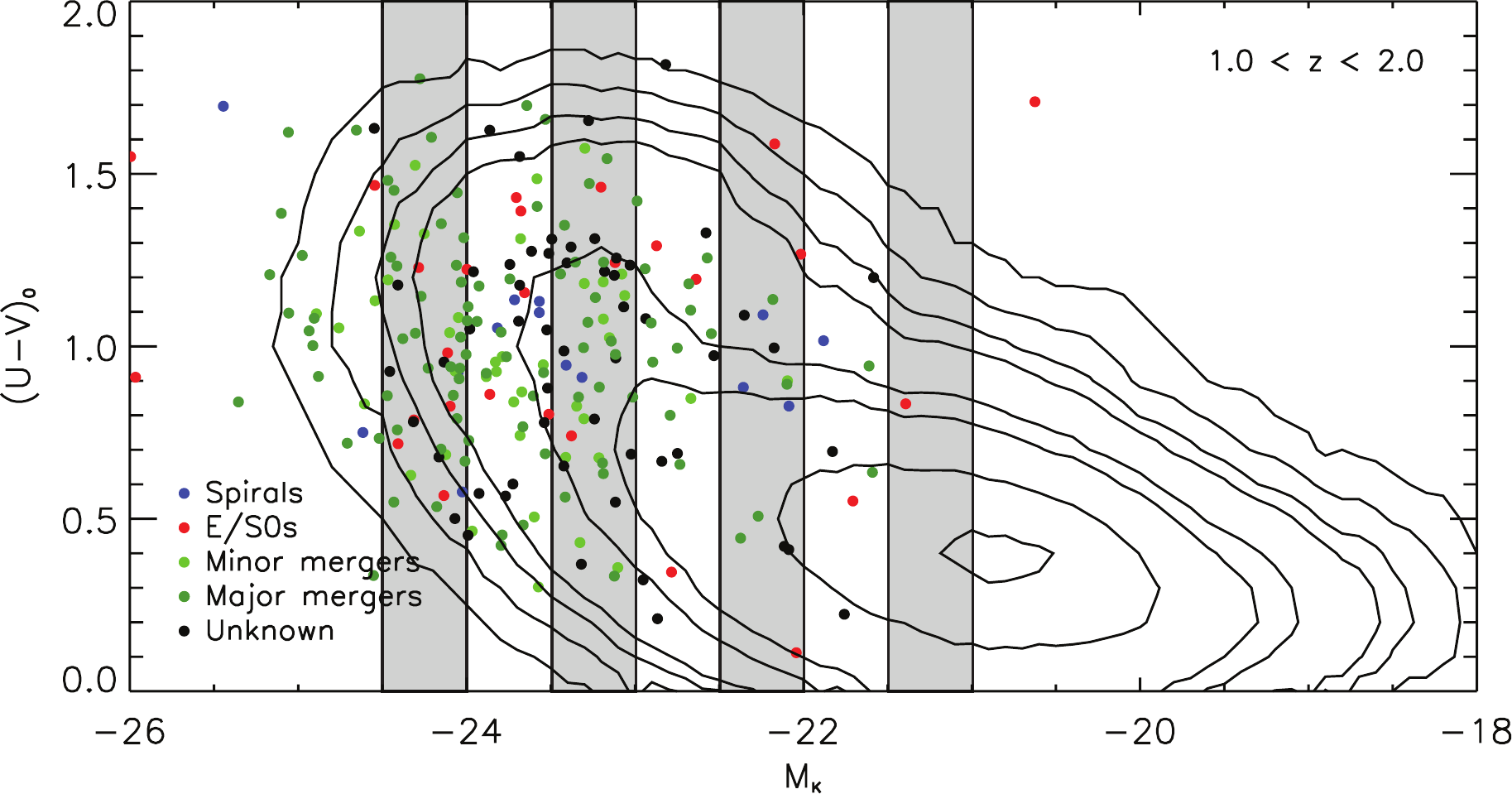}
\epsscale{1.1}
\hspace*{-0.6in}
\plotone{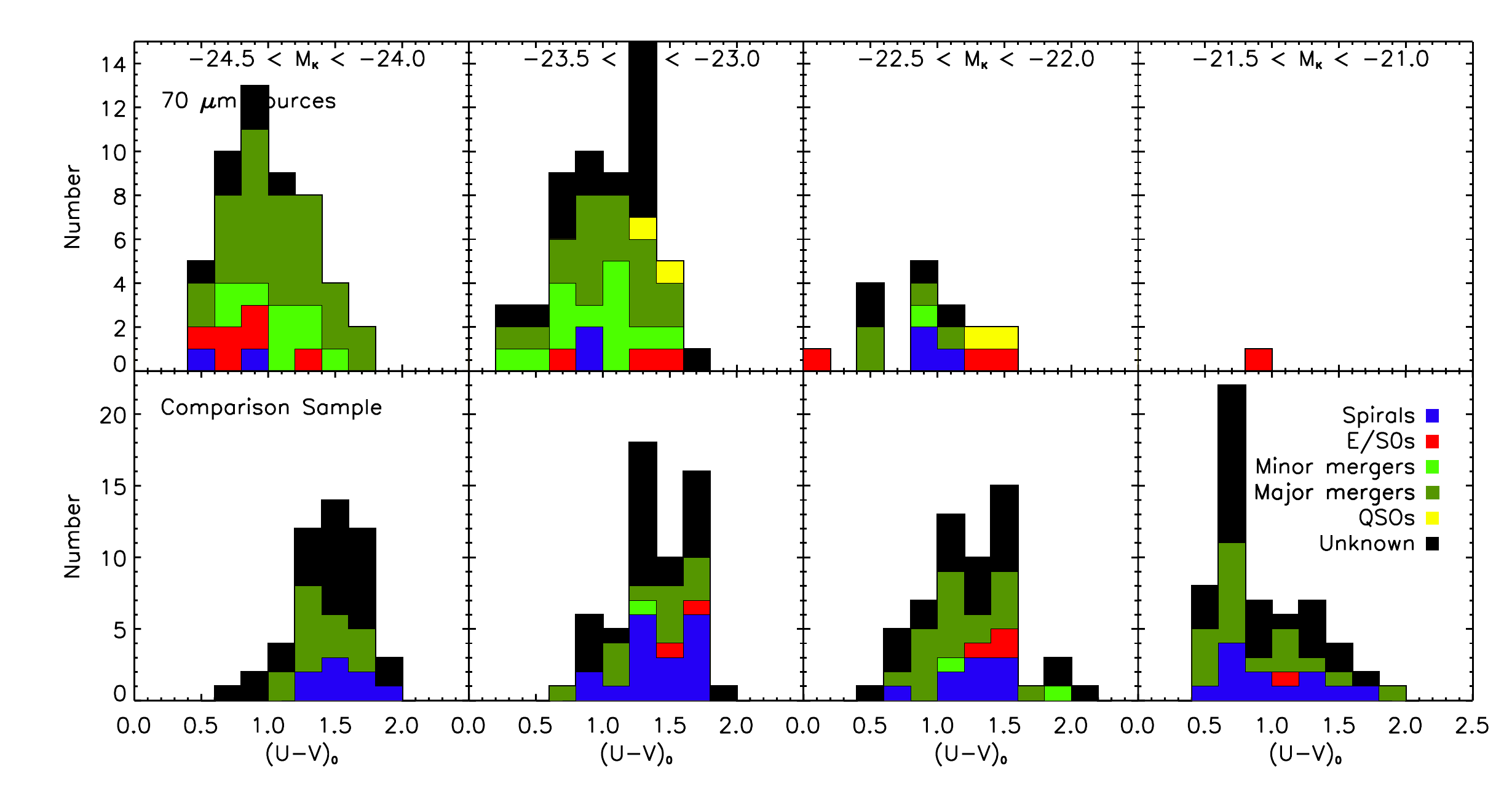}
\caption{Top: Color magnitude diagram for the $1.0<z<2.0$ redshift bin (as in Fig.~\ref{UV_MK}) with the 70\ts$\mu$m selected sources color coded by their visual morphological classification. The shaded regions highlight the four $M_{K}$ slices presented in the bottom panel. Bottom: $U-V$ histogram for the 70\ts$\mu$m sample (top panel) and a comparison sample of $\sim200$ randomly chosen optically selected galaxies, 50 in each $M_{K}$ slice (bottom panel), in the $1.0<z<2.0$ redshift bin in four different $M_{K}$ slices color coded by morphology.}
\label{UV_comp3}
\end{figure*}
\subsection{Relation to Morphology}

In order to investigate the relationship between $U-V$ color and morphology we compared the properties of our 70\ts$\mu$m sources with our comparison sample (described in \S3.2) in three different redshift bins. To ensure that we were comparing similar galaxies, we chose four slices in $M_{K}$ (illustrated in Figures~\ref{UV_comp1}--\ref{UV_comp2}) and compared objects within each slice. The results are shown in Figure~\ref{UV_comp1} for the $0.4<z<0.6$ redshift bin ($<L_{\rm IR}>=11.3$), Figure~\ref{UV_comp2} for the $0.8<z<1.0$ redshift bin ($<L_{\rm IR}>=11.8$), and Figure~\ref{UV_comp3} for the $1.0<z<2.0$ redshift bin  ($<L_{\rm IR}>=12.4$). The top row in each figure shows the $U-V$ distribution for the 70\ts$\mu$m sources in each $M_{K}$ slice color coded by morphology and the same is shown in the bottom row for the comparison sample. The typical color dichotomy can be seen in the comparison sample in each of the low redshift ($z<1$) bins with one peak in the red and one peak in the blue. In general the galaxies classified as spirals tend to be bluer and those classified as ellipticals tend to be redder, as expected. The few galaxy mergers in the comparison sample span a wide range in color but tend to be bluer on average. This color dichotomy is not seen in the 70\ts$\mu$m sample. The few ellipticals in this sample are not all red, but instead span the full range in color as do the galaxy mergers. There are not as many galaxies classified as ellipticals in the $1<z<2$ comparison sample, as expected, and the few there are tend to have red colors. The spirals in this bin span a wide range of in color, just like the mergers. The 70\ts$\mu$m sample at this redshift is composed of mostly mergers that span a wide range in color. The few ellipticals are much bluer than their optically selected counterparts.

\begin{figure*}
\hspace*{-0.6in}
\epsscale{1.0}
\plotone{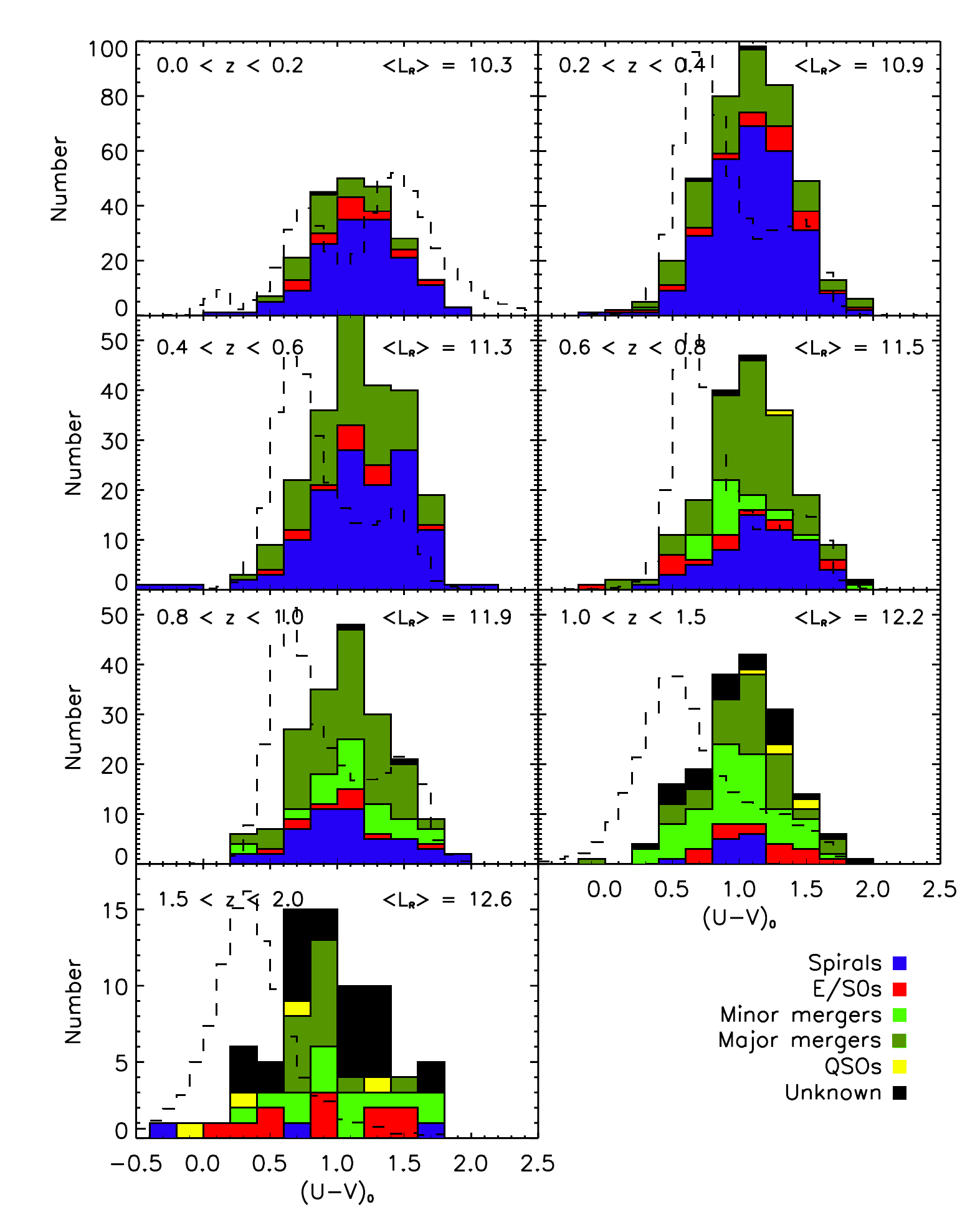}
\caption{Rest frame $U-V$ histogram for all 70\ts$\mu$m sources in the seven different redshift bins color coded by morphology. The $U-V$ distribution for the entire optically selected sample over the same redshift and mass range is shown as the dashed line. The median $L_{\rm IR}$ is given for reference.}
\label{UV_bin1}
\end{figure*}

\begin{figure}
\epsscale{1.2}
\plotone{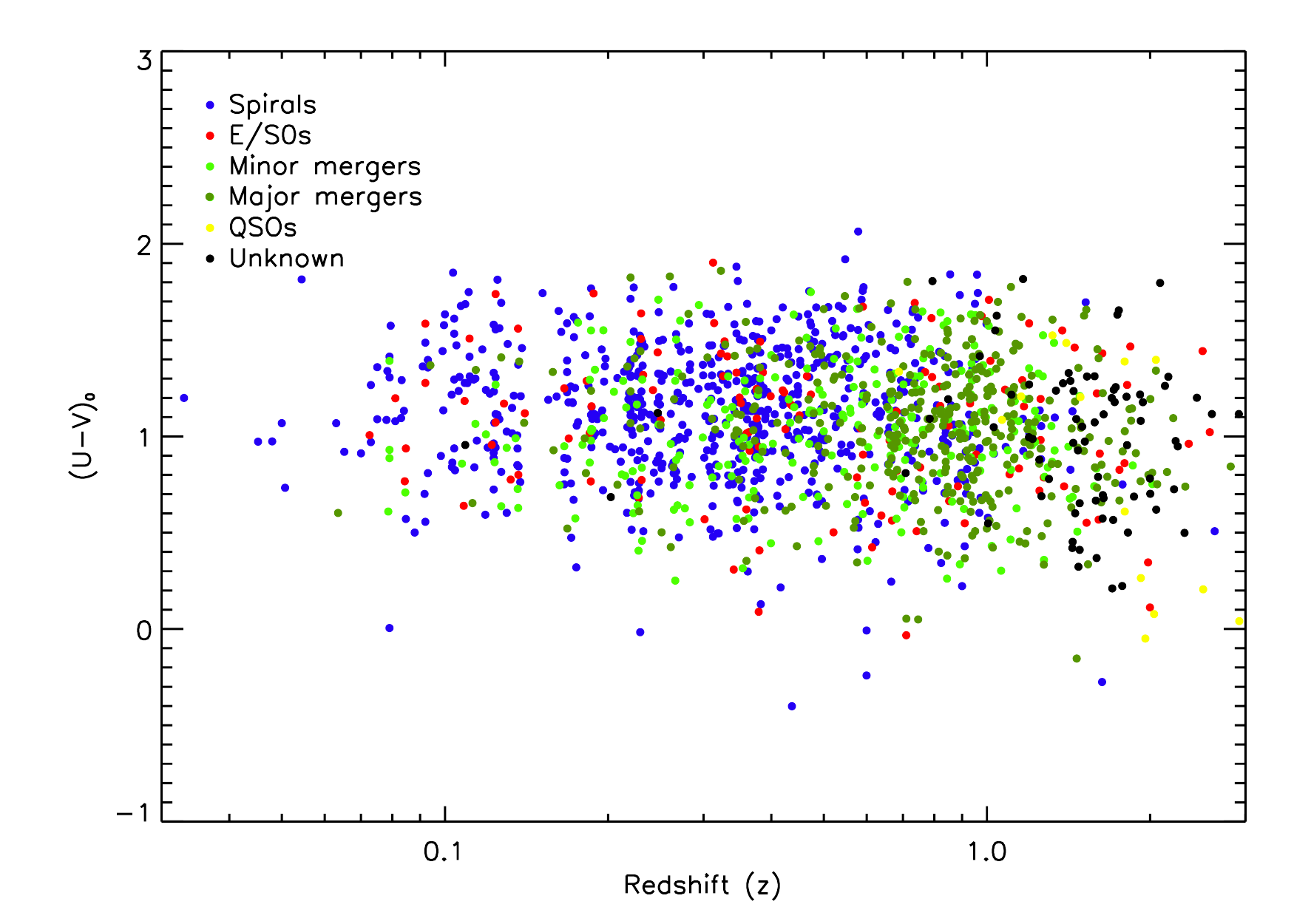}
\plotone{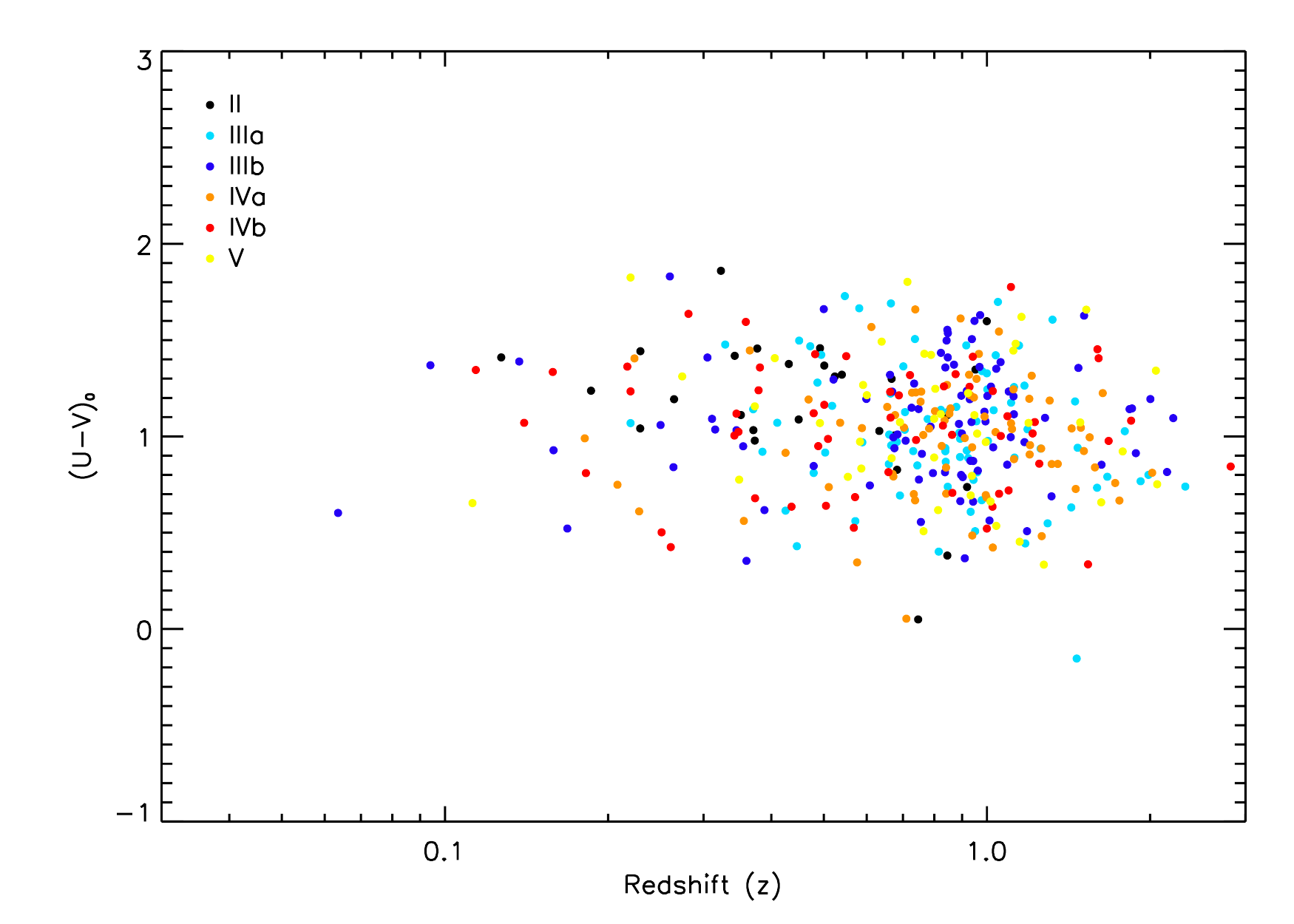}
\caption{Rest frame $U-V$ colors as a function of redshift color coded by morphology (top) and by interaction class for all major mergers (bottom). Note that the QSOs make up the blue tail of the distribution and that the spirals and ellipticals both span the entire range of colors. The major mergers are mostly in the green valley. }
\label{UV_z}
\end{figure}

Histograms for several redshift bins (including sources at all $M_{K}$ values) are shown in Figure~\ref{UV_bin1}, again color coded by morphology and with the $U-V$ distribution of the optical sample in the same redshift range normalized and plotted as the dashed line. The same general results are true at all redshifts. The 70\ts${\mu}$m selected galaxies do not follow the same color dichotomy as optically selected galaxies. Major mergers span a wide range in color but generally peak in the green valley ($U-V\sim 1.1$). Galaxies classified as unknown become significant at high redshift and they also span the entire color space. Galaxies where the QSO completely dominates (colored yellow on the histograms) tend to be the bluest objects in our sample and account for the very blue tail in ULIRGs and HyLIRGs. We plot the $U-V$ color as a function of redshift in Figure~\ref{UV_z} for all galaxies color coded by their morphology in the top panel and for all major mergers color coded by interaction class in the bottom panel. This figure clearly shows that spirals and ellipticals each span the full range in color and do not separate in color space by their morphology ($<U-V>=1.18$ and 1.16, respectively). The major mergers (colored dark green, $<U-V>=1.0$) have a sharper peak than the rest. The QSOs in the sample clearly separate toward the bottom high redshift portion of the plot ($<U-V>=0.05$). On average the galaxies appear to be bluer at high redshift ($z>1$) but the same trend is seen in the optically selected sample as well (see Fig.~\ref{UV_bin1}), which is consistent with observations of the evolution of colors with redshift \citep[e.g.,][]{Bell:2004p4626}. When separated by interaction class (bottom panel of Fig.~\ref{UV_z}), class V objects are on average slightly redder ($<U-V>=1.05$) than the class III ($<U-V>=1.0$) or class IV ($<U-V>=0.92$) objects.

It is also worth considering that many of the galaxies classified as major mergers are pairs of galaxies, some of which are blended in the optical catalogs. This means that for some objects the colors represent a single galaxy and for others the combination of both objects. To test how much of an effect this has on our results we combined the photometry and derived new colors for the separated pairs and compared the resulting distribution to that obtained for the individual objects. We find that the colors for the individual objects does not differ greatly from the combined color of the pair ($\Delta (U-V)<0.1$). This implies that there are very few mixed mergers in our sample (i.e., a merger between a spiral and an elliptical) and that the gas-rich nature of the progenitor galaxies is necessary for the increased infrared luminosity.

\section{Discussion}
\subsection{Comparison with Local (U)LIRGs}

Early results of morphological studies of infrared galaxies in the local universe from the BGS \citep{Sanders:1988p1635,Sanders:1996p1630,Ishida:2004p3661} found that the merger fraction increases systematically with $L_{\rm IR}$, from $\sim 10\%$ for objects with $10^{10.5}<L_{\rm IR}<10^{11}\ts L_{\odot}$ to 100\% for ULIRGs. We also see a systematic trend of an increased fraction of mergers in our higher redshift sample of COSMOS 70\ts$\mu$m selected sources, though the magnitude of this trend is not as high. Studies of local ULIRGs in particular agree that galaxy mergers are ubiquitous \citep{Sanders:1988p1635,Kim:1995p3875,Murphy:1996p4660,Veilleux:2002p920} and all but 1 of the 118 ULIRGs in the IRAS 1\ts Jy ULIRG sample show clear evidence of merging \citep{Veilleux:2002p920}. For our sample of 305 70\ts$\mu$m selected ULIRGs, 41\% are major mergers. Another 18\% are classified as unknown and are also possibly major mergers as indicated by the automated techniques discussed in \S3.5. Even if these objects are included, a maximum of 59\% of these high redshift ULIRGs are mergers. This fraction actually drops for the most extreme luminosities as only 32\% of HyLIRGs are mergers (as illustrated in Fig.~\ref{fraction}), though this is most likely due to the increase of QSOs (23\%).

LIRGs in the local universe show a wide range of morphologies. \cite{Ishida:2004p3661} found that for a complete sample of 56 BGS LIRGs, all objects with $L_{\rm IR}>10^{11.5}\ts L_{\odot}$ showed merger signatures while lower luminosity LIRGs were a mix of mergers (74\%) and single isolated disk galaxies (26\%).  This result generally agrees with that of \cite{Wang:2006p5068} who studied a sample of 159 local ($z<0.1$) LIRGs identified by cross referencing IRAS detected sources with the SDSS and found that 48\% of their sample were clearly interacting and merging systems with another 6\% possibly merging. Their sample is dominated by low luminosity LIRGs and the fraction of mergers goes up to $\sim 90\%$ above $L_{\rm IR}=4\times 10^{11}\ts L_{\odot}$. For our high redshift LIRGs we see a similar difference between low and high luminosity objects where 20\% of objects with $10^{11.0}<L_{\rm IR}<10^{11.5}\ts L_{\odot}$ are mergers while $\sim$ 40\% of objects with $10^{11.5}<L_{\rm IR}<10^{12.0}\ts L_{\odot}$ are. This is accompanied by a similar drop in the fraction of spirals and minor mergers. At the low luminosity end ($10^{10}<L_{\rm IR}<10^{11}\ts L_{\odot}$), $\sim 6\%$ of objects are major mergers, similar to objects locally \citep{Sanders:1996p1630}. However, it should be noted that it is very difficult to disentangle trends with $L_{\rm IR}$ from trends with redshift. The low luminosity objects in the sample are all at $z<0.3$ so it is perhaps not surprising that the fraction of mergers agrees with that locally.

If we consider the objects in a single redshift bin (to minimize the effects of redshift evolution) we find that for the 184 objects in the $0.8<z<1.0$ bin (see Fig.~\ref{lir} for the luminosity distribution) the fraction of major mergers rises from 47\% for $10^{11.5}<L_{\rm IR}<10^{12.0}\ts L_{\odot}$ to 52\% for $10^{12.0}<L_{\rm IR}<10^{12.5}\ts L_{\odot}$. These merger fractions are more reliable than the fractions for the entire sample since there are very few ``unknowns" in this redshift bin. The fraction of minor mergers remains roughly constant. There are only 7 galaxies with $L_{\rm IR}>10^{12.5}\ts L_{\odot}$ and all but one ($\sim 85\%$) of these are major mergers.  The lower redshift bin ($0.4<z<0.6$) probes objects at $10^{10.5}<L_{\rm IR}<10^{12.0}\ts L_{\odot}$ (though very few at the highest luminosities) and shows a rise in major mergers from 9\% to 19\% to 24\% for $10^{10.5}<L_{\rm IR}<10^{11.0}\ts L_{\odot}$, $10^{11.0}<L_{\rm IR}<10^{11.5}\ts L_{\odot}$, and $10^{11.5}<L_{\rm IR}<10^{12.0}\ts L_{\odot}$, respectively. At $1.2<z<1.4$ major mergers make up $\sim 60\%$ of ULIRGs.

A comparison of our results to the theoretical predictions of \cite{Hopkins:2010p7149} show general agreement. They find that at all redshifts, the high-luminosity population of galaxies becomes increasingly dominated by mergers but that the transition luminosity between normal galaxies and mergers changes between $z\sim 0$ and $z\sim 2$ from $\sim 10^{12}\ts L_{\odot}$ to $\sim 10^{13}\ts L_{\odot}$. While our data at $z\sim 2$ are sparse and the morphology is difficult to determine in our images at these high redshifts, our results at $z\sim 1$ would be consistent with such an evolution. Further work  with detailed morphologies covering wider range of luminosities at each redshift is needed to pin down if such an evolution is truly present.

Studies in the local universe also find a trend in merger phase with $L_{\rm IR}$ (and nuclear separation). \cite{Veilleux:2002p920} found that 56\% of ULIRGs in the 1-Jy sample are late stage mergers with a single disturbed nucleus (class IV and V) while 39\% had two nuclei but with tidal tails, indicating they were past first passage (class III). They found none at the earliest stages of merging (classes I and II). Pairs and double nuclei amongst local ULIRGs have a mean projected nuclear separation of $2-3$\ts kpc though they generally range from $<0.3$ to 10\ts kpc with a tail out to $20-40$\ts kpc \citep{Sanders:1996p1630,Murphy:1996p4660, Veilleux:2002p920}. \cite{Ishida:2004p3661} found that LIRGs span a wide range in nuclear separations from 5 to 50 kpc with the mean separation increasing while $L_{\rm IR}$ decreases. The results presented in Table~\ref{mergers} indicates that a large fraction of COSMOS ULIRGs are also late stage mergers (classes IIIb, IV, and V) but a significant fraction ($\sim 40\%$) are pairs at separations of $> 10$\ts kpc (i.e., class IIIa). 

Numerical simulations of equal mass mergers \citep{Mihos:1996p4245} show that there are two episodes of star formation activity tied to the phase of interaction that can lead to an elevation in infrared luminosity. The relative strengths of the two bursts depend on several factors including the orbital geometry of the encounter and the morphological structure of the progenitor galaxies. In models that include a disk, bulge, and halo, the peak of star formation activity occurs in the late stages of a galaxy merger just before and after final coalescence (nuclei at this stage would be within a few kpc of each other), $\sim 10^{9}$\ts yr after the first encounter. This is qualitatively consistent with the (U)LIRGs observed in interaction classes IV and V in our sample as well as most of the local ULIRGs. A peak in star formation activity also occurs shortly after the first close encounter and is actually larger in models involving a disk and halo only. At this stage, $\sim 1-2\times 10^{8}$ yr after the first passage, the two galaxies can actually be separated by tens of kpc. It is possible that the majority of (U)LIRGs in our sample, those with separations greater than a few kpc, are at this stage of the merger with this burst of star formation enhancing the infrared luminosity. This could also apply to the minority of local ULIRGs (and most LIRGs) with large separations (as suggested by \citeauthor{Veilleux:2002p920}). However, the difference in the fraction of early stage mergers between low and high redshift still remains unexplained.  One possibility is that the higher gas content of high redshift gas-rich spirals enables the first peak in star formation activity to trigger the ULIRG stage whereas locally it tends to happen during the second peak (with the first peak triggering the LIRG stage). The drop in the fraction of class IIIa mergers and the increase in the fraction of late stage (class IIIb, IV, and V) mergers for HyLIRGs could be an indication that these extreme luminosities are only reached at final coalescence.

It is also possible that multiple mergers can play a role in triggering ULIRG activity. Approximately 6\% of the ULIRGs and 3\% of the LIRGs in our sample appear to be part of a small group. It is possible that previous interactions with other group members, or past mergers, could have lead to an increase in infrared activity.  \cite{Ishida:2004p3661} finds some evidence for minor mergers for low luminosity LIRGs and our findings indicate that 16\% of objects with $10^{11.0}<L_{\rm IR}<10^{11.5}\ts L_{\odot}$ are minor mergers.  There is a significantly higher fraction  of LIRGs that are minor mergers (24\%) compared to optically selected galaxies in our comparison sample (13\%) in the $0.4<z<0.6$ redshift bin, suggesting that minor mergers could have a real effect at these luminosities. Finally it is worth noting that the numerical simulations mentioned here do not take into account the role of AGN activity in contributing to the total infrared luminosity.  We know that the fraction of objects hosting an AGN increases very strongly with $L_{\rm IR}$, both locally and at high redshift  (e.g., \citealt{Veilleux:1995p2081} and Paper I). The role of AGN in the sample will be discussed in more detail in \S5.3.

\subsection{Previous High Redshift Studies}

Our results show that LIRGs at intermediate redshifts ($0.2<z<1.2$) are 20\% major mergers at $L_{\rm IR}<10^{11.5}\ts L_{\odot}$ and 40\% major mergers at  $L_{\rm IR}>10^{11.5}\ts L_{\odot}$. Minor mergers also play a role (20\%). This result is fairly consistent with previous LIRG studies at these redshifts.  These studies have found a range in merger fractions from $\sim 50\%$ (\citealt{Zheng:2004p4983}; \citealt{Melbourne:2005p1175} (at the low-z end); \citealt{Shi:2006p4394,Shi:2009p4334}) to fairly low fractions (37\%: \citealt{Flores:1999p4815}; 28\%: \citealt{Bell:2005p4631}; 30\%: \citealt{Melbourne:2005p1175} (at the high-z end); 26\%: \citealt{Bridge:2007p1511}; 30\%: \citealt{Melbourne:2008p944}; 15\%: \citealt{Lotz:2008p347}). There are several possible explanations for the seemingly discrepant results among these studies. The first is that while they all focus on LIRGs, some actually probe slightly different $L_{\rm IR}$ ranges. For example, \cite{Bell:2005p4631} looks at all galaxies with $L_{\rm IR}> 6\times 10^{10}\ts L_{\odot}$ and finds 28\% are mergers, however, they also find that this fraction increases as a function of $L_{\rm IR}$ though the numbers are not given. \cite{Bridge:2007p1511} also obtain a merger fraction, selected morphologically, using a wider range of infrared luminosities (lower limit of their sample is $5\times 10^{10}\ts L_{\odot}$). They also find 25\% of their objects are in pairs, though the overlap with the morphologically selected mergers is unclear, so the actual fraction of mergers in the sample may be higher than 26\%. The range of luminosities probed by these studies is likely exacerbated by the selection at 24\ts$\mu$m. As discussed in Paper I, 24\ts$\mu$m selection introduces a larger error to the estimate of $L_{\rm IR}$ than the 70\ts$\mu$m selection used here. Many of these studies use small samples of a few tens of galaxies (with the exception of \citealt{Melbourne:2005p1175}, \citealt{Bridge:2007p1511}, and \citealt{Shi:2009p4334}) so small number statistics may also play a role. The result of \cite{Lotz:2008p347} stands out as the lowest merger fraction ($\sim 15\%$) among (U)LIRGs in the literature. They use the Gini/$M_{20}$ criteria described in \S3.5 to identify potential mergers. As discussed in \cite{Lotz:2008p4477}, this technique only selects galaxies at particular times during the merger process (i.e., at first passage and at final coalescence) so, in our opinion, it cannot be used to obtain a complete census of galaxy mergers at all stages.

The fraction of identifiable major mergers among our high-redshift ULIRGs ($z\sim 1-2$) is 30--40\%. Other high redshift ULIRG studies find similar results \citep[e.g.,][]{Dasyra:2008p1540}. In a survey of 31 dust obscured galaxies (DOGs, \citealt{Dey:2008p2965}) from the Bo\"{o}tes Field, \cite{Bussmann:2009p5135} find that they are more relaxed than local ULIRGs and are possibly at a post-merger stage. They also find that many of the objects in their sample have power-law SEDs, consistent with the presence of an AGN. \cite{Melbourne:2009p5148} obtained Keck adaptive optics imaging of 15 DOGs and found that only two show evidence for possible mergers. 

In Paper I we determined which of our 70\ts$\mu$m selected sources had AGN signatures using several different methods (X-ray detection, IRAC color selection, power-law SED shape, high infrared to optical ratios, and radio excess). In the next section we discuss the morphological properties of our sample in the context of the presence of an AGN.

\subsection{Merger Fraction Among AGN}

\begin{figure}
\epsscale{1.2}
\plotone{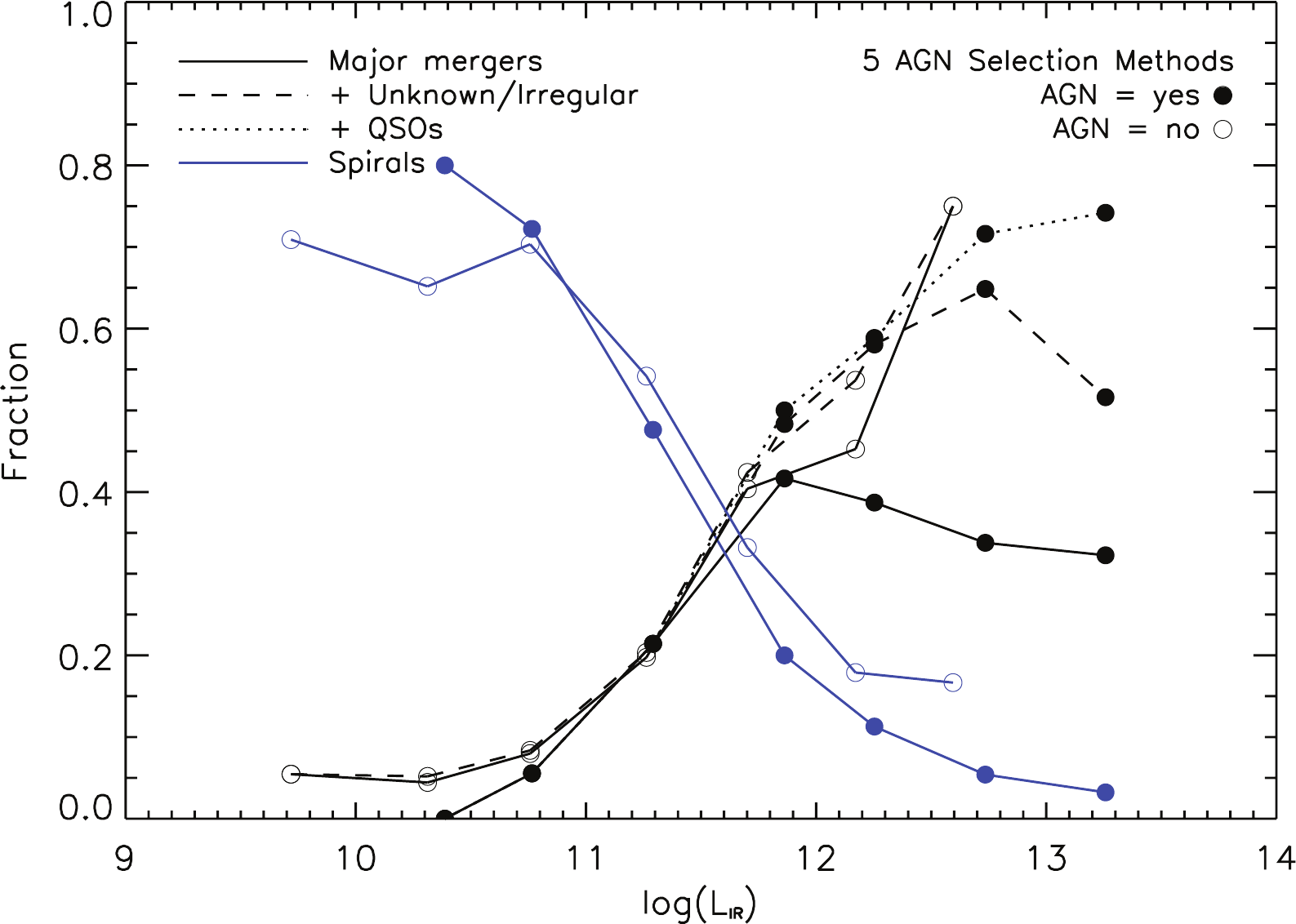}
\caption{Same as Fig.~\ref{fraction} except divided into two sub-samples: sources that are detected as possible AGN using one of five different selection methods and those that are not (see text for details). The plot shows the fraction of sources with (filled circle) or without (open circles) any of these AGN signatures that are morphologically classified as spirals (blue solid line) or major mergers (black solid line), as well as the cumulative fraction of sources that are either a major merger or unknown/irregular (dashed line) or  either a major merger, unknown/irregular, or a QSO (dotted line). None of the sources without an AGN signature fall into the QSO classification. The sources classified as ellipticals and minor mergers are not shown and make up the remaining fraction of sources (hence the fractions in a given luminosity bin do not sum up to 100\%). A strong trend with $L_{\rm IR}$ can be seen for both sub-samples, in particular, the fraction of spirals decreases and the fraction of major mergers increases for both. For ULIRGs, it appears that a higher fraction of the objects without a signature of an AGN are major mergers, however, a larger fraction of AGN are classified as unknown and as QSOs. All of the HyLIRGs in the sample have AGN signatures and 20\% of them are QSOs.}
\label{fraction_agn}
\end{figure}

Figure~\ref{fraction_agn} shows the fraction of galaxies in the 70\ts$\mu$m sample that are spirals (blue) and major mergers (black) split by whether (filled circles) or not (open circles) they exhibit AGN signatures using any one of the five selection methods described in Paper I.  Also shown is the cumulative contribution from galaxies classified as unknown or QSOs. Both sub-samples show a strong trend of an increase in the fraction of major mergers and decrease in the fraction of spirals with $L_{\rm IR}$. In fact, ULIRGs without an AGN signature have a higher fraction of major mergers (up to $\sim 75\%$) than those with an AGN ($\sim 40\%$). However, the sources with an AGN have a much larger fraction of galaxies classified as unknown ($20-30\%$ versus 5\%) and also have a significant fraction that are QSOs ($>5\%$ of ULIRGs and 23\% of HyLIRGs). Since all of the HyLIRGs in our sample show AGN signatures it is possible that the presence of an AGN (and in many cases a QSO) is required for a galaxy to reach these extreme IR luminosities.

\subsection{The Green Valley}

Our results presented in \S4 have shown that our sample of 70\ts$\mu$m sources peaks strongly in the green valley, in between the blue cloud of star forming galaxies and the quiescent red sequence. However, there are several possible interpretations for the cause of the reddened galaxy colors in the green valley. The first is that the truncation of star formation in these galaxies indicates that they are truly transition objects between strongly star forming blue cloud galaxies and ``red and dead" red sequence galaxies. In such a scenario it is also possible that the color can vary over the galaxy (e.g., red cores surrounded by blue outer regions). Our colors represent a combined color for the entire galaxy and the reduced resolution at high redshift makes looking at the spatial color variations difficult. However, local samples of galaxies in the green valley should provide an excellent way to probe this question. Studies of local (U)LIRGs have found that they are as blue as the general blue cloud populations (with a few exceptions) when looking at a galaxy's total integrated light, but that there are dramatic color gradients within the galaxy (\citealt{Chen:2010p7052}; Larson \etal in preparation). 

Dust extinction can also have a large effect on the galaxy colors and most of the galaxies in the green valley may simply be highly reddened star forming galaxies. Indeed, \cite{Salim:2009p4834} found that when correcting for dust extinction among a sample of infrared galaxies ($10^{8}<L_{\rm IR} < 10^{12}\ts L_{\odot}$) from the Far-Infrared Deep Extragalactic Legacy Survey (FIDEL) that most were actually star forming blue cloud galaxies. However, some were intrinsically red even after the dust correction. Similarly, \cite{Kaviraj:2009p4849} found that almost all of the objects in a sample of 561 low redshift ($z<0.2$) LIRGs were in the blue cloud but that those with AGN were on average redder while they were not any dustier than those without AGN. These studies show that dust extinction plays a role but cannot be the only factor. 

\begin{figure}
\epsscale{1.2}
\plotone{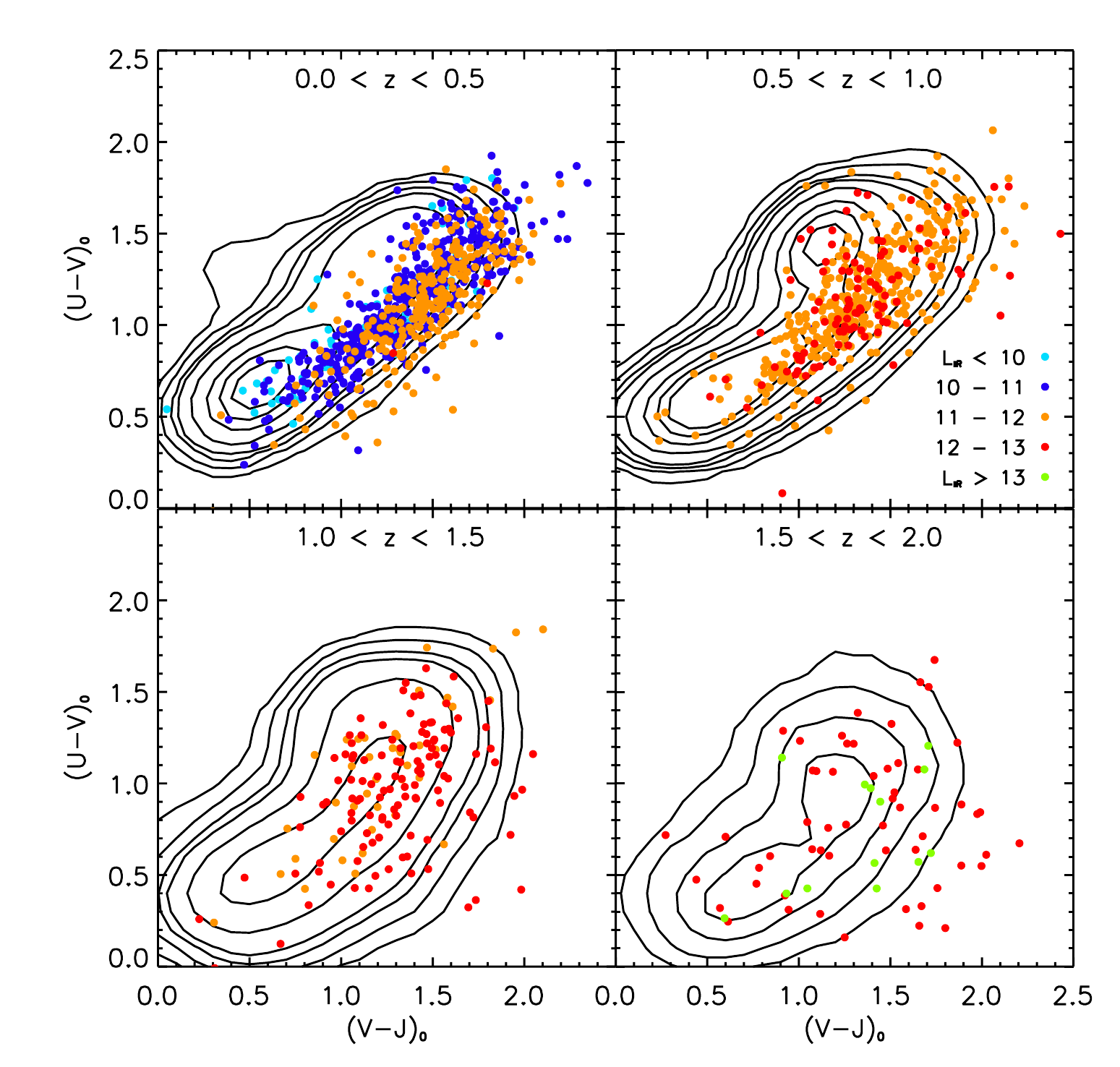}
\caption{Color-color diagram ($U-V$ versus $V-J$) illustrating the position of the 70\ts$\mu$m selected sources relative to the optically selected population in four different redshift bins. The 70\ts$\mu$m sources are color coded by their $L_{\rm IR}$ and the optically selected galaxies are represented by contours. The contour levels represent the number of objects in $U-V$ and $V-J$ bins of 0.1 with levels of 10, 20, 35, 50, 100, 200, 300, and 400 objects. Many of the 70\ts$\mu$m sources follow the star forming sequence with scatter increasing with $L_{\rm IR}$ (and redshfit).}
\label{UVJ_LIR}
\end{figure}

To explore the role of dust extinction, we plot the position of the 70\ts$\mu$m sources on a color-color diagram ($U-V$ vs $V-J$) in Figure~\ref{UVJ_LIR}. Star forming galaxies on this diagram follow a diagonal sequence where objects with increasing $V-J$ are redder due to dust \citep[e.g.][]{Wuyts:2007p7057,Wuyts:2009p7059,Williams:2009p7114}. Quiescent galaxies lie in a clump above this sequence with red $U-V$ colors but bluer $V-J$ colors. The contours in Figure~\ref{UVJ_LIR} illustrate the colors of the optically selected population in each of the four redshift bins. While the star forming sequence can be easily seen in each redshift bin, the quiescent clump can only be clearly seen in the $0.5<z<1.0$ bin. Many of the 70\ts$\mu$m sources, color coded by their $L_{\rm IR}$, appear to follow the star forming sequence, indicating that they are indeed reddened by dust and dustier than most of the normal blue cloud galaxies. The presence of so many galaxy mergers with apparently more dust than the blue cloud (where the progenitor galaxies presumably resided) suggest that the merger itself may be responsible for kicking up the dust responsible for the redder colors. While very few of the 70\ts$\mu$m sources lie in the quiescent area of the diagram (a few can be seen in the $0.5<z<1.0$ redshift bin), there is a lot of scatter that increases with $L_{\rm IR}$ (and redshift). At high redshift, there are many extremely blue sources -- almost all of these are AGN. The increased scatter toward redder $U-V$ colors above the star forming sequence may be an indication that these sources are transitioning toward the red sequence. Further work is needed to estimate the dust corrected star formation rates of these galaxies to determine the cause of the increased scatter.

Most studies of galaxies in the green valley have focused on AGN host galaxies because many have been found to reside there \citep[e.g.,][] {Sanchez:2004p4852,Schawinski:2007p4883,Salim:2007p4832,Nandra:2007p4895,Westoby:2007p4897,  Georgakakis:2008p4900,Silverman:2008p4951,Silverman:2009p2150,Hickox:2009p4151, Schawinski:2009p4884,Smolcic:2009p4902}. This result is often cited as evidence for AGN feedback as a mechanism for truncating star formation. Indeed, \cite{Westoby:2007p4897} find that Type II AGN identified spectroscopically in the SDSS reside predominantly in the red sequence with a tail that stretches through the green valley to the edge of the blue cloud while AGN/star forming composite galaxies peak on the blue edge of the red sequence with a significant number extending into the green valley indicating that composite galaxies represent a transition stage between the blue cloud and the red sequence. Likewise, \cite{Smolcic:2009p4902} finds evidence for a transition class of objects among high excitation radio AGN. 

\begin{figure}
\epsscale{1.2}
\plotone{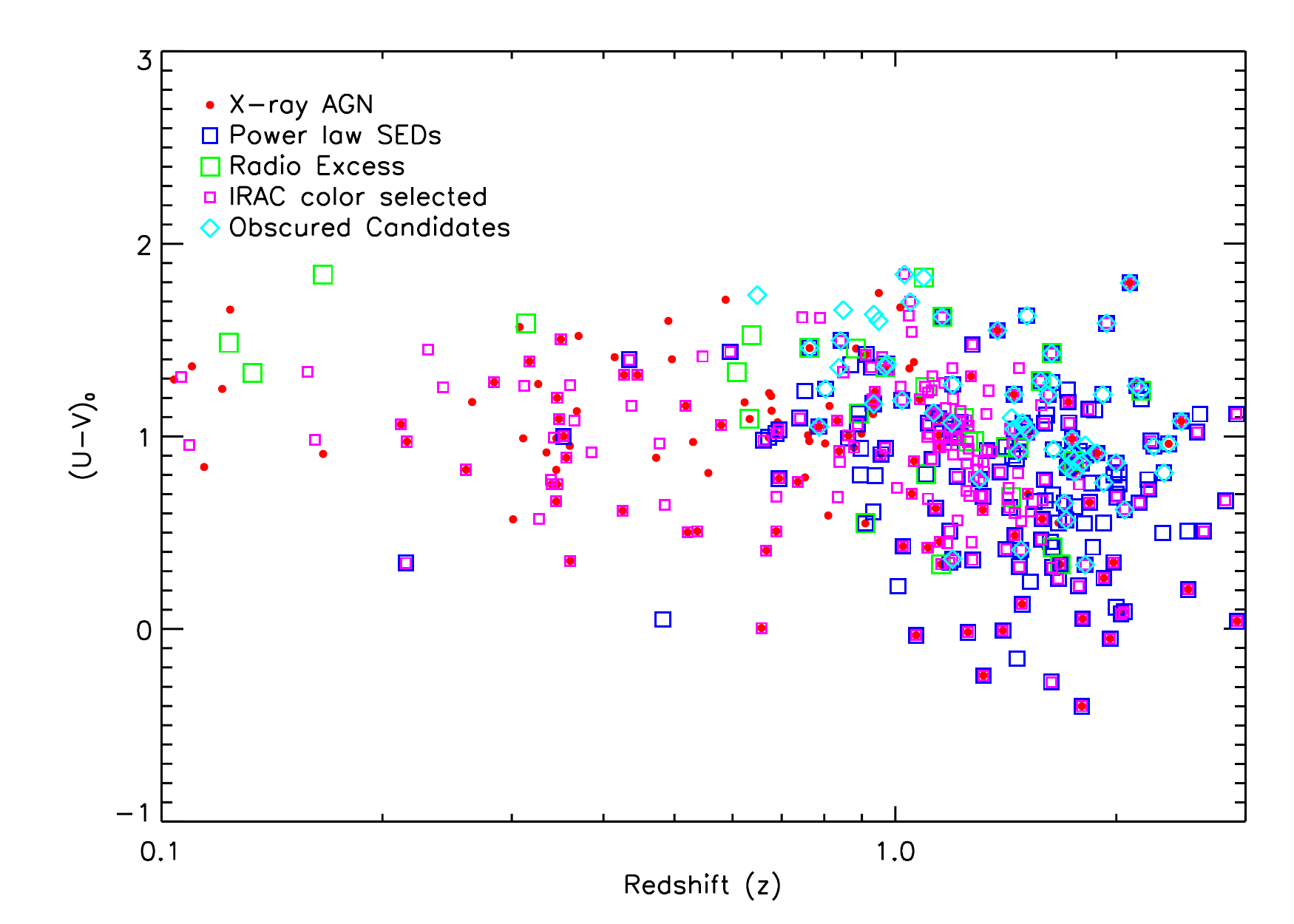}
\caption{Same as Fig.~\ref{UV_z} but only for the 355 AGN identified in the sample coded by their selection technique. The radio excess AGN and the obscured AGN candidates are on average redder than the rest of the AGN in the sample. This is consistent with these objects being red-sequence radio AGN and and highly dust obscured, respectively. }
\label{UV_agn}
\end{figure}

In Figure~\ref{UV_agn} we once again plot the $U-V$ colors as a function of redshift, but this time only for the various AGN in our sample of 70\ts$\mu$m sources color coded by their selection techniques (see Paper I for a complete description of the selection). The AGN appear to peak even more strongly in the green valley than the galaxies without AGN signatures in the sample do. The radio excess AGN (green) and the obscured AGN candidates (cyan, $\rm F(24\ts {\mu} m)/F(R) > 10^3$) are redder than the rest of the AGN sample, consistent with them being red sequence radio AGN and highly dust obscured, respectively. There is no evidence to suggest that the rest of the AGN in our sample are dustier than those without AGN signatures and indeed, the typical $E(B-V)$ for both sub-samples are practically identical ($\sim 0.3$).

\begin{figure}
\epsscale{1.2}
\plotone{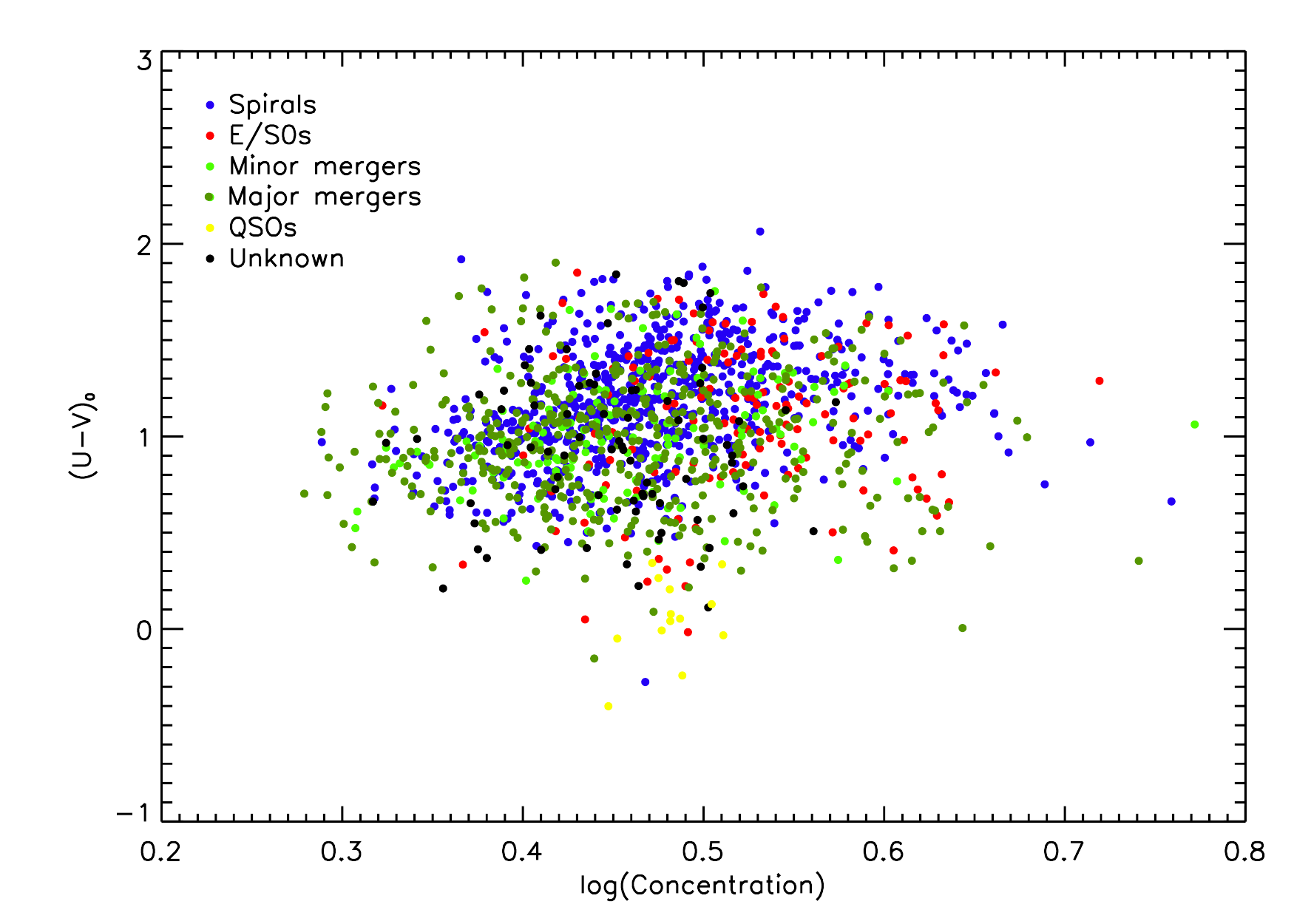}
\plotone{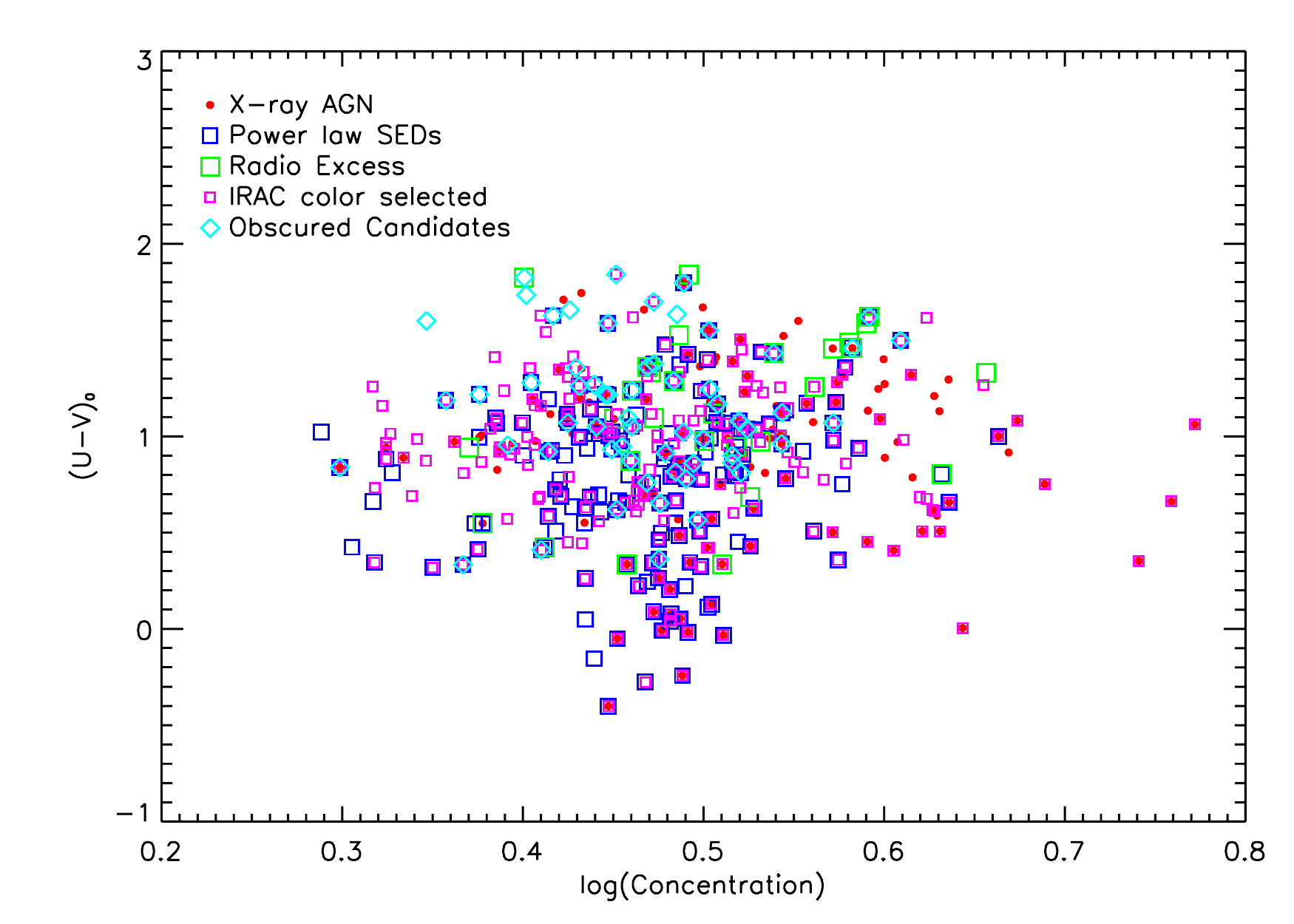}
\caption{Rest frame $U-V$ colors as a function of the concentration morphological parameter for all 1503 70\ts$\mu$m selected sources color coded by morphology (top) and for all 355 AGN in the sample color coded by their selection method (bottom). Note that there is a slight trend for more concentrated galaxies to be redder in color.}
\label{UV_con}
\end{figure}

We further explore the evolution of these objects by plotting their $U-V$ color as a function of the morphological concentration parameter (Fig.~\ref{UV_con}). We find that there is a slight trend for more highly concentrated galaxies to be redder in color as expected for galaxies that are on the way to becoming red elliptical galaxies. This suggests the possibility that the galaxies classified as ellipticals in our sample are actually merger remnants with either some remaining star formation activity or an AGN  providing the energy for the observed infrared emission. The bottom panel shows the AGN in the sample with the same color codes as Figure~\ref{UV_agn}. The trend of redder colors with more concentrated objects appears strongest for the power law selected AGN and the X-ray detected AGN have the highest concentrations on average.

\subsection{Elliptical Galaxy Formation}

Several studies have found that the stellar mass in the red sequence has doubled since $z\sim1$ \citep{Bell:2004p4626,Faber:2007p4912} and this growth of the red sequence has become a subject of much research and debate. It has become clear that passive evolution of disk galaxies is not likely to be the explanation since most galaxies on the red sequence have spheroidal morphologies \citep{Ilbert:2010p6983}. Major mergers have been shown to be capable of producing spheroidals \citep{Toomre:1972p4262,Barnes:1992p4505} but the question remains whether or not there are enough mergers to form the red sequence. It has been suggested however, that while gas-rich mergers may account for the formation of the red sequence at the low mass end there are not enough high luminosity gas-rich systems to form the present day high mass ($\gtrsim L^{\star}$, not including the largest cD ellipticals which likely gain mass through the accretion of many smaller objects over time) end of the red sequence \citep[e.g.,][]{Bell:2004p4626,Weiner:2005p4632,Willmer:2006p4920,Faber:2007p4912}. In this case, it becomes necessary to invoke a contribution from gas-poor mergers (so-called ``dry mergers") to explain the formation of the red sequence \citep[e.g.,][]{vanDokkum:2001p4924,Bell:2006p4630}.

Gas-rich mergers that are massive enough to produce high mass ellipticals should be bright in the infrared and therefore should be detected by infrared surveys. Our large sample of 70\ts$\mu$m sources allows us to uniquely explore the question of whether there are enough gas-rich mergers taking place to form the observed red sequence.  Let us consider if it is possible that gas-rich mergers could have formed the red-sequence galaxies observed in a single redshift and $M_{K}$ bin. For example, in the $0.8<z<1.0$ bin, and the interval $-23.5<M_{K}<-23.0$, there are $\sim 780$ red sequence galaxies. In the same redshift and $M_{K}$ bins we observe $\sim 50$ (U)LIRGs with $L_{\rm IR}>10^{11.5}\ts L_{\odot}$. The timescale on which a single major merger with a typical nuclear separation of 7\ts kpc would be observed with  $L_{\rm IR}>10^{11.5}\ts L_{\odot}$ is $\sim 1.4\times 10^8$ years given that $\Delta V\sim 100$\ts km\ts s$^{-1}$ between the two nuclei (typical velocity observed between close galaxy pairs in our sample) and assuming it takes the same amount of time for the source to reach its peak luminosity and then fade to below  $L_{\rm IR}=10^{11.5}\ts L_{\odot}$. In order to have formed all 780 ellipticals currently observed, it would have taken $780/50 \times (1.4\times 10^8)\ts \rm yr \sim 2.2\times 10^9\ts \rm yr$.  This is certainly plausible if all of the ellipticals observed in this bin formed between $z\sim1.5$ and 1. This assumes that the number of (U)LIRGs is constant over this redshift interval. If one takes into account the evolution of the (U)LIRG luminosity function ($\sim (1+z)^{3-5}$, e.g., \citealt{Kim:1998p3280,LeFloch:2005p2544,Caputi:2007p2597,Magnelli:2009p2619}) the timescale to have formed the observed massive ellipticals would be even shorter.  This simple argument implies that enough gas-rich mergers take place to form the observed red sequence at stellar masses $< 10^{12}\ts M_{\odot}$, consistent with what one would expected based on \cite{Kormendy:2009p7116} and \cite{vanderWel:2009p7121}.

\cite{Ilbert:2010p6983} have found that the high mass end of the red sequence is already in place by $z\sim1$ and that a transition between the dominance of star forming and quiescent galaxies takes place around this redshift. If dry mergers were responsible for the formation of the high mass end of the red sequence then we would expect there to be significant evolution in the high mass end between $z=1$ to the present day, which we do not see. The observed luminosity function for infrared galaxies implies that (U)LIRGs are much more common at high redshifts and with LIRGs dominating beyond $z=0.7$ \citep{LeFloch:2005p2544} and ULIRGs by $z\sim 2$ \citep{Caputi:2007p2597,Magnelli:2009p2619}. At these redshifts there are enough gas-rich mergers taking place to form the red sequence at the peak of massive elliptical galaxy formation.

\section{Summary}

We have presented visual morphological classifications for each object in our complete sample of 1503 70\ts$\mu$m selected sources in the COSMOS field and have compared them to 1) comparison galaxies selected in the optical, 2) automated classification techniques, and 3) morphologies of (U)LIRGs in the local universe. We have also presented the color distribution for the sample. From the analysis of these properties, the following conclusions can be drawn:

\begin{enumerate}

\item{The sample of 70\ts$\mu$m sources span the redshift range $0.01<z<3.5$ with a median redshift of 0.5 and an infrared luminosity range of $10^{8} < L_{\rm IR} < 10^{14}\ts L_{\odot}$ with a median luminosity of $10^{11.4}\ts L_{\odot}$. These sources are all massive, spanning a stellar mass range of $\sim 10^{10}-10^{12}\ts \rm M_{\odot}$ and luminous with $-25 < M_{\rm K} < -20$\ts.}

\item{The fraction of major mergers increases systematically with increasing $L_{\rm IR}$ while the fraction of spirals shows a systematic decline with increasing $L_{\rm IR}$. At the highest luminosities ($>10^{12}\ts L_{\odot}$, where the majority of sources are at $z>1$), sources still identifiable as major mergers  represent $\sim50\%$ of the sample and spirals comprise $<10\%$.  Additionally, a high fraction of sources at these high luminosities and redshifts are classified as unknown ($\sim 20\%$) or QSOs ($\sim20\%$) and could possibly be mergers as well. Minor mergers represent a significant fraction of sources at low luminosity ($L_{\rm IR}<10^{11.5}\ts L_{\odot}$). Sources identified as AGN constitute a large fraction ($>70\%$) of sources at high luminosity ($L_{\rm IR}>10^{12}\ts L_{\odot}$).}

\item{A comparison of our visual classification to several automated classification techniques commonly used in the literature (e.g., concentration, asymmetry, clumpiness, Gini, and $M_{20}$) shows that while there is some broad agreement between visual and automated morphological classes there is still substantial overlap between adjacent classifications. In particular, none of the automated techniques is sensitive to major mergers at all phases, and therefore any selection using these automated methods will be incomplete. Visual classification, while time consuming, is still the most robust method for identifying merger signatures.}

\item{The $U-V$ colors of the 70\ts$\mu$m sources peaks in the ``green valley" ($<U-V>= 1.1$), but with a broad spread of colors extending into the ``blue cloud" ($<U-V>\sim 0.7$) and ``red sequence" ($<U-V>\sim1.5$). We find that the galaxies classified as major mergers and the sub-sample classified as AGN peak more strongly in the green valley than the rest of the morphological classes.}

\item{We have shown that given the relatively short timescale ($\lesssim 2\times10^{8}$\ts yr) of the (U)LIRG phase of a major merger, the number of observed gas-rich major mergers is sufficient to account for the formation of the red-sequence ($< 10^{12}\ts M_{\odot}$) without the need to invoke dry mergers. This argument becomes even more plausible once the evolution in the number density of (U)LIRGs over the redshift range $z=1-3$ is taken into account, which has been shown to evolve quite strongly for these luminous sources.}

\end{enumerate}

\acknowledgments

Support for this work was provided in part by NASA through contracts 1282612, 1298213 and 1344920 issued by the Jet Propulsion Laboratory. This work was supported in part by a grant from The City University of New York PSC-CUNY Research Award Program. We would also like to recognize the contributions from all of the members of the COSMOS Team who helped in obtaining and reducing the large amount of multi-wavelength data that are now publicly available through the NASA Infrared Science Archive (IRSA) at http://irsa.ipac.caltech.edu/Missions/cosmos.html. The analysis pipeline used to reduce the DEIMOS data was developed at UC Berkeley with support from NSF grant AST-0071048. This research has made use of the NASA/IPAC Extragalactic Database (NED) which is operated by the Jet Propulsion Laboratory, California Institute of Technology, under contract with the National Aeronautics and Space Administration. This research has also made use of data from the Sloan Digital Sky Survey (SDSS-DR7). Funding for the SDSS and SDSS-II has been provided by the Alfred P. Sloan Foundation, the Participating Institutions, the National Science Foundation, the U.S. Department of Energy, the National Aeronautics and Space Administration, the Japanese Monbukagakusho, the Max Planck Society, and the Higher Education Funding Council for England. The SDSS Web Site is http://www.sdss.org/. We thank the anonymous referee for their thorough comments, which greatly improved the paper. We also thank Marco Barden and Knud Jhanke for the use of the FERENGI redshifting code as well as their help with using it.

\clearpage

%\begin{landscape}
\begin{deluxetable*}{lccccccccccccccccc}
 \tabletypesize{\footnotesize}
  \tablecaption{Morphological catalog of COSMOS 70\ts$\mu$m Sources \label{catalog}}
 \tablewidth{0pt}
\tablehead{
\colhead{ID} & 
\colhead{ \hspace*{0.1in} Name} &  
\colhead{$L_{\rm IR}$\tablenotemark{a}} & 
\colhead{Redshift} & 
\colhead{Flag\tablenotemark{b}} & 
\colhead{Morphological\tablenotemark{c}} & 
\colhead{Nuclear\tablenotemark{d}} & 
\colhead{Mass\tablenotemark{e}} & 
\colhead{Mass\tablenotemark{f}} & 
\colhead{M$_{\ts K}$\tablenotemark{g}} & 
\colhead{$U-V$\tablenotemark{h}} &
\colhead{C\tablenotemark{i}} & 
\colhead{A\tablenotemark{j}} & 
\colhead{Clumpiness} & 
\colhead{Gini} &
\colhead{M$_{20}$} & 
\colhead{Auto\tablenotemark{k}} \\
\colhead{} & \colhead{} & \colhead{} & \colhead{} & \colhead{} & \colhead{Class} & \colhead{Separation} & \colhead{Ratio} & \colhead{} & \colhead{} & \colhead{color}}
\startdata
 1017 & \hspace*{-1.1in} COSMOS J100046.07+013439.89 &  $11.16\pm0.25$ 	& 0.518 	& I 	& MIIIa  	& 15.4 	& 1.9 	& 10.72 	&  -22.13	& 1.16	& 2.57 & 0.26 & -0.03 & 0.43 & -1.60 & 3\\
 1077 & \hspace*{-1.1in} COSMOS J100034.22+013543.18  & $10.87\pm0.22$ 	& 0.23	& P 	& S, m	& \nodata 	& \nodata	& 10.74 	& -21.98 	& 0.68	& 2.79 & 0.11 & -0.07 & 0.49 & -1.90 & 2\\
 1081& \hspace*{-1.1in} COSMOS J100133.37+013547.59  &  $10.44\pm0.22 $ 	& 0.23	& P 	& MIIIa, G 	& 6.4 	& 5.5		& 10.07 	& -20.32 	& 0.41	& 3.41 & 0.11 & -0.05 & 0.50 & -1.81 & 2\\
 1082 & \hspace*{-1.1in} COSMOS J100041.55+013552.68  & $11.26\pm0.23 $	& 0.37	& P 	& MIVb 	& \nodata 	& \nodata	& 11.03 	& -22.81 	& 0.68 	& 3.55 & 0.22 & -0.08 & 0.55 & -1.94 & 3\\
 1083 & \hspace*{-1.1in} COSMOS J100241.78+013550.07  &  $12.62\pm0.32 $ 	& 1.20	& P	& U	 	& \nodata 	& \nodata 	& 10.55 	& -22.17 	& 0.99 	& 2.70 & 0.33 &  0.37 & 0.38 & -1.58 & 3
\enddata

\tablecomments{Table~\ref{catalog} is published in its entirety in the online edition of the article. A portion is shown here for guidance regarding its form and content.}

\tablenotetext{a}{$\rm log(L_{\rm IR}/L_{\odot})$: total infrared luminosity derived from SED template fitting described in Paper I}
\tablenotetext{b}{Redshift Flag --- D: DEIMOS spectroscopy on Keck II, Z: zCOSMOS spectroscopy from VLT/VIMOS, I: IMACS spectroscopy from Magellan, S: SDSS spectroscopy, 2dF: 2dFGRS spectroscopy, F: FORS1 spectroscopy, M: Hectospec spectroscopy from MMT, P: Photometric redshift from \cite{Ilbert:2009p2146} or \cite{Salvato:2009p2142}}
\tablenotetext{c}{Morphological classification and interaction class --- S: spiral, E: elliptical,  Q: QSO, m: minor merger, M: Major merger, U: unknown,  G: group, I--V: interaction classes. See text for details}
\tablenotetext{d}{Nuclear separation between galaxy pairs and double nuclei in kpc.}
\tablenotetext{e}{Ratio of the galaxy stellar masses for galaxy pairs with resolved optical photometry.}
\tablenotetext{f}{$\rm log$(stellar mass$/M_{\odot}$) determined from the photometry. For galaxy pairs that are resolved in the optical photometry catalog, this mass refers to the optical counterpart whose coordinates are given. For those that are not resolved, the mass is the combined mass of the system. The combined mass for the resolved pairs can be obtained using the mass ratio given.}
\tablenotetext{g}{Rest frame absolute K-band magnitude}
\tablenotetext{h}{Rest-frame $U-V$ color}
\tablenotetext{i}{Concentration parameter}
\tablenotetext{j}{Rotational asymmetry parameter}
\tablenotetext{k}{Automated classification from Cassata \etal (2010) --- 1: early type, 2: late type, 3: irregular}
 \end{deluxetable*} 
 \clearpage
% \end{landscape}

\begin{deluxetable*}{lcccccccc}
  \tablewidth{0pt}
  \tabletypesize{\scriptsize}
  \tablecaption{Percentage of COSMOS 70\ts$\mu$m Sources in Each Morphological Class \label{morphology}}
  \setlength{\tabcolsep}{0.05in}
\tablehead{\colhead{log$(L_{\rm IR}/L_{\odot})$\tablenotemark{a}} & \colhead{Total\tablenotemark{b}} &  \colhead{med(z)\tablenotemark{c}} & \colhead{Spirals\tablenotemark{d}} & \colhead{E/S0\tablenotemark{d}} & \colhead{Minor Mergers} & \colhead{Major Mergers\tablenotemark{e}} & \colhead{QSOs\tablenotemark{f}} & \colhead{Unknown\tablenotemark{g}}} 
\startdata
% lir         	     Total	  z                  S       		    E     		   m     		   M       		   Q     		  u   
$<10.0$        & 60      & 0.09	& 70$\pm$11 	& 13$\pm$5 	& 10$\pm$4   	&  5$\pm$3	&  0  			& 2$\pm$2   \\
$10.0-10.5$ & 140	& 0.17	& 66$\pm$7 	& 11$\pm$3 	& 18 $\pm$4	& 4$\pm$2	&  0  			& 0.7$\pm$0.7   \\
$10.5-11.0$ & 281	& 0.27	& 70$\pm$5 	&   8$\pm$2 	& 13$\pm$2 	& 8$\pm$2 	&  0	  		& 0.4$\pm$0.4  \\
$11.0-11.5$ & 376	& 0.47	& 53$\pm$4 	&   6$\pm$1	& 20$\pm$2 	& 20$\pm$2 	& 0   			& 0.5$\pm$0.4  \\
$11.5-12.0$ & 310	& 0.77	& 31$\pm$3 	&   5$\pm$1 	& 20$\pm$3 	& 41$\pm$4 	&  0.3$\pm$0.3 & 3$\pm$1    \\
$12.0-12.5$ & 219	&1.05	& 14$\pm$3 	& 11$\pm$2	& 18$\pm$3 	& 42$\pm$4 	&  0.5$\pm$0.5 & 15$\pm$3  \\ 
$12.5-13.0$ & 86	&1.48	& 7$\pm$3   	&   9$\pm$3 	& 12$\pm$4 	& 40$\pm$7 	&  6$\pm$3  	& 27$\pm$6  \\
$>13$           & 31	& 2.00	&  3$\pm$3  	& 16$\pm$7 	&   7$\pm$5 	& 32$\pm$10 	& 23$\pm$9   	& 19$\pm$8  \\
\hline									
All $L_{\rm IR}$    & 1503 & 0.50	& 44$\pm$2 	& 8$\pm$1 & 17$\pm$1 	& 24$\pm$1 	& 1$\pm$0.2	& 5$\pm$0.6  	
\enddata

    \tablenotetext{a}{The total infrared luminosity for each galaxy in the sample was derived from SED template fitting as described in Paper I}
    \tablenotetext{b}{Total number of galaxies in each luminosity bin}
    \tablenotetext{c}{Median redshift for each luminosity bin}
    \tablenotetext{d}{Percentage of spiral and elliptical/S0 galaxies}
    \tablenotetext{e}{Major mergers from 4 interaction classes (II--V) as defined in \cite{Veilleux:2002p920}. Class I objects are left out of this category since they can be placed into one of the other categories first (i.e., into the spiral or elliptical class).}
    \tablenotetext{f}{Percentage of sources with a central point source that overwhelms the light of the host galaxy}
    \tablenotetext{g}{Percentage of galaxies that are too faint, small, or irregular to classify}
\end{deluxetable*}

\begin{deluxetable*}{lccccccccc}
  \tablewidth{0pt}
  \tabletypesize{\small}
  \tablecaption{Percentage of COSMOS 70\ts$\mu$m Sources in Each Major Merger Interaction Class \label{mergers}}
\tablehead{\colhead{log$(L_{\rm IR}/L_{\odot})$\tablenotemark{a}} & \colhead{Total\tablenotemark{b}} & 
 \colhead{II\tablenotemark{c}} & \colhead{IIIa\tablenotemark{d}} &
\colhead{IIIb\tablenotemark{d}} & \colhead{IVa\tablenotemark{e}} & \colhead{IVb\tablenotemark{e}} &
\colhead{V\tablenotemark{f}} & \colhead{$<NS>$\tablenotemark{g}}}

\startdata
% LIR               #               II              	IIIa            		 IIIb           		  IVa  		IVb    		  V  	NS
$<10.0$     &      3   	&  0$\pm$	 	& 0	  		& 67$\pm$47	& 0			& 0   			& 33$\pm$33	& 6.7 \\
$10.0-10.5$ &  6  	&   17$\pm$17	& 0	  		& 24$\pm$24	& 17$\pm$17	& 33$\pm$24   	&  0 			& 8.4 \\
$10.5-11.0$ &  22	&  27$\pm$11	& 5$\pm$5	& 9$\pm$6 	& 14$\pm$8	& 36$\pm$13	&  9$\pm$6	& 9.3 \\
$11.0-11.5$ & 75	&  12$\pm$4	& 23$\pm$5	& 19$\pm$5	& 16$\pm$5 	& 19$\pm$5	&  12$\pm$4	& 13.9  \\
$11.5-12.0$ & 126	&   6$\pm$2	& 16$\pm$5	& 23$\pm$4	& 18$\pm$4 	& 11$\pm$3	&  11$\pm$3	& 11.5  \\
$12.0-12.5$ & 91	&  3$\pm$2	& 18$\pm$4	& 27$\pm$5	& 24$\pm$5 	& 16$\pm$4	&  12$\pm$4	& 9.3 \\ 
$12.5-13.0$ & 34	&  0			& 19$\pm$7	& 24$\pm$8	& 32$\pm$10 	& 6$\pm$ 4	&  21$\pm$8	& 8.8 \\
$>13$           &  10	&    0			&  20$\pm$14	& 50$\pm$22	& 20$\pm$14	& 10$\pm$10	& 0			& 9.0 \\
\hline									
All $L_{\rm IR}$     & 367  	& 7$\pm$1& 22$\pm$2	& 24$\pm$3	& 20$\pm$2	& 15$\pm$2	& 12$\pm$2	& 10.0
\enddata

    \tablenotetext{a}{The total infrared luminosity for each galaxy in the sample was derived from SED template fitting as described in Paper I}
    \tablenotetext{b}{Total number of major mergers in each luminosity bin}
   \tablenotetext{c}{Percentage of major mergers in interaction class II: first contact }
	\tablenotetext{d}{Percentage of major mergers in interaction class III: pre-merger }
	\tablenotetext{e}{Percentage of major mergers in interaction class IV: merger }
	\tablenotetext{f}{Percentage of major mergers in interaction class V: old merger; See text for a complete description of each interaction class. }
	\tablenotetext{g}{Mean nuclear separation in kpc between galaxy pairs and double nuclei (i.e., objects in classes II, IIIa, and IIIb). }
\end{deluxetable*}

\begin{deluxetable*}{lccccccccc}
  \tablewidth{0pt}
  \tabletypesize{\scriptsize}
  \tablecaption{Percentage of COSMOS 70\ts$\mu$m Sources in Each Class as a Function of Redshift and Luminosity \label{morphology}}
  \setlength{\tabcolsep}{0.05in}
\tablehead{\colhead{Redshift} &  \colhead{N\tablenotemark{a}} & \colhead{$9.0-10.0$\tablenotemark{b}} 	 & \colhead{$10.0-10.5$} 	& \colhead{$10.5-11.0$}	& \colhead{$11.0-11.5$} 	& \colhead{$11.5-12.0$} 	& \colhead{$12.0-12.5$} 	& \colhead{$12.5-13.0$} 	& \colhead{$>13$ }} 
\startdata
    \sidehead{\it Major Mergers}
$0.0-0.5$ 		&        751		&	5 		&      4 		&      8  		&    15 		&     23		&      50		&       \nodata	&       \nodata \\
$0.5-1.0$ 		&       477		&	\nodata 	&      \nodata	&       0	&      26		&      43		&      47		&      88		&       \nodata \\
$1.0-1.5$ 		&      	172		&	 \nodata     &	\nodata	&       \nodata	&       \nodata	&      53		&      39		&      38		&       \nodata \\
$1.5-2.0$ 		&      	73		&	 \nodata     &      \nodata	&       \nodata	&       \nodata	&       \nodata	&      34		&      31		&      33 \\
$2.0-3.0$ 		&       29		&	\nodata	&       \nodata	&       \nodata	&       \nodata	&       \nodata	&      25		&      30		&      27 \\
\hline									
    \sidehead{\it Minor Mergers}
$0.0-0.5$ 		&        751		&	9 		&      18 		&      13  		&    17 		&     17		&      50	&       \nodata	&       \nodata \\
$0.5-1.0$ 		&      477		&	 \nodata 	&      \nodata	&       0	&      24		&     21		&     18		&      0		&       \nodata \\
$1.0-1.5$ 		&       172		&	\nodata     &	\nodata	&       \nodata	&       \nodata	&      23		&      21		&      26		&       \nodata \\
$1.5-2.0$ 		&       73		&	\nodata     &      \nodata	&       \nodata	&       \nodata	&       \nodata	&      7		&      0		&      7 \\
$2.0-3.0$ 		&       29		&	\nodata	&       \nodata	&       \nodata	&       \nodata	&       \nodata	&      0		&      0		&      7 \\
\hline									
    \sidehead{\it Spirals}
$0.0-0.5$ 		&        751		&	70 		&      66 		&      70  		&    60 		&     58		&      0	&       \nodata	&       \nodata \\
$0.5-1.0$ 		&      477		&	 \nodata 	&      \nodata	&       100		&    44		&     29		&     24		&      0		&       \nodata \\
$1.0-1.5$ 		&       172		&	\nodata     &	\nodata	&       \nodata	&       \nodata	&      2		&      8		&      10		&       \nodata \\
$1.5-2.0$ 		&       73		&	\nodata     &      \nodata	&       \nodata	&       \nodata	&       \nodata	&      3		&      3		&      7 \\
$2.0-3.0$ 		&       29		&	\nodata	&       \nodata	&       \nodata	&       \nodata	&       \nodata	&      0		&      10		&      0 \\
\hline									
    \sidehead{\it Ellipticals}
$0.0-0.5$ 		&        751		&	14 		&      11 		&      8  		&    7 		&     2		&      0	&       \nodata	&       \nodata \\
$0.5-1.0$ 		&      477		&	 \nodata 	&      \nodata	&       0		&    4			&     6		&     11		&      12		&       \nodata \\
$1.0-1.5$ 		&       172		&	\nodata     &	\nodata	&       \nodata	&       \nodata	&      7		&      12		&      5		&       \nodata \\
$1.5-2.0$ 		&       73		&	\nodata     &      \nodata	&       \nodata	&       \nodata	&       \nodata	&      14		&      14		&      20 \\
$2.0-3.0$ 		&       29		&	\nodata	&       \nodata	&       \nodata	&       \nodata	&       \nodata	&      0		&      10		&      13 \\
\hline									
    \sidehead{\it QSOs}
$0.0-0.5$ 		&        751		&	0 		&      0 		&      0  		&    0 		&     0		&      0	&       \nodata	&       \nodata \\
$0.5-1.0$ 		&      477		&	 \nodata 	&      \nodata	&       0		&    0			&     0		&     0		&      0		&       \nodata \\
$1.0-1.5$ 		&       172		&	\nodata     &	\nodata	&       \nodata	&       \nodata	&      0		&      1		&      10		&       \nodata \\
$1.5-2.0$ 		&       73		&	\nodata     &      \nodata	&       \nodata	&       \nodata	&       \nodata	&      0		&      3		&      20 \\
$2.0-3.0$ 		&       29		&	\nodata	&       \nodata	&       \nodata	&       \nodata	&       \nodata	&      0		&      0		&      27 \\
\hline									
    \sidehead{\it Unknown}
$0.0-0.5$ 		&        751		&	2 		&      1 		&      0  		&    0 		&     0		&      0	&       \nodata	&       \nodata \\
$0.5-1.0$ 		&      477		&	 \nodata 	&      \nodata	&       0		&    1			&     1		&     0		&      0		&       \nodata \\
$1.0-1.5$ 		&       172		&	\nodata     &	\nodata	&       \nodata	&       \nodata	&      14		&      19		&      10		&       \nodata \\
$1.5-2.0$ 		&       73		&	\nodata     &      \nodata	&       \nodata	&       \nodata	&       \nodata	&      41		&      48		&      13 \\
$2.0-3.0$ 		&       29		&	\nodata	&       \nodata	&       \nodata	&       \nodata	&       \nodata	&      75		&      50		&      27 

\enddata
    \tablenotetext{a}{The total number of sources in each redshift bin}
    \tablenotetext{b}{Total infrared luminosity (log$(L_{\rm IR}/L_{\odot})$) bins}
\end{deluxetable*}

\begin{deluxetable*}{lccccccccc}
  \tablewidth{0pt}
  \tabletypesize{\footnotesize}
  \tablecaption{Comparison Between 70 $\mu$m Sources and Optically Selected Comparison Sample \label{comparison}}
  \setlength{\tabcolsep}{0.05in}
  \tablehead{
  \colhead{Morphological} & 
  \multicolumn{2}{c}{$0.4 < z < 0.6$} &
   \multicolumn{2}{c}{$0.8 < z < 1.0$} &
   \multicolumn{2}{c}{$1.0 < z < 2.0$} &
   \multicolumn{2}{c}{$2.0 < z < 3.0$} &\\
 \colhead{Classification} &
 \colhead{70 $\mu$m} & \colhead{Optical} &
 \colhead{70 $\mu$m} & \colhead{Optical} &
 \colhead{70 $\mu$m} & \colhead{Optical} &
 \colhead{70 $\mu$m} & \colhead{Optical}}
 \startdata
 
 Spirals             &  134 (58) 	&  96 (44) 		& 49 (26) 		&  83 (38)		& 15 (6)	& 47 (21)	& 1 (3)	& 23 (10) \\ 
 Ellipticals         &  15 (6) 		&  37 (17)     	& 10 (5) 		&   52 (24)		&  27 (11)	& 6 (3)	& 3 (10)	& 3 (1)\\
 QSOs                & 0  			&    4  (1.8)     	& 0  			&   0      		& 9 (4)	& 0		& 4 (14)	& 0 \\
 Unknown         &  0  			&    4 (1.8)     	& 2 (1) 		&  15 (6.9)       	& 55 (22)	& 100 (46)	& 12 (41)	& 131 (59) \\
 Minor Merger  & 62 (27)  		&  29 (13)   	& 38 (21)  		&  23 (10.6)	& 44 (18)	& 3 (1)	& 1 (3)	& 0 \\
 Major: I             & 14 (6)  		&  18 (8.2)  	& 4 (2) 		&  23 (10.6)	& 3 (1)	& 1 (0.5)	& 0		& 0   \\
 Major: II            & 6 (3)  		&    2  (0.9)  	& 5 (3) 		&   5   (2.3)	& 0		&  0		& 0		& 0   \\
 Major: III           & 18 (8)  		&   13 (5.9) 	& 50 (27) 		&  11 (5.1)		& 45 (18)	& 25 (11)	& 4 (14)	& 26 (12)      \\ 
 Major: IV           & 16 (7)  		&     9 (4.1)  	& 28 (15) 		&   7  (3.2) 	& 40 (16)	& 36 (16)	& 2 (7)	& 36 (16)  \\ 
 Major: V           & 6 (3)  		&     5 (2.3) 	&  9  (5) 		&   1  (0.5)		& 12 (5)	& 2 (1)	& 2 (7)	& 2 (1)  \\
 Group               & 13 (5.8)  	&     6  (2.7) 	&  11 (6) 		&    5   (2.3)	& 19 (7.8)	& 7 (3.2)	& 4 (14)	& 4 (1.8)   \\
 \hline
 Total \#   &    225  &   220   &   184   &   217   & 245 & 219 & 29 & 221 \\
 $<\rm L_{IR}>$ & 11.33 & \nodata &11.88 & \nodata & 12.35 & \nodata & 13.00 & \nodata \\
 $<\rm mass>$ & 11.06 & 11.02    & 11.17 & 10.92 & 11.14  & 10.78 & 11.07 & 10.54 
 
 \enddata
\tablecomments{Values represent the total number in each morphological class with the percentage of the total in each redshift bin given in parentheses. Note that the percentages do not add up to 100\% since a single object can be placed into more than one classification.}
 \end{deluxetable*}

\begin{deluxetable*}{lccccccccccccc}
  \tablewidth{0pt}
  \tabletypesize{\small}
  \tablecaption{Automated Morphological Parameters for 70 $\mu$m Sources \label{auto_table}}
  \setlength{\tabcolsep}{0.05in}
\tablehead{\colhead{$log(\rm L_{IR}/\rm L_{\odot})$} & 
\multicolumn{2}{c}{Concentration} &
\multicolumn{2}{c}{Asymmetry} &
\multicolumn{2}{c}{Clumpiness} &
\multicolumn{2}{c}{Gini} &
\multicolumn{2}{c}{M$_{20}$} & \multicolumn{3}{c}{Class [\%] }\\
\colhead{ } & \colhead{median} & \colhead{$\sigma$} &
\colhead{median} & \colhead{$\sigma$} &
\colhead{median} & \colhead{$\sigma$} &
\colhead{median} & \colhead{$\sigma$} &
\colhead{median} & \colhead{$\sigma$} & 
\colhead {E/S0} & \colhead{S} & \colhead{Irr}}
\startdata
$<10.0$     &  2.83   &   0.53   &   0.11  &    0.16  &  -0.05  &   0.08   &   0.51  &    1.51  &    -1.82  &    0.38 & 29 & 47 & 24 \\
$10.0-10.5$ &   2.91  &    0.57   &   0.12  &    0.09  &  -0.09  &    0.06   &   0.51  &     3.23   &   -1.85  &    0.33 & 25 & 60 & 15 \\
$10.5-11.0$ &  2.97   &   0.50   &   0.12   &  0.10  &  -0.08  &   0.10   &   0.50    &   3.28  &    -1.85  &    0.33  & 23 & 63 & 14\\
$11.0-11.5$ &  2.97    &  0.53   &   0.15   &   0.11 &   -0.07  &    0.09   &   0.50   &   0.70  &    -1.79  &    0.32  & 15 & 62 & 23 \\
$11.5-12.0$ &  2.92   &   0.54   &   0.20   &   0.15  &  -0.03  &    0.19   &   0.49   &  0.07   &   -1.69   &   0.36 & 12 & 46 & 43 \\
$12.0-12.5$ &   2.93   &   0.53  &    0.23  &    0.14  &   0.06  &    0.24   &   0.47   &  0.08   &   -1.63   &   0.37 & 14 & 33 & 53\\ 
$12.5-13.0$ &  2.80    &  0.50   &   0.26    &  0.15  &   0.14   &   0.23   &   0.44    & 0.07   &   -1.58   &   0.29 & 17 & 22 & 62 \\
$>13$           &   2.99   &   0.41  &    0.16   &   0.16  &   0.10   &   0.27  &    0.49  &   0.08  &    -1.55   &   0.25 & 32 & 36 & 32\\
\hline									
All $L_{\rm IR}$   &  2.94  &    0.53  &    0.15  &    0.13  &  -0.05  &    0.18  &    0.50    &   1.82  &    -1.74  &    0.35  & 18 & 51 & 31	
\enddata
\end{deluxetable*}

\end{document}